\documentclass[
		twoside,openright,titlepage,numbers=noenddot,headinclude,
                footinclude=true,cleardoublepage=empty,
                BCOR=5mm,paper=a4,fontsize=11pt, 
                ngerman,american, 
                ]{scrreprt} 
                
\PassOptionsToPackage{eulerchapternumbers,listings, pdfspacing, subfig,beramono,eulermath,parts}{classicthesis}

\newcommand{\myTitle}{Non Common Path Aberrations Correction \xspace}
\newcommand{\mySubtitle}{Application of Electric Field Conjugation on SPHERE and use of Phase Diversity on the AOF.\xspace}
\newcommand{\myDegree}{Master Thesis\xspace}
\newcommand{\myName}{Jean-Baptiste Ruffio\xspace}
\newcommand{\myProf}{S\'ebastien Massenot\xspace}

\newcommand{\mySupervisor}{Markus Kasper\xspace}
\newcommand{\myFaculty}{\xspace}
\newcommand{\myDepartment}{Adaptive Optics Department, ESO\xspace}
\newcommand{\myUni}{\label{myUni}Institut Sup\'erieur de l'a\'eronautique et de l'espace, ISAE-Supa\'ero, Toulouse, France\xspace}
\newcommand{\myOtherUni}{Universit\'e Paul Sabatier, Toulouse III, France\xspace}
\newcommand{\myLocation}{Garching, Germany\xspace}
\newcommand{\myTime}{august 2014\xspace}
\newcommand{\myVersion}{version 1.0\xspace}

\newcounter{dummy} 
\providecommand{\mLyX}{L\kern-.1667em\lower.25em\hbox{Y}\kern-.125emX\@}

\usepackage{lipsum} 

\PassOptionsToPackage{latin9}{inputenc} 
\usepackage{inputenc}
 
\usepackage{babel}

\PassOptionsToPackage{square,numbers}{natbib}
 \usepackage{natbib}
 
\PassOptionsToPackage{fleqn}{amsmath} 
 \usepackage{amsmath}
 
\PassOptionsToPackage{T1}{fontenc} 
\usepackage{fontenc}

\usepackage{xspace} 

\usepackage{mparhack} 

\usepackage{fixltx2e} 

\PassOptionsToPackage{smaller}{acronym} 
\usepackage{acronym} 

\PassOptionsToPackage{pdftex}{graphicx}
\usepackage{graphicx} 

\usepackage{tabularx} 
\newcommand{\tableheadline}[1]{\multicolumn{1}{c}{\spacedlowsmallcaps{#1}}}
\newcommand{\myfloatalign}{\centering} 
\usepackage{caption}
\usepackage{subfig}  

\usepackage{listings} 
\PassOptionsToPackage{pdftex,hyperfootnotes=false,pdfpagelabels}{hyperref}
\usepackage{hyperref}  
\hypersetup{
colorlinks=true, linktocpage=true, pdfstartpage=3, pdfstartview=FitV,
breaklinks=true, pdfpagemode=UseNone, pageanchor=true, pdfpagemode=UseOutlines,
plainpages=false, bookmarksnumbered, bookmarksopen=true, bookmarksopenlevel=1,
hypertexnames=true, pdfhighlight=/O, urlcolor=webbrown, linkcolor=RoyalBlue, citecolor=webgreen,
pdftitle={\myTitle},
pdfauthor={\textcopyright\ \myName, \myUni, \myFaculty},
pdfsubject={},
pdfkeywords={},
pdfcreator={pdfLaTeX},
pdfproducer={LaTeX with hyperref and classicthesis}
}   

\usepackage{ifthen} 
\newboolean{enable-backrefs} 
\setboolean{enable-backrefs}{false} 

\newcommand{\backrefnotcitedstring}{\relax} 
\newcommand{\backrefcitedsinglestring}[1]{(Cited on page~#1.)}
\newcommand{\backrefcitedmultistring}[1]{(Cited on pages~#1.)}
\ifthenelse{\boolean{enable-backrefs}} 
{
\PassOptionsToPackage{hyperpageref}{backref}
\usepackage{backref} 
\renewcommand*{\backref}[1]{}  
\renewcommand*{\backrefalt}[4]{
\ifcase #1 
\backrefnotcitedstring
\or
\backrefcitedsinglestring{#2}
\else
\backrefcitedmultistring{#2}
\fi}
}{\relax} 

\makeatletter
\@ifpackageloaded{babel}
{
\addto\extrasamerican{

}
\addto\extrasngerman{

}
}{\relax}
\makeatother

\usepackage{classicthesis} 

\usepackage{comment} 

 \usepackage{chapterbib}
 \usepackage{placeins}
 \usepackage{scrextend}
 
\usepackage{mathtools}
\DeclarePairedDelimiter\abs{\lvert}{\rvert}%
\DeclarePairedDelimiter\mean{\langle}{\rangle}%
\newcommand{\diag}[1]{ \text{diag}\left( #1 \right) }
\newcommand{\vecTwoD}[2]{ \left[ \begin{array}{c} #1 \\ #2 \end{array} \right] }
\newcommand{\matTwoD}[4]{ \left[ \begin{array}{cc} #1&#2 \\ #3&#4  \end{array} \right] }

\graphicspath{{D:/latex/EFC/}{D:/latex/phaseDiversity/}{D:/latex/optics_others/}}

\newcommand{\relChapter}[1] {\chapter{#1}}
\newcommand{\relSection}[1] {\section{#1}}
\newcommand{\relSubsection}[1] {\subsection{#1}}
\newcommand{\relSubsubsection}[1] {\subsubsection{#1}}

\begin{document}

\frenchspacing 

\raggedbottom 

\selectlanguage{american} 


\pagenumbering{roman} 

\pagestyle{plain} 



\begin{titlepage}

\begin{addmargin}[-1cm]{-3cm}
\begin{center}
\large

\hfill
\vfill

\begingroup
\color{Maroon}\spacedallcaps{\myTitle} \\ \bigskip 
\endgroup

\spacedlowsmallcaps{\myName}\footnote{\myUni}\footnote{\myOtherUni} 

\vfill

\includegraphics[width=\linewidth]{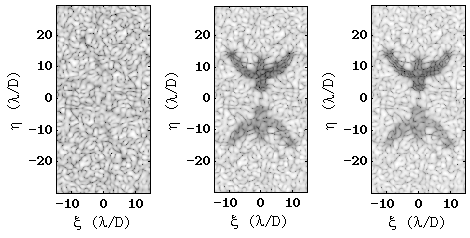} \\ \medskip 

\vfill

\mySubtitle \\ 
\vfill
\spacedlowsmallcaps{\myDegree} \\
\myDepartment, \myLocation \\ \medskip

Scientific Supervisor: \mySupervisor\footnote{\myDepartment, \myLocation} \\
University Supervisor: \myProf\footref{myUni} \\ \bigskip

\vfill
\myTime\ -- \myVersion 

\includegraphics[width=0.75\linewidth]{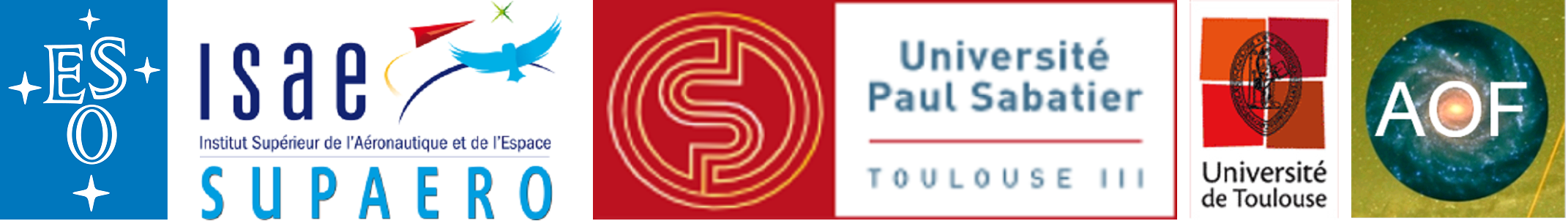}

\end{center}
\end{addmargin}

\end{titlepage} 


\thispagestyle{empty}

\hfill

\vfill

\noindent\myName: \textit{\myTitle,} \mySubtitle, 
\textcopyright\ \myTime










\cleardoublepage

\pdfbookmark[1]{Abstract}{Abstract} 

\begingroup
\let\clearpage\relax
\let\cleardoublepage\relax
\let\cleardoublepage\relax

\chapter*{Abstract} 

The primary goal of this thesis was the correction of Non-Common-Path-Aberrations in the SPHERE instrument for helping it meeting its contrast requirements. SPHERE's purpose is the search and characterization of giant exo-planets around nearby stars. The author implemented a method called Electric Field Conjugation that he tested in simulation as well as on the real system. A full week was booked in SPHERE schedule a few days before the second commissioning in June 2014. It gave the opportunity to the author to travel to the VLT in Chile and experiment directly on the system. The contrast gain objective of another order of magnitude in a medium-sized area has successfully been reached bringing SPHERE raw speckle contrast from about $10^{-6}$ to $10^{-7}$. The algorithm has therefore proven its value and will be further investigated and hopefully automated by the SPHERE team based on the codes developed by the author. However it is important to keep in mind that Electric Field Conjugation is more effective for follow-up studies in order to improve the quality of the observations. Indeed the area for a good correction is very limited. It can't be used for exo-planets discovery unless the corrected area is made big enough but the performance will be less.

Electric Field Conjugation is only one of the many algorithms for correcting speckles and that is why the author got interested in another project. ESO's Adaptive Optics Facility is currently being tested on the ASSIST test bench at ESO headquarters in Garching, Germany. This bench suffers from a bad quality Point Spread Function and the goal of the author was to understand why an existing method called Phase Diversity couldn't improve the Strehl Ratio. Therefore he implemented a new and more flexible code and performed a series of tests. The ASSIST test bench was indeed booked during a week in Mai 2014. After validating the code and exploring different parameters the AOF team got convinced that the problem was not coming from low order phase aberration that could be corrected using the Deformable Mirror and Phase Diversity. In average the performance of the author's algorithm on ASSIST was an improvement of the Strehl Ratio of about $10\%$ from $60$ to $70\%$ with a peak at almost $80\%$ at a wavelength of $1.6\mu m$.

\endgroup			

\vfill 


\cleardoublepage

\pdfbookmark[1]{Acknowledgements}{Acknowledgements} 

\begin{flushright}{\slshape    
In this world there is room for everyone. And the good earth is rich and can provide for everyone. The way of life can be free and beautiful, but we have lost the way. Greed has poisoned men's souls, has barricaded the world with hate, has goose-stepped us into misery and bloodshed. We have developed speed, but we have shut ourselves in. Machinery that gives abundance has left us in want. Our knowledge has made us cynical. Our cleverness, hard and unkind. We think too much and feel too little. More than machinery we need humanity. More than cleverness we need kindness and gentleness. Without these qualities, life will be violent and all will be lost.} \\ \medskip
\href{http://www.youtube.com/watch?v=45BotfjfGpE}{Hinkel The Great Dictator - Final Speech of the Barber},\marginpar{\texttt{Youtube} link.} \\ ---  Charlie Chaplin.
\end{flushright}

\bigskip


\begingroup

\let\clearpage\relax
\let\cleardoublepage\relax
\let\cleardoublepage\relax

\chapter*{Acknowledgements} 

First I would like to thank my supervisor Markus Kasper for this opportunity to work in a great environment such as ESO. Few are the interns having the opportunity to travel to Chile and visit le Very Large Telescope. By the way thank you to Norbert Hubin who agreed to finance this trip. Despite the distance Markus made himself available for answering any of my questions and who knows how annoying I am. \\
Then big thanks to Johann Kolb who accepted my contribution to his project when I was getting low in tasks to do. \\
Thank you to Christophe V\'{e}rinaud for welcoming me in Grenoble and helping me understand Electric Field Conjugation. \\
Thank you to Bruce Macintosh for recommending me to Markus Kasper and therefore making this internship possible. \\
And I have to thank Anthony Berdeu for the skype-chat preventing me from working. Mh\dots Actually I might be the one preventing him from working!

Finally thank you to Andr\'e Miede for developing and making available the template \texttt{classicthesis} used for this document. A real postcard has been sent to him as he wished.

\endgroup 

\pagestyle{scrheadings} 

\cleardoublepage

\refstepcounter{dummy}

\pdfbookmark[1]{\contentsname}{tableofcontents} 

\setcounter{tocdepth}{2} 

\setcounter{secnumdepth}{3} 

\manualmark
\markboth{\spacedlowsmallcaps{\contentsname}}{\spacedlowsmallcaps{\contentsname}}
\tableofcontents 
\automark[section]{chapter}
\renewcommand{\chaptermark}[1]{\markboth{\spacedlowsmallcaps{#1}}{\spacedlowsmallcaps{#1}}}
\renewcommand{\sectionmark}[1]{\markright{\thesection\enspace\spacedlowsmallcaps{#1}}}

\clearpage

\begingroup 
\let\clearpage\relax
\let\cleardoublepage\relax
\let\cleardoublepage\relax


\refstepcounter{dummy}
\pdfbookmark[1]{\listfigurename}{lof} 

\listoffigures

\vspace*{8ex}
\newpage


\refstepcounter{dummy}
\pdfbookmark[1]{\listtablename}{lot} 

\listoftables
        
\vspace*{8ex}
\newpage

\refstepcounter{dummy}
\pdfbookmark[1]{Acronyms}{acronyms} 

\markboth{\spacedlowsmallcaps{Acronyms}}{\spacedlowsmallcaps{Acronyms}}

\chapter*{Acronyms}

\begin{acronym}[UML]
\acro{AOF}{Adaptive Optics Facility}
\acro{ASSIST}{Adaptive Secondary Setup and Instrument Simulator}
\acro{COFFEE}{COronagraphic PHase diversitY}
\acro{DOTF}{Differential Optical Transfer Function}
\acro{EFC}{Electric Field Conjugation}
\acro{EPICS}{Exoplanet Imaging Camera and Spectrograph}
\acro{FFREE}{Fresnel-FRee Experiment for EPICS}
\acro{ESO}{European Southern Observatory}
\acro{GALACSI}{Ground Atmospheric Layer Adaptive Optics for Spectroscopic Imaging}
\acro{GRAAL}{GRound-layer Adaptive optics Assisted by Laser}
\acro{HOT}{High Order Testbench}
\acro{IM}{Interaction Matrix}
\acro{IPAG}{Institut de Plan\'{e}tologie et d'Astrophysique de Grenoble}
\acro{IRDIS}{Infra-Ref Dual-beam Imager and Spectrograph}
\acro{ONERA}{Office National d'\'Etudes et de Recherches A\'erospatiales}
\acro{PD}{Phase Diversity}
\acro{PSF}{Point Spread Function} 
\acro{PSI}{Phase Sorting Technique}
\acro{RMS}{Root-Mean-Square}
\acro{SPHERE}{Spectro-Polarimetric High-contrast Exoplanet REsearch}
\acro{UT3}{Unit Telescope 3}
\acro{VLT}{Very Large Telescope}
\end{acronym}

\endgroup 

\cleardoublepage

\pagenumbering{arabic} 

\cleardoublepage 


\chapter{Introduction} 
The Spectro-Polarimetric High-contrast Exoplanet REsearch (SPHERE) instrument has been installed at the Very Large Telescope (VLT) in spring 2014. SPHERE includes a very powerful Adaptive Optics for correcting the wavefront distortion due to the atmospheric turbulence. However even the best Adaptive Optics systems still suffer apparition of noise called speckles originating from defects inside the optical train and after the Adaptive Optics system. Methods have therefore been developed to measure the wavefront distortion using the science camera. A shape of the deformable mirror is inferred to suppress the speckles. One of this method is called Electric Field Conjugation (EFC). The goal of the Master Thesis is to implement it for SPHERE using Matlab. In addition of this main goal another phase aberration correction techniques was implement for the Adaptive Optics Facility which is a project of secondary Deformable Mirror for the VLT. This method is called Phase diversity and aims at improving the quality of the Point Spread Function by removing low order aberrations. This document first emphasizes the general context and the aims of the thesis in an introductory chapter. Then it is followed by one part for each the Phase Diversity and the Electric Field Conjugation method. These two parts are built in the same manner: first basic elements of theory are given, then simulated respectively experimental results are exposed. Besides there is one bibliographic section called State of the Art and a list of references per part. To finish almost every chapter has a paragraph emphasizing the author's contributions.

The demonstrations of the methods even though very interesting have been pushed to the appendix. A week of tests for each instrument SPHERE and the AOF has been performed. The very essential of these tests has been exposed in the following. The complete set of tests for SPHERE is presented in the appendix while the complete tests for the AOF are available in a separate technical report.

\paragraph{Warning} The author acknowledge that the $30$ pages requirement of \myOtherUni is not formally fulfilled as the table of contents states $47$. It is explained by the layout of this document which is very light with big margins and a lot of white pages for clarity. However it doesn't contain more than $12$ thousand words which gives $26$ pages using a standard $500$ words pages with $11\text{pt}$ according to \url{http://www.wordstopages.com/}.

\relSection{Context}

Extrasolar planets or exo-planets are very faint objects orbiting very close to their host star. They are therefore very difficult to observe but thanks to technological and data processing techniques improvements it became one of the most dynamic research field in Astronomy. Nowadays the most common way to detect exo-planets is to use indirect methods. They are indirect because they consist in observing the different perturbations of the host star due to the presence of a close-orbiting body\marginpar{For instance transits or radial velocity for the common ones}. However for the past ten years a couple of exo-planets around nearby stars have been directly imaged opening a new era for detecting and characterizing exo-planets. The development of eXtreme \marginpar{Extreme means very efficient.} Adaptive Optics systems have triggered the development of a new generation of instrument dedicated to exo-planets imaging. Direct imaging is briefly described in the following. The other methods short descriptions as well as their advantages and limitations can be found in \autoref{app:ExoDetecMeth}.

 Direct imaging allows one to get the light of the planet itself. Its great advantage is the possibility of characterizing the planet by studying the spectrum of its atmosphere for instance. Indirect methods will only provide with basic parameters like mass, size and orbital parameters.\marginpar{Sometimes also a very low resolution spectrum\dots} Besides direct imaging is not constraint by the limitations inherent to indirect methods which need significant induced perturbations on the host star. The direct detection is limited by the separation power or resolution of the telescope and the brightness of the planet which determines a required contrast to achieve. The contrast is the ratio between the intensity of the star and the intensity of the planet. It is easier to detect bright planets far from the star. The ultimate goal of high contrast imaging is to observe an earth-like planet with evidence of life but it is not even sure that the next generation of 40 meters telescopes will suffice. Direct imaging instruments are usually coronagraphs with eXtreme Adaptive Optics. Coronagraphs are described in \autoref{app:APLCdef} and Adaptive Optics system in \autoref{app:AO}. Two such instruments have recently been built both on a eight meter class telescopes : SPHERE on the Very Large Telescope (VLT) and GPI on GEMINI-South.\marginpar{SPHERE is European and GPI is American-Canadian}
 
 \paragraph{SPHERE}
 SPHERE stands for Spectro-Polarimetric High-contrast Exo-planet Research described in \citeauthor{beuzitSPHERE2008} \citep{beuzitSPHERE2008}. It is being commissioned at the VLT during this summer 2014. It includes three science paths: a differential infrared imager and spectrograph called IRDIS, an integral field spectrograph and a visible light polarimeter.
 
\begin{figure}[tb]
  \begin{center}
    \includegraphics[width=0.4\linewidth]{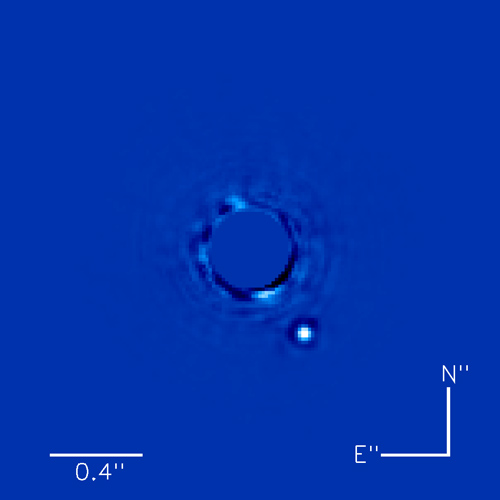}
    \caption[GPI First Light.]{Gemini Planet Imager's first light image of Beta Pictoris b, a planet orbiting the star Beta Pictoris. Source: \url{www.gemini.edu}.}
    \label{fig:GPI_first_light}
  \end{center}
\end{figure}

 \paragraph{Gemini Planet Imager (GPI)}
 GPI described in \citeauthor{macintoshGPI2006} \citep{macintoshGPI2006} was a few months ahead and it has already begun its science phase at the Gemini south telescope. It is very similar to SPHERE but it possesses a single scientific instrument: a infrared Integral Field Spectrometer with a polarimetric mode. Figure \ref{fig:GPI_first_light} shows the first light image of an already known planet Beta Pictoris b.
 
\relSection{Master Thesis}
\relSubsection{Motivations and Objectives}
The purpose of the master thesis was to correct Non-Common-Path-Aberrations of optical instruments. The Non-Common-Path-Aberrations\marginpar{For a description of the non-common-path see \autoref{app:NCPA}.} are phase aberrations due to the optics defects but not seen by the wavefront sensor of the Adaptive Optics system.
More particularly it was meant to be done on SPHERE. The method chosen for correcting these aberrations is Electric Field Conjugation and it will be described later on. However the author quickly realized that this would not keep him occupied for five months. At the mean time the Adaptive Optics Facility (AOF) was being tested on a bench called ASSIST in the laboratory of ESO in Garching. This bench suffered from a bad quality and unstable Point Spread Function measured by a relatively low Strehl Ratio\marginpar{Low means $50-80\%$. For more on Strehl Ratio see \autoref{sec:strehlRatio}.}. A solution to improve Point Spread Functions is to apply a method called Phase Diversity. It is an algorithm for correcting Non-Common-Path-Aberrations. The algorithm used by ESO was developed by ONERA\footnote{Office National d'\'Etudes et de Recherches A\'erospatiales (ONERA)} using IDL\footnote{Interactive Data Language (IDL)} and it didn't give the expected results. The author took that opportunity to work on this other algorithm. The goal of the author was to help better understanding the ASSIST bench and try to improve the image quality or at least identify the causes. It went through coding a new algorithm on Matlab.

However the main goal of the thesis was still to work on SPHERE. In the framework of high-contrast imaging what determines the accuracy of an instrument is the contrast it is able to achieve. The contrast is indeed limited by the existence of random speckles in the image and it is not possible to detect planets fainter that the speckles themselves. The different causes for the speckles are first the atmospheric turbulence and then the optical aberrations. The Adaptive Optics deals with the main part of the atmospheric turbulence. The residual atmospheric speckles are short lived and randomly distributed so that a long exposure time makes them negligible. Indeed they average themselves with time forming a uniform background tending to zero relatively to the planet which doesn't move and therefore strengthen its flux. At the end only what are called the quasi-static speckles are left. This speckles are constant on long time scales\marginpar{Here long time scale means the exposure time which is the most critical. Longer time scales can be otherwise calibrated and shorted time scales are averaged like with the atmospheric speckles.} so that long exposure makes them brighter in the same way as the planet. These are the speckles caused by optical aberrations in the instrument. The goal of Electric Field Conjugation is to correct for these aberrations and therefore improving the contrast. However the goal of the internship was less to reach a given contrast than to give preliminary results for validating the principles and therefore justifying the need of further investigations.

\relSubsection{Methodology}
The methodology was the same in both cases for the Phase Diversity as well as for the Electric Field Conjugation algorithm. The first step has been to get a deep understanding of the theory behind each method. In order to do so the author read the associated reference papers and made the demonstrations all over again\marginpar{The demonstration was even improved for EFC.}. The demonstration can be found in Appendix. Then simulation codes on Matlab were implemented to verify the overall principles and to perform some sensitivity studies. Both algorithms are based on a single Fourier Optics library of functions developed by the author for this occasion. Afterwards experiments were done on real systems which are respectively the AOF and SPHERE. To finish all the results were recorded in the present report and other documentations.

\relSubsection{Means}
The AOF is a future upgrade of the VLT UT4 which aims to create a adaptive telescope by replacing the secondary mirror by a deformable one. The AOF also includes several instruments optimized for this configuration. The AOF is currently being tested for two years 2014-2015 on a bench called ASSIST\footnote{Adaptive Secondary Setup and Instrument Simulator (ASSIST)} in Garching, Germany. ASSIST includes the Secondary Deformable Mirror and one of the two Adaptive Optics modules GRAAL\footnote{GRound-layer Adaptive Optics Assisted by Laser (GRAAL)} or GALACSI\footnote{Ground Atmospheric Layer Adaptive Optics for Spectroscopic Imaging (GALACSI)}. The author was present when the first Adaptive Optics module GRAAL was tested. A full week from June $2^{th}$ to $7^{th}$ was dedicated to Phase Diversity tests with the author's code.

Electric Field Conjugation was applied on the instrument IRDIS of SPHERE at UT3 of the VLT in Paranal, Chile. IRDIS means Infra-Red Dual-beam Imager and Spectrograph. IRDIS is one of the three instruments of SPHERE. The author had the opportunity to spend a week at the summit of Paranal between the $24^{th}$ and the $29^{th}$ of june 2014 for testing the method. This took place a week before the second commissioning of SPHERE. \marginpar{The overall stay in Chile was from June, 23 and July, 2.} Besides the author spent a week from the April $19^{th}$ to $26^{th}$ at the Institut de Plan\'{e}tologie et d'Astrophysique de Grenoble (IPAG), Grenoble, France for preparing the experiment on a test bench called FFREE. FFREE stands for Fresnel-FREE Experiment for EPICS.
However a week was to short for the author to really experiment on the bench. This visit was still a good opportunity to talk about the method with Christophe V\'erinaud who actually already applied the method on FFREE.

\relSection{State of the Art}
There are a lot of different methods with different names for correcting non-common-path phase aberrations in regular or coronagraphic environments: EFC, COFFEE, PSI, DOTF\dots. However they are based on the same principle which is to create diversity in the phase for estimating the Electric Field from intensity measurements. The diversity indeed works as a sort of interferometer. Using the science camera to do the measurements allows one to correct for the Non-Common-Path-Aberrations as the same optical path is used for the science and for the correction. A list of references is given in \autoref{app:aberrCorrRefs}\marginpar{These references is in Appendix because the author read only the abstracts and don't feel confortable about them.}. Besides another bibliographical section is available for each part one Phase Diversity and one for Electric Field Conjugation.

The methods above should be applied before the measurements are made. However two post-processing methods are worth mentioning here because they are very efficient and widely used. Both methods are usually used in combination.

\begin{figure}[tb]
  \begin{center}
    \includegraphics[width=\linewidth]{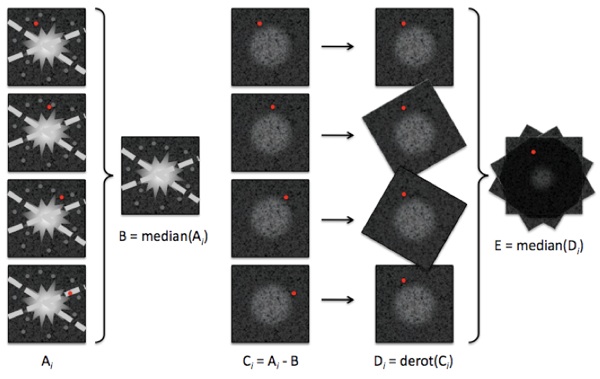}
    \caption[Principle of Angular Differential Imaging.]{Description of the working principle of Angular Differential Imaging. Source: C. Thalmann \url{http://www.cinga.ch/Academic/ADI.html}.}
    \label{fig:ADI}
  \end{center}
\end{figure}

The first one is called Angular Differential Imaging (ADI) \citeauthor{Marois2006} \citep{Marois2006}. The principle of Angular Differential Imaging is to build a model of the speckles using a few sky-rotated images as shown in \autoref{fig:ADI}. In practice the images are taken after disabling the sky rotation compensation of the telescope. The planet will then rotate around the star as the sky is rotating during the night while the speckles will remain constant in the image. The speckles model is then subtracted from the images and finally the rotated images are combined.

The second method called Spectral Differential Imaging also builds a model of the speckles. However this time it uses their wavelength dependency.\marginpar{But the position of the planet is not wavelength dependent.} The speckles indeed follow the behaviour of the Point Spread Function which widen as the wavelength gets lower. The speckles can then be retrieved from radially scaled images taken at different wavelength. If the spectral range of the images is to limited for excluding the planet from the model it can also make use of spectral features in the planet not present in the star. For example the planets usually include methane in their atmosphere making them very faint compared to the star in the methane absorption band. Therefore one can subtract a scaled image at such wavelengths from another image where there isn't methane dimming.

\cleardoublepage

\label{app:bibliography} 


\addcontentsline{toc}{chapter}{\tocEntry{\bibname}}

\bibliographystyle{plainnat}

\bibliography{D:/latex/ESO_PFE/masterThesis/bibliography/bib_intro} 

\ctparttext{Phase Diversity is a method to improve the quality of a Point Spread Function or its Strehl Ratio by correcting the low order quasi-static phase aberrations. The first chapter develops the theory of the method including the principles for applying the algorithm. In a second chapter are shown some simulated and experimental results. The experiments were performed on the ASSIST test bench for the AOF in ESO Headquarters, Garching, Germany. The third and  last chapter is a discussion about the results in the context of the AOF.} 

\part{Phase Diversity} 

\relChapter{Theory} 

This part will develop the theory behind Phase Diversity. It is mainly based on the article by \citeauthor{blancJOSAA2003} \citep{blancJOSAA2003}. The purpose of this part is for the author to have a reference of his current understanding of the method but there is otherwise nothing new to science here.

\paragraph{Authors's Contributions} The author didn't make any contribution to Phase Diversity theory. However he made sure to demonstrate by himself almost all the results.\\

Phase Diversity is the name of a method for estimating phase static-aberration in non coronographic optical systems from the science detector images. It is simply a Maximum \textit{a posteriori} estimator where the estimated parameters are the modes amplitudes and the extended source shape also called object. The modes \marginpar{Modes are usually Zernike polynomials\dots} are the base on which the pupil phase is expanded. The possibility to estimate at the same time the phase perturbations and the extent of the light source is the interesting particularity of Phase Diversity. The principle of any estimator is to estimate parameters from measurements using a model to link them both. For Phase Diversity the measurements are focused and defocused images and the model for reconstructing the data from the parameters is simple Fourier optics. The defocus is the diversity which removes the indetermination caused by the fact that we are measuring intensities and not directly complex amplitudes\marginpar{It is still not clear to the author if there is really an indetermination in the case of estimating only a few modes but whatever. The more data the better!}. The difference between Maximum \textit{A Posteriori} and Maximum Likelihood is the use of \textit{a priori} knowledge on the estimated parameters. It is also called regularization in the frame of ill-posed inverse problems but it is the same. \\

Once the aberrations are estimated it can be corrected directly in the optical path thanks to a Deformable Mirror. Another way to use Phase Diversity is to restore an image by use of deconvolution. Indeed the estimated object would be the original image before the convolution with a bad Point Spread Function. \\

Two Maximum \textit{A Posteriori} estimators are developed in \citep{blancJOSAA2003}. The first one is a joint estimator where the extended source and the aberrations are estimated together while the second estimator integrates the object out of the likelihood. The second estimator has better mathematical properties however only the first one has been implemented successfully in Matlab by the author. Therefore only the joint estimator is included in the main body of this document and the second estimator can be found in \autoref{app:MAP}.

This section is organized as follow: First a short bibliography, then the definition of the problem, afterwards the description of the Joint Maximum \textit{A Posteriori} estimator and to finish a description of how to apply Phase Diversity to a real system.

\relSection{State of the Art}
\marginpar{There are many more references but the author didn't take the time to read them so they are not cited here.}Phase Diversity was first proposed by \citeauthor{gonsalves1982} \citep{gonsalves1982}.
 \citeauthor{blancJOSAA2003} \citep{blancJOSAA2003} developped a new estimator where the object is integrated out of the problem drastically reducing the number of unknowns. This document is mainly based on this paper. However the idea was already presented at a SPIE conference with \citeauthor{blanc2000} \citep{blanc2000}. Phase Diversity was extensively developed and studied by \citeauthor{blancPhD2002} during her PhD \citep{blancPhD2002}. \citeauthor{JohannPhD2005} also applied Phase Diversity for his PhD \citep{JohannPhD2005}. His contribution was to measure the Projection Matrix instead of simulating it. \marginpar{See \autoref{sec:projMat} for details on the Projection Matrix}

\relSection{Problem Definition}
\label{sec:JMAPprobDef}
The detailed definition of the model is given in \autoref{app:Model}. The estimation problem is expressed as follow:

\begin{description}

\item[The Unknowns] The $M$ Zernike coefficients of the phase,
\begin{equation*}
A=(a_{4},\dots,a_{M+3})^\mathsf{T},
\end{equation*}
and the object with $N^2$ pixels $o_{ij}$. The object is the shape of the extended light source.
\begin{equation*}
O=(o_{11},o_{12},\dots,o_{21},\dots,o_{NN})^\mathsf{T}.
\end{equation*}
The aberration phase is expanded on Zernike polynomials $\mathcal{Z}_{k}$ as,\marginpar{See \autoref{app:EFCzerPol}}
\begin{equation}
\phi(x,y) = \sum_{k=4}^{M+3}a_{k}\mathcal{Z}_{k}(x,y).
\end{equation}

\item[The (noisy) Measurements] The focused and defocused images with $N^2$ pixels $i^{f,d}_{ij}$ each.
\begin{align*}
I^{f} &= (i^{f}_{11},i^{f}_{12},\dots,i^{f}_{21},\dots,i^{f}_{NN})^\mathsf{T}, \\
I^{d} &= (i^{d}_{11},i^{d}_{12},\dots,i^{d}_{21},\dots,i^{d}_{NN})^\mathsf{T}. \\
\end{align*}

\item[The Model] with $N^{f}$ and $N^{d}$ the random Gaussian vectors with mean and standard deviation equal to $(0,\sigma^2)$.
\begin{equation*}
  \begin{dcases*}
        I^{f} = H^{f}(A)\times O + N^{f} & focused image,\\
        I^{d} = H^{d}(A)\times O + N^{d}& defocused image.
  \end{dcases*}
\end{equation*}
$H^{f}$ and $H^{d}$ are matrices allowing the convolution of the simulated Point Spread Function with the object $O$. The simulated Point Spread Function depends on the aberrations $A$ and so does $H^{f}$ and $H^{d}$. The expression of the convolution matrices is given in \autoref{app:convoMat}.

\end{description}

\relSection{Joint Maximum \textit{A Posteriori}}
\label{sec:JMAP}

\relSubsection{Bayesian Approach}
In this case the Maximum \textit{A Posteriori} estimator corresponds to the following minimization problem,
\begin{equation}
\left(\tilde{A},\tilde{O}\right) = \underset{A,O}{\text{argmin}}\underbrace{-\text{Ln}\left(f\left(A,O,I^{f},I^{d}\right)\right)}_{L_{JMAP}}.
\end{equation}

The complete expression of $J_{JMAP}$ is given by,\marginpar{The expression for the Maximum Likelihood is simply the third line if ignoring the constants}
\begin{align}
L_{JMAP} &= \frac{N^2}{2}\text{Ln}2\pi + \frac{N^2}{2}\text{Ln}2\pi + \frac{M}{2}\text{Ln}2\pi + \frac{N}{2}\text{Ln}2\pi & \text{Big constant} \nonumber \\
 &+ N^2\text{Ln}\sigma^2 + \frac{1}{2}\text{Ln}\abs*{R_a} + \frac{1}{2}\text{Ln}\abs*{R_o} & \text{Uncertainties} \nonumber \\
 &+ \frac{1}{2\sigma^{2}}(I^{f}-H^{f}O)^\mathsf{T}(I^{f}-H^{f}O) & \dots \nonumber \\
 &+  \frac{1}{2\sigma^{2}}(I^{d}-H^{d}O)^\mathsf{T}(I^{d}-H^{d}O) & \dots\text{Model} \nonumber \\
 &+ \frac{1}{2}A^\mathsf{T}R_{a}^{-1}A + \frac{1}{2}(O-O_m)^\mathsf{T}R_{o}^{-1}(O-O_m)) & \text{\textit{A priori}}
 \label{eq:LJMAP}
\end{align}
$R_a$ and $R_o$ are the covariance matrices of the a priori knowledge of the aberration vector respectively the object.
The full demonstration is available in \autoref{app:JMAPBaysApproach}. Only the last two lines depend on the unknowns.

\relSubsection{Object Estimation}
The minimization of $L_{JMAP}$ \eqref{eq:LJMAP} for the variable $O$ only is a least square problem. Therefore it is possible to find a closed-form expression of $O$ depending on the aberration vector $A$ and the other parameters. This closed-form expression is given by,
\begin{equation}
\tilde{O} = \left(H^{f\mathsf{T}}H^{f} + H^{d\mathsf{T}}H^{d} + \sigma^2 R_o^{-1} \right)^{-1} \left(  H^{f\mathsf{T}}I^{f} + H^{d\mathsf{T}}I^{d} + \sigma^2 R_o^{-1}O_m \right).
\end{equation}
The demonstration is given in \autoref{app:JMAPobjEsti}.

\relSubsection{Fourier Space}
The problem can then be expressed in Fourier space according to the demonstration in \autoref{app:JMAPfour}. The criterion becomes,
\begin{align}
L_{JMAP}(O,A) &= \text{cst}+ N^2\text{Ln}\sigma^2 + \frac{1}{2}\text{Ln}\abs*{R_a} + \frac{1}{2}\text{Ln}\abs*{R_o} \nonumber \\
&+ \sum_{k,l=1}^{N} \frac{1}{2\sigma^{2}}  \abs*{\widehat{i^f}_{kl}-\widehat{h^f}_{kl}\widehat{o}_{kl}}^2  \nonumber \\
&+ \sum_{k,l=1}^{N} \frac{1}{2\sigma^{2}}  \abs*{\widehat{i^d}_{kl}-\widehat{h^d}_{kl}\widehat{o}_{kl}}^2  \nonumber \\
&+ \sum_{k,l=1}^{N} \frac{1}{2\sigma^{2}}  \frac{\abs*{\widehat{o}_{kl}-\widehat{o_m}_{kl}}^2}{s_{o,kl}}  \nonumber \\
&+ \frac{1}{2}A^\mathsf{T}R_{a}^{-1}A,
\end{align}
With $\widehat{x}$ representing the Fourier Transform of $x$ and $s_{o,kl}$ are the eigen-values of $R_o$.
The estimation of the object is given by,
\begin{equation}
\widehat{\tilde{o}}_{kl} = \frac{\widehat{h^f}_{kl}^{\ast}\widehat{i^f}_{kl} + \widehat{h^d}_{kl}^{\ast}\widehat{i^d}_{kl} + \sigma^2\frac{\widehat{o_m}_{kl}}{s_{o,kl}}}{\abs*{\widehat{h^f}_{kl}}^2+\abs*{\widehat{h^d}_{kl}}^2+\frac{\sigma^2}{s_{o,kl}}}.
\end{equation}

\relSection{Application}

\begin{figure}[h]
	\centering	
   	\includegraphics[width=0.8\linewidth]{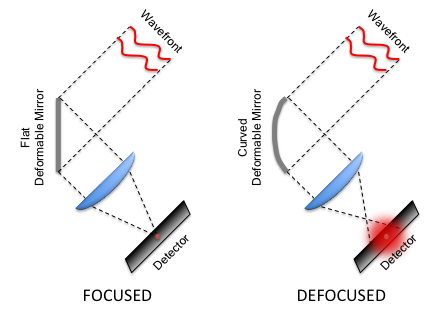}
   	\caption[Optical Configuration for Phase Diversity.]{\label{fig:defocConfig} Optical Configuration for acquiring focus and defocus images for Phase Diversity with a Deformable Mirror.}
\end{figure}

\relSubsection{Measurements}
\label{sec:measurements}
Phase Diversity needs a pair of focused and defocused images acquired with the science camera for the estimation of the phase aberration. There are different ways to introduce the defocus in the system for instance moving directly the detector or the source, introducing a beam splitter or using the Deformable Mirror. The principle is to shape it as a converging or diverging mirror creating the defocus as shown in \autoref{fig:defocConfig}. The defocus function is the fourth Zernike polynomial when using a single indexation. In the case of polynomials normalized by their Root-Mean-Square value and according to simulations the best defocus amplitude depending on the wavelength is given by $\delta_{rms,nm} = 0.3 \lambda_{nm}$. The other options for applying a defocus are describing in \autoref{app:PDappMeas}.

\relSubsection{Projection Matrix}
\label{sec:projMat}
The estimation with Phase Diversity is based on a model of the optics with Fourier Transforms. However the orientation of the modelled pupil might not match the real pupil which could cause the estimation to be irrelevant as one doesn't know how to apply it to the Deformable Mirror. In addition the real system doesn't reproduce exactly theoretical modes. Tuning the simulation so that it solves all these uncertainties would be complicated and really time-consuming. This is the reason why \citeauthor{JohannPhD2005} \citep{JohannPhD2005} decided to measure a matrix transforming estimated coefficients into the coefficients for feeding the system. This matrix would be the inverse of a so called Projection Matrix.
The column vectors of the Projection Matrix are the estimation of the pupil phase when a mode is applied on the Deformable Mirror. In order to fill out the matrix one needs to apply each mode, then takes focus and defocused images with that mode and then run the estimation. The Projection Matrix is built by concatenating all the estimations.
Doing like this would works assuming there is no existing aberrations in the optics. A way to remove the contribution of the aberrations is to apply the mode positively and negatively. On one hand the half subtraction of the two will isolate the sole contribution of the mode. On the other hand the mean isolate the sole aberrations.\marginpar{The phase in the pupil is indeed equal to the sum of the phase of the mode and the aberrations. When the sign of the mode is flipped the sign of the aberrations doesn't change.} An example of Projection Matrix and the corresponding isolated aberrations can be found in \autoref{fig:PDAOFiterPDPM}.

\relSubsection{Iterations}
In theory the application of the method is straightforward. 
\begin{enumerate}
\item Acquisition of a focused and defocused images as described in \autoref{sec:measurements}. The images are then cropped to a small stamp around the center of Point Spread Function and the stamps are normalized by their mean.
\item Estimation of the phase aberration using Phase Diversity. This is done using an iterative optimization algorithm.
\item Inversion of the Projection Matrix and multiplication of the estimated aberrations vector by the inverse for obtaining the correction vector. This vector is a set of mode amplitudes that should be applied on the Deformable Mirror.
\item Application of the correction vector to the Deformable Mirror.
\end{enumerate}
In practice it might need a few iterations to converge toward a nice correction. One just needs to apply several times the steps above.

\relChapter{Results} 
This chapter exposes the simulated or experimental results of Phase diversity applied on the ASSIST bench for testing the AOF during the June $2^{th}$-$7^{th}$ week.

\paragraph{Authors's Contributions} The author reimplemented a Phase Diversity code from scratch on Matlab and used it for the simulations and the experiments reported in this chapter.\marginpar{Implementing a new code allowed the author to vary many parameters.} The code produces consistent results in simulation and experiments and reduces the optical aberrations in ASSIST. \\

The full set of tests is available in the technical report \citeauthor{ruffioAOFPDtest} \citep{ruffioAOFPDtest}.\marginpar{They are not in Appendix like for EFC.} Only the essential tests are given here.

\relSection{Simulations}
The simulations were used to validate the estimation algorithm and check that it was at least able to recover artificially introduced aberrations. Besides it was used to run some sensitivity studies.

The code has been validated on simulated data built with the same model as the one used inside the phase diversity algorithm. First an aberration vector of the first $33$ Zernike modes is randomly created. The amplitude is defined by a normal law with standard deviation equal to $50 \text{nm rms}$\marginpar{The unit of the coefficients is the nm rms because the Zernike polynomials are normalized to a unit RMS value.} and then multiplied by an inverse square function in order to artificially lower down the highest orders. Then two $32\times32$ pixels images with one focused and the other one defocused are built using a Fast Fourier Transform method. The array used for the Fourier Transform is bigger than the image and in this case $128\times128$ pixels. A simple Gaussian noise is then added so that the signal to noise ratio be $500$. To finish images are normalized to a unit mean value.

\begin{figure}[bt]
\centering
\includegraphics[width=\linewidth]{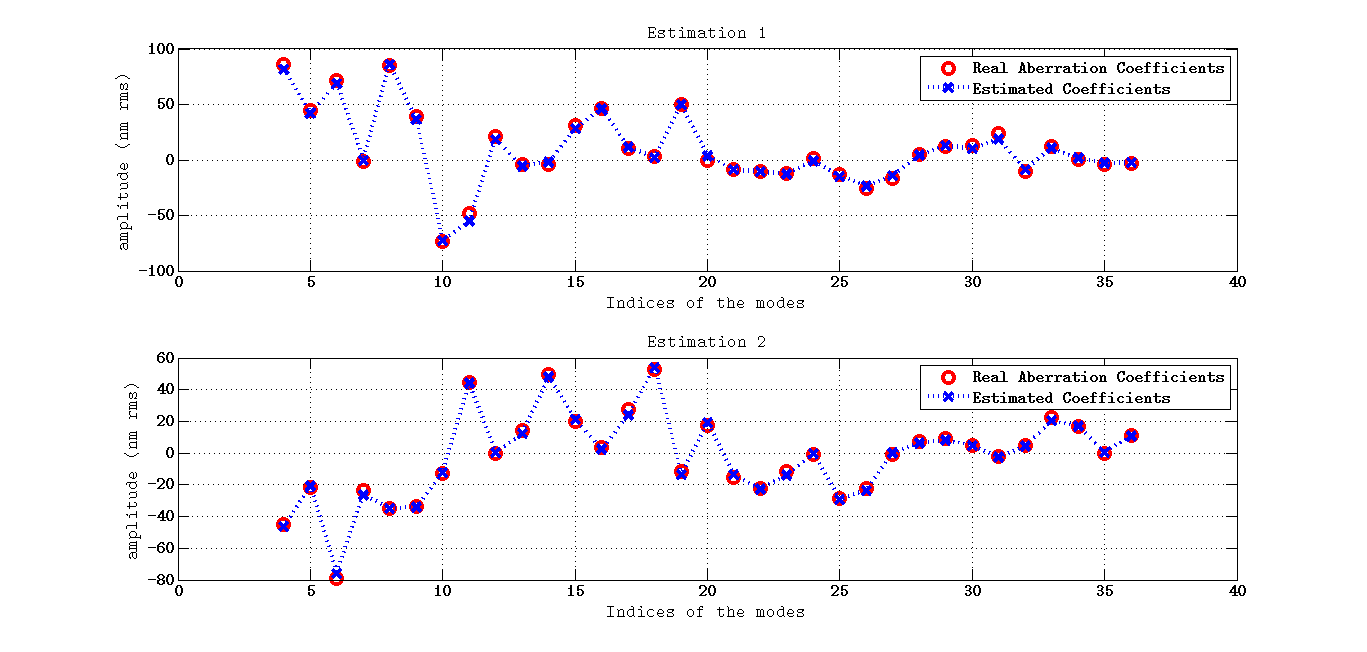}
\caption[Phase Diversity estimation of synthetic random aberrations.]{Estimation in blue of synthetic phase aberrations in red circle using $33$ Zernike coefficients. The first mode indexed at $4$ is the defocus. The Root-Mean-Square error for the first estimation is $2.4\text{nm rms}$ and $1.5\text{nm rms}$ for the second.}
\label{fig:PDcodeValid}
\end{figure}

\relSubsection{Zernike Estimation}
\autoref{fig:PDcodeValid} shows two examples of Phase Diversity estimation on $33$ Zernike coefficients with synthetic aberrations. The mean performance over $150$ simulations of Phase Diversity is given in \autoref{tab:PDperfo}.

\begin{table}
\myfloatalign
\begin{tabularx}{\textwidth}{>{\centering\arraybackslash}X>{\centering\arraybackslash}X>{\centering\arraybackslash}X}
\toprule
\tableheadline{Infinity Norm} & \tableheadline{RMS} & \tableheadline{$2$-Norm} \\ \midrule
$5.1$ & $2.0$ &  $11.7$ \\
\bottomrule
\end{tabularx}
\caption[Phase Diversity simulated mean performance.]{Mean performance of Phase Diversity over 150 simulations. A simulation includes a new aberration vector, a new Gaussian noise pattern for the image and a new estimation. The performance is expressed as the Infinity Norm of the error which is the maximum absolute value, the Root-Mean-Square and finally the $2$-Norm which is the square root of the sum of the squares.}  
\label{tab:PDperfo}
\end{table}

\begin{figure}[bt]
\myfloatalign
\subfloat[Noisy Synthetic PSF.]
{\label{fig:PDimRecons-a} \includegraphics[width=0.66\linewidth]{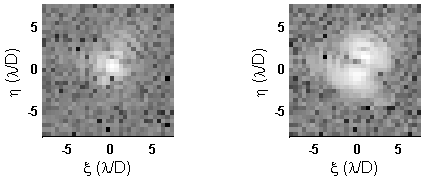}} \\
\subfloat[Reconstructed PSF.]
{\label{fig:PDimRecons-b} \includegraphics[width=0.66\linewidth]{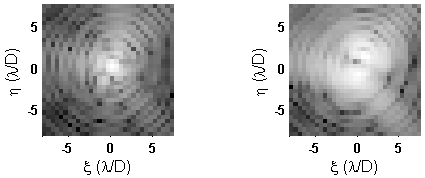}} \\
\subfloat[Reconstructed Noisy PSF.]
{\label{fig:PDimRecons-c} \includegraphics[width=0.66\linewidth]{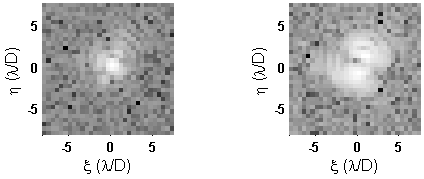}}
\caption[Phase Diversity PSF Reconstruction.]{Reconstruction of Point Spread Function with Phase Diversity. The focused images are on the left and the defocused on the right. \autoref{fig:PDimRecons-a} shows the input synthetic images. \autoref{fig:PDimRecons-b} shows the reconstruction of the Point Spread Function with the estimated Zernikes coefficients. \autoref{fig:PDimRecons-c} is the same with an additional noise of the same amplitude as the input images in order to ease the comparison. These images don't include the convolution with the extended source however there is no difference visually.}
\label{fig:PDimRecons}
\end{figure}

\relSubsection{Image Reconstruction}
Once the coefficients are estimated it is possible to reconstruct simulated images like in \autoref{fig:PDimRecons}.

\begin{figure}[bth]
\myfloatalign
\subfloat[Object.]
{\label{fig:PDobjRecons-a} \includegraphics[width=0.45\linewidth]{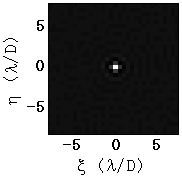}} \quad
\subfloat[Fourier Transform.]
{\label{fig:PDobjRecons-b} \includegraphics[width=0.45\linewidth]{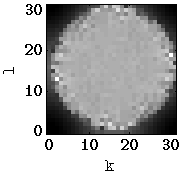}}
\caption[Phase Diversity Object Reconstruction.]{Reconstruction of the Object using Phase Diversity. The Object in \autoref{fig:PDobjRecons-a} is the shape of the extended light source. \autoref{fig:PDobjRecons-b} is the Discrete Fourier Transform of \autoref{fig:PDobjRecons-a}. The input images were computed with a point like source.}.
\label{fig:PDobjRecons}
\end{figure}

\relSubsection{Object Reconstruction}
Phase Diversity estimates the object as well. \autoref{fig:PDobjRecons} shows the estimation of a point source in simulation.

\relSection{Adaptive Optics Facility}
The team for the tests included Johann Kolb and the author himself.

Results presented here are only two of the many tests performed on the ASSIST test bench for the AOF. The complete set of result is available in the technical report \citeauthor{ruffioAOFPDtest} \citep{ruffioAOFPDtest}. An entire week was booked for these tests. In total $16$ iterations tests were performed as well as code validation tests like the linearity check briefly mentioned below. Another important experiment was the definition of spiders\marginpar{The spiders are the structural elements fixing the secondary mirror from the edges of the primary mirror.} but it has been skipped for this document.

\begin{figure}[bt]
\centering
\includegraphics[width=0.66\linewidth]{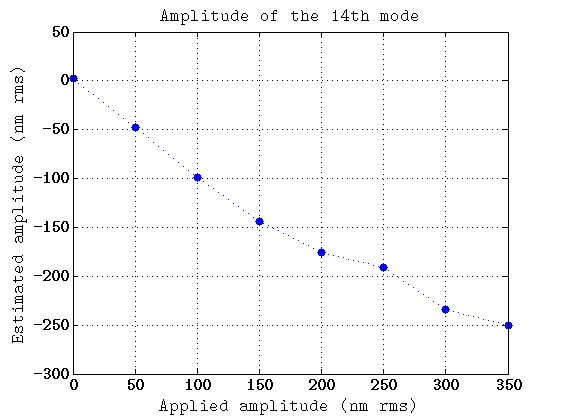}
\caption[Phase Diversity linearity test on the AOF.]{ Linearity test of Phase diversity applied on the AOF. The $14^{th}$ Zernike mode is applied with different amplitudes going from $0$ to $350\text{nm rms}$ with a step of $50\text{nm rms}$.  The figure shows the estimated amplitude of this $14^{th}$ modes.}
\label{fig:PDAOFlintest}
\end{figure}

\relSubsection{Linearity Test}
The linearity test is one of the first test that the author performed on the AOF. It is a good way to verify the behaviour of the algorithm.
To do so we applied different amplitudes to the $14^{th}$ Zernike mode\marginpar{The defocus is the $4^{th}$} on the AOF, recorded images and applied Phase Diversity. Eight different amplitudes have been tested from $0$ to $350\text{nm rms}$ with a step of $50\text{nm rms}$. This mode was selected because it is usually well detected by the algorithm with low residuals on the other modes. For every coefficient we apply a positive and a negative offset for reconstructing the sole contribution of the mode by difference and the sole aberration by addition.
The short result of the test is shown in \autoref{fig:PDAOFlintest} and the complete results can be found in \citeauthor{ruffioAOFPDtest} \citep{ruffioAOFPDtest}.

One can see that the linearity is very good until a saturation of the amplitude of the estimated coefficient at $200\text{nm rms}$ even if it begins to deviate at $150\text{nm rms}$. Besides the other modes are perturbed when the saturation is reached because the algorithm try to use other modes to reconstruct the aberration. \marginpar{It is very likely to be the $2\pi$ phase aberration limit indicated by ONERA.}

\begin{figure}[bth]
\myfloatalign
\subfloat[Initial PSF.]
{\label{fig:PDAOFiterBeforeAfter-a} \includegraphics[width=0.66\linewidth]{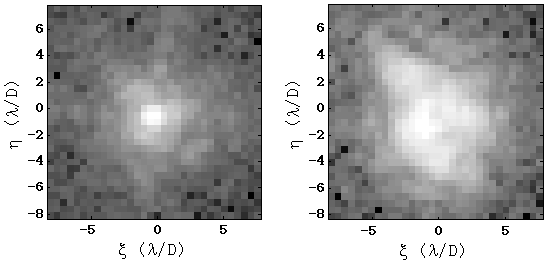}} \\
\subfloat[Corrected PSF.]
{\label{fig:PDAOFiterBeforeAfter-b} \includegraphics[width=0.66\linewidth]{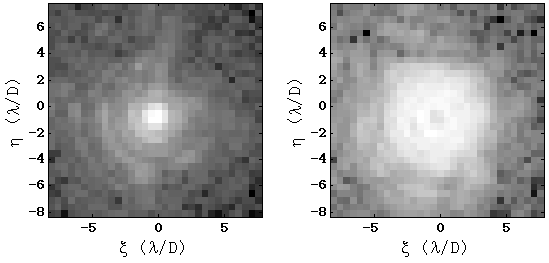}}
\caption[AOF PSF corrected by Phase Diversity.]{The leftmost images are the Point Spread Function before \autoref{fig:PDAOFiterBeforeAfter-a} and after \autoref{fig:PDAOFiterBeforeAfter-b} correction using two iterations of Phase Diversity on the AOF. The rightmost images are the same but defocused.}.
\label{fig:PDAOFiterBeforeAfter}
\end{figure}

\begin{figure}[bth]
\myfloatalign
\subfloat[Estimated Coefficients.]
{\label{fig:PDAOFiter-a} \includegraphics[width=0.45\linewidth]{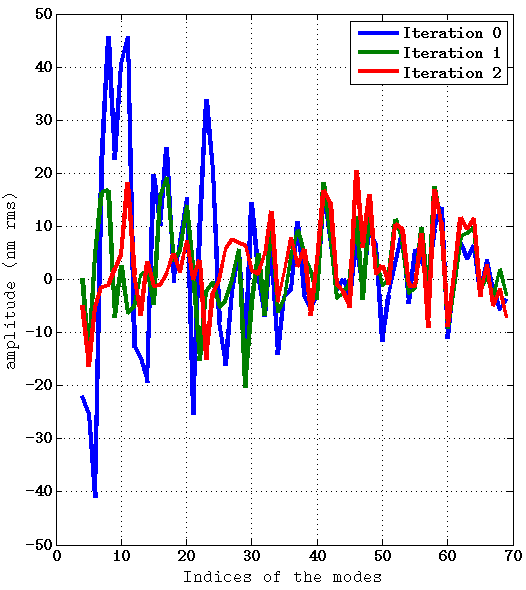}} \quad
\subfloat[RMS of the Coefficients.]
{\label{fig:PDAOFiter-b} \includegraphics[width=0.45\linewidth]{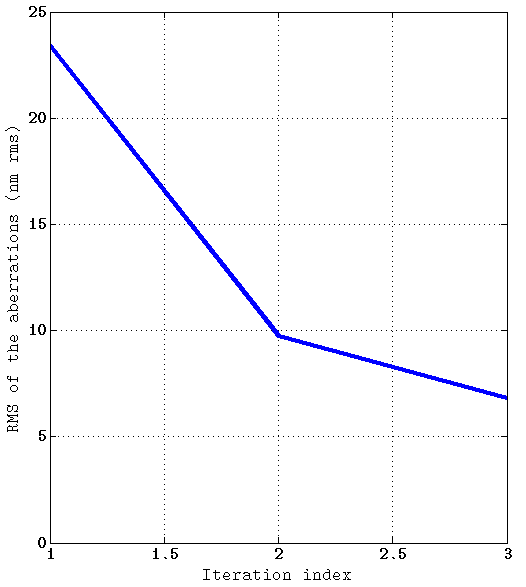}} \\
\subfloat[Strehl Ratio of the PSF $\lambda=1.6\mu m$.]
{\label{fig:PDAOFiter-c} \includegraphics[width=0.45\linewidth]{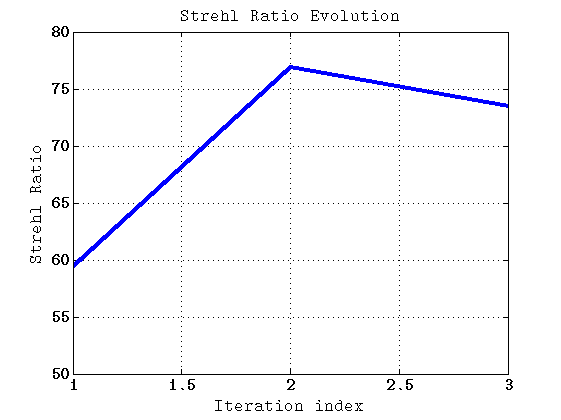}} 
\caption[Phase Diversity Iterations on the AOF.]{Phase Diversity iterations performed on the AOF. \autoref{fig:PDAOFiter-a} shows the $66$ estimated Zernike coefficients estimated for each iterations. \autoref{fig:PDAOFiter-b} shows the convergence of the algorithm with the Root-Mean-Square value of the first $27$ corrected coefficients. Finally \autoref{fig:PDAOFiter-c} shows the improvement in the Strehl Ratio. For this iterations the optimized model of the spiders was used, the images dimension was $32\times32$ pixels, no sub-pixel sampling has been used, the amplitude used for the Projection Matrix was $150\text{nm rms}$ and the standard deviation of the noise was estimated to $0.15$ in normalized data number.}.
\label{fig:PDAOFiter}
\end{figure}

\relSubsection{Iterations}
\autoref{fig:PDAOFiterBeforeAfter} shows the Point Spread Function of the AOF after and before Phase Diversity correction. Visually the image becomes clearly more symmetrical and portions of rings appears. The analysis of the two iterations is presented in \autoref{fig:PDAOFiter}. The Projection Matrix used is available in \autoref{app:PDresAOFit} \autoref{fig:PDAOFiterPDPM}. One can see that the Strehl Ratio jump of $17\%$ at the first iteration and then go down a little. However the estimated aberrations are always converging toward zero. One should remember that it is only one of the $16$ iterations test performed on the AOF.\marginpar{For the complete set of results see \citeauthor{ruffioAOFPDtest} \citep{ruffioAOFPDtest}.} $17\%$ is the biggest jump that the author could reach in one iteration. Otherwise the highest Strehl reached was close to $80\%$. However the Strehl is still quite unstable and the improvement is globally chaotic when iterating\marginpar{Chaotic means that the Strehl Ratio went up and down.}. See the discussion in \autoref{cha:PDdiscussion} for more comments.

\relChapter{Discussion}
\label{cha:PDdiscussion}

During the first half of 2014 the AOF was in Maintenance and Commissioning mode with GRAAL in ESO's laboratory in Garching. There is therefore no immediate scientific goal to the current experiments on ASSIST.\marginpar{Even if some performance requirements are tested.} The main purpose is to better understand the system in order to be able to identify possible problems. One of the problems AOF's people faced was a very poor quality Point Spread Function. The usual method to improve the Strehl Ratio is to apply Phase Diversity. However it assumes that the problem comes from phase error in the non-common optical path. But the existing algorithm didn't succeed. Different explanations were suggested but the main one was the existence of spiders in the pupil while the algorithm didn't take account for them. The new author's code actually proved using simulations and experimental tests\marginpar{These tests are shown in \citeauthor{ruffioAOFPDtest} \citep{ruffioAOFPDtest}.} that it had not a noticeable impact on the result. Other paths to improve the estimation were explored but without success\marginpar{For example a better sampled pupil has been tried or different hyper-parameters.}. The author's algorithm has proven itself reliable as it passed the linear test and it visually succeeds in removing the very low order aberrations. However the new code didn't give better Strehl Ratio\marginpar{The Strehl was stuck below $80\%$} and it is still very unstable so it convinced the team that the problem is not caused by low order phase aberrations that can be solved with the Deformable Mirror. The Strehl Ratio is also something complex to estimate and the existing algorithm are not very stable and robust. One could therefore partially question these values. Other explanation could include amplitude errors but they have not been estimated yet.\marginpar{The pupil amplitude variations should indeed turn around $20-25\%$.} The existence of a half ring in the left image of \autoref{fig:PDAOFiterBeforeAfter-b} strongly points toward this explanation. Indeed a Deformable Mirror can correct amplitude aberrations in only one half of the Point Spread Function.

\cleardoublepage

\label{app:bibliography} 


\addcontentsline{toc}{chapter}{\tocEntry{\bibname}}

\bibliographystyle{plainnat}

\bibliography{D:/latex/phaseDiversity/bibliography/bib_phase_diversty} 

\ctparttext{Electric Field Conjugation is a method for removing Quasi Static Speckles in a predefined area of the detector. The first chapter develops the theory of the method including instructions for applying the algorithm. In a second chapter are shown some simulated and experimental results. The experiments were performed on Infra-Red Dual-beam Imager and Spectrograph (IRDIS) which is one of the instruments of the Spectro-Polarimetric High-contrast Exoplanet REsearch (SPHERE) project at the VLT UT3, Paranal, Chile. The third and last chapter is a discussion about the results including some perspectives relative to the scientific context of direct exo-planet imaging.} 

\part{Electric Field Conjugation} 

\relChapter{Theory} 

This chapter develops the theory behind Electric Field Conjugation. The essential of this part is inspired from \citeauthor{giveon2011} \citep{giveon2011}. The core method is something that has been used extensively in laboratory however it is not yet very common on-sky. 

Electric Field Conjugation is a method to darken a predefined area in a coronagraphic image using a Deformable Mirror and the science camera. The nice side of this method is its simplicity. Indeed there is no estimation theory involved here and it is conceived so that everything is linear. The only real mathematical tool used is the Singular Value Decomposition but basic knowledge on Pseudo-Inverse is sufficient. The idea of Electric Field Conjugation is to record images while applying so called probes positively and negatively on the Deformable Mirror. The probes are shapes of the Deformable Mirror which result in adding a known complex amplitude to the speckles in the focal plane. Based on some assumptions the subtraction of the two images is linearly proportional to the electric field in the focal plane. Besides using Fourier Optics the electric field itself is linearly related to the phase in the pupil at least to the first order and therefore to the shape of the Deformable Mirror. Altogether it means that the relation between the shape of the Deformable Mirror and its effect on the intensity in the detector is linear. It means that one is able to define a shape of the Deformable Mirror for cancelling out the speckles. This shape is in practice expanded on a base of modes. For example the effect of Fourier modes is to enlighten a spot at a position defined by the frequency of the sine in the pupil. 

\paragraph{Author's Contributions} One contribution has been to make the mathematical demonstration cleaner. The Taylor expansions were indeed not rigorously made which requires some not well justified approximations. The author solved the problem in the case of coronagraphic images. The author also suggested to use singular modes instead of Fourier modes which should be in theory more efficient. \marginpar{The definition of a probe and a mode will be detailed in \autoref{sec:EFCdarkholearea}.} \\

The section begins first with a short bibliography of Electric Field Conjugation. Then it develops the intensity expression in the focal plane.\marginpar{Because of number of pages requirement most of the interesting demonstrations are in Appendix.} Using this model it explains how the electric field can be estimated in the focal plane using intensity measurements. Then it develops how the speckle correction can be performed using probes and modes themselves described in the following section. To finish a more practical description of Electric Field Conjugation is given for whoever wants to apply it in real life.

\relSection{State of the Art}
Many different types of correction algorithm have been developed based on the estimation of the complex electric field in the focal plane.\marginpar{The author didn't read all the papers so he will only cite the ones he knows.}
During his master thesis at the European Southern Observatory (ESO) the author based his work on the formulation of the problem by \citeauthor{giveon2011} in \citep{giveon2011} and \citep{giveon2007}.
Others similar methods have been tried elsewhere for example \citeauthor{martinacheSUBARUEFC} \citep{martinacheSUBARUEFC} with a Speckle Nulling technique on-sky at the Subaru telescope. However the algorithm presented in this document seems more straightforward. \citeauthor{thomas2010} in \citep{thomas2010} shows results of Electric Field Conjugation at the Laboratory of Adaptive Optics in Santa Cruz. She reached an additional order of magnitude in a $5\times 5 (\lambda/d)$ rectangle area at a distance of $6.5 (\lambda/d)$. However \citeauthor{thomas2010} kept separated the estimation of the electric field and the correction in her algorithm. Besides she based the estimation on a simulation of the bench which requires a very good knowledge of it. The method developed by the author and Markus Kasper is simpler to implement because all the calibrations are measured and there is no estimation of the electric field.

A previous master student \citeauthor{jankowsky2012} worked on Electric Field Conjugation in 2012 at the ESO for his master thesis \citep{jankowsky2012}. He tried to apply Electric Field Conjugation on the High Order Testbench (HOT)\marginpar{The High Order Testbench purpose was to test Extreme Adaptive Optics systems.} from ESO in the same way it was applied on a bench at the Institut de Plan\'{e}tologie et d'Astrophysique de Grenoble (IPAG) called Fresnel-FRee Experiment for EPICS\footnote{EPICS means Exoplanet Imaging Camera and Spectrograph} (FFREE). However the speckles on FFREE were artificially added using a phase screen before trying to suppress them. This experience reached a maximum contrast improvement of $40$ in term of the Root-Mean-Square of the speckles in a rectangle area. The results can be found in \citeauthor{verinaudFREEEFC} \citep{verinaudFREEEFC}.

\relSection{Model}
\label{sec:EFCModel}

The complex amplitude in the pupil plane is defined by,

\begin{equation}
   \mathcal{P}(x,y) = \mathcal{P}_m(x,y) e^{i(\phi(x,y)+\psi(x,y)+\omega(x,y))} ,
\end{equation}
 With $P_{m}(x,y)$ the pupil mask.\marginpar{Note that the model doesn't include amplitude aberrations.} $x$ and $y$ are the spatial coordinates in the pupil normalized by the pupil mask diameter. $\phi(x,y)$, $\psi(x,y)$ and $\omega(x,y)$ are the phase respectively of the aberrations, the probes and the correction phase. The correction phase is later expanded on modes. 
 
Electric Field Conjugation is dealing with coronagraphic images so that the complex amplitude in the focal plane can't be computed using simply one Fourier Transform. However it would use a series of Fourier Transforms to take into account the succession of pupil and focal planes in which are the apodizer, the focal-mask and the Lyot-stop.\marginpar{For more about the Apodized-Lyot-Coronagraph see \autoref{app:APLCdef}.} This transformation is a combination of linear operators so that it is also linear and it is noted $\mathcal{C}$ standing for Coronagraph. Therefore the complex amplitude in the focal plane is given by,

\begin{equation}
  \mathcal{D}(\xi,\eta)=\mathcal{C}\left( \mathcal{P} \right)=\mathcal{C}\left( \mathcal{P}_m e^{i(\phi+\psi+\omega)} \right) ,
  \label{eq:EFCpup2foc}
\end{equation}
With $D(\xi,\eta)$ the complex amplitude in the focal plane. $\xi$ and $\eta$ are the spatial coordinates in the focal plane normalized by $\lambda / d$\marginpar{The unit of $\xi$ and $\eta$ is sometimes called resel.}.

The phase $\omega$ is ignored for now and $\phi$ and $\psi$ are assumed small compared to unity. In order to make the notation more compact the operator $\mathcal{C}$ will now be noted with an upper bar. \marginpar{The upper bar is sometimes used for conjugation but here the conjugate of $z$ is noted $z^\ast$.} Expanding the expression of the intensity $\mathcal{I} = \mathcal{D} \mathcal{D}^{\ast}$ with a second Taylor expansion gives,
\begin{align}
 \mathcal{I} &= \abs*{\overline{\mathcal{P}_m}}^2 & 0\, \text{Order,} \nonumber \\
 &+ 2\text{Re} \left[ i \left(\overline{\mathcal{P}_m}\right)^{\ast} \left( \overline{\mathcal{P}_m \phi}+ \overline{\mathcal{P}_m \psi} \right) \right] & 1^{st}\, \text{Order,} \nonumber \\
 &- 2\text{Re} \left[ \overline{\mathcal{P}_m} \left( \overline{\mathcal{P}_m \frac{\phi^2}{2}} + \overline{\mathcal{P}_m \frac{\psi^2}{2}} + \overline{\mathcal{P}_m \psi\phi} \right)^{\ast} \right] & 2^{nd}\, \text{Order,} \nonumber \\
 &+ \abs*{\overline{\mathcal{P}_m \phi}}^2 + \abs*{\overline{\mathcal{P}_m \psi}}^2 + 2\text{Re} \left[ \left(\overline{\mathcal{P}_m \phi}\right)^{\ast}\overline{\mathcal{P}_m \psi} \right]& \dots
 \label{eq:EFCintens}
\end{align}
The complete demonstration is given in Appendix \autoref{app:EFCAppModel}.

\relSection{Electric Field Estimation}
\label{sec:EFCesti}
The principle of Electric Field Conjugation is to apply the probe $\psi$ positively and negatively. As it is demonstrated in \autoref{app:EFCAppEFEst} and assuming a perfect coronagraph the difference between the two resulting images is linearly proportional to the electric field of the speckles. The only terms remaining of the difference in \eqref{eq:EFCintens} are indeed the one proportional to $\psi$. Using vector and matrix representation of the variables the problem takes the form,\marginpar{The function in the pupil or in the focal plane are discretized and shaped in vector form.}
 \begin{equation}
 \underbrace{\vecTwoD{\frac{1}{2}(I_1^+ - I_1^-)}{\frac{1}{2}(I_2^+ - I_2^-)}}_{\delta I_\Phi}= 2
 \underbrace{\left[ \begin{array}{cc}
 \text{Re} \left[C \Psi_1\right] & \text{Im} \left[C \Psi_1\right] \\
 \text{Re} \left[C \Psi_2\right] & \text{Im} \left[C \Psi_2\right] \\
 \end{array} \right]}_{E}
 \vecTwoD{\text{Re} \left[C \Phi \right]}{\text{Im} \left[C \Phi \right]},
 \label{eq:EFCesti}
 \end{equation}
With $\Psi_1$ and $\Psi_2$ two different probe vectors and $I_1$ respectively $I_2$ the vector of the pixel intensities for the respective probes. $I_k^+$ and $I_k^-$ correspond to the use of the positive and the negative offset of the probe. \\
\eqref{eq:EFCesti} is the core of Electric Field conjugation

\relSection{Correction}
\label{sec:EFCcorr}
The goal of Electric Field Conjugation is to cancel the term $\phi$ using the correction phase $\omega$ in,
\begin{equation}
  \mathcal{P}= \mathcal{P}_m e^{i(\phi+\omega)} ,
\end{equation}
so that the speckles disappear. $\omega$ is created using the Deformable Mirror so that it can be defined as a vector of commands $A$ for the actuators. The demonstration in \autoref{app:EFCAppCorr} shows that $A$ can be obtained from,
\begin{equation}
GA=-\delta I_\Phi \quad \text{or} \quad A=-R\delta I_\Phi,
\label{eq:EFCproblem}
\end{equation}
With $G$ a so called Interaction Matrix and $R$ its inverse called the Reconstruction Matrix. \marginpar{The expression of $G$ is given in \autoref{app:EFCAppCorr}.} The Interaction Matrix $G$ should be either computed through simulations or directly measured. \marginpar{It is quite common to simulate the Interaction Matrix however measurements prevent one from hidden parameters, biases or more generally wild assumptions.}

\relSection{Dark Hole Definition}
\label{sec:EFCdarkholearea}

\relSubsection{Modes}
\label{sec:EFCmodes}

The modes are a base of functions on which the correction phase $\omega(x,y)$ is expanded. The base should be complete enough so that every pixel of the area can be corrected. The number of independent modes is limited by the number of actuators. \marginpar{In the experiences done by the author on SPHERE it seems like there is no further improvement when using more than $\simeq150$ modes.}

The problem as defined in \eqref{eq:EFCproblem} was presented with $A=[a_1\,a_2\,\dots\, a_K]^\mathsf{T}$ being the actuators commands however it is actually more generally the coordinate vector of the phase $\Omega$ in the base of the modes. 
\begin{equation}
\Omega = \sum_{k=1}^{K} a_k Z_k = ZA,
\end{equation}
With $Z_k$ the modes as vectors containing the pupil phase values and $Z$ the matrix formed with the column vectors $Z_k$. The method to obtain the commands vector from the modes for the actuators is given in \autoref{app:EFCAppModesVolt}.

 The contribution of a mode to the electric field in the focal plane is to the first order equal to the Fourier Transform of the phase times the pupil mask.\marginpar{It comes from a first order Taylor expansion and the linearity of $\mathcal{C}$ in \autoref{eq:EFCpup2foc}.} So one can roughly infer the effect of a mode directly from the phase instead of from the complex amplitude.
 
 It is then straightforward to define sine and cosine functions which are called Fourier modes. Their Fourier Transform being a pair of Dirac functions the effect on the detector will be two symmetric non coronagraphic Point Spread Function\marginpar{If they are not on axis the coronagraph is negligible.} centred on a pixel defined by the frequency of the wave. A detailed description of the Fourier modes with closed form expression and example figure is given in \autoref{app:EFCAppModesFour} and \autoref{fig:modeSin}.
 
 More efficient modes can be defined from the Singular Value Decomposition of the Interaction Matrix. The author called them Singular Modes. The detailed explanation is given in \autoref{app:EFCAppModesSing}.

\relSubsection{Probes Definition}
The probes noted $\psi$ in the previous sections are phase function produced by the Deformable Mirror for introducing diversity into the measurement and then retrieving the electric field from intensities. Their effect in the focal plane should cover all the area of interest\marginpar{Even if it is also possible to patch probes.}. Two probes should have different phase for any point in the focal plane falling in the area. The best probes would have orthogonal vectors which means a difference in phase of $\frac{\pi}{2}$ in the focal plane.
\citeauthor{giveon2011} \citep{giveon2011} gives an example of probes for making a rectangle in the detector. Their closed form expression, their explanation and a figure is available in appendix \autoref{app:EFCAppProbes} and \autoref{fig:probeSin}. \marginpar{New formula for rotated probes is also available.}

\relSubsection{Area}
The area of interest also called dark hole can have any shape. However Electric Field Conjugation cannot correct for both phase and amplitude aberrations if the area overlap with itself when mirrored. It means for instance that as long as the area remains in one half of the image the algorithm is able to correct amplitude aberration as well. This is due to the fact that the Deformable Mirror can only act on the phase. \marginpar{However the author doesn't know yet a proper explanation for this\dots}

 Besides the spatial frequency in the pupil is limited by the distance between two actuators. The limitation in spatial frequency in the pupil plane results in the impossibility of correcting speckles too far from the center. For example if the Deformable Mirror has an array of $40\times40$ actuators it can produce a maximum of $40/2=20$ periods on its surface and therefore the limit of correction would be at $20 (\lambda/d)$. Even inside the boundaries the efficiency of the correction is not constant with the distance to the center. Some frequencies are indeed not well reproduced by the discrete array of actuators. In practice one can neither correct too close of the center because of the residual leaking light of the coronagraph.  Because of all these constraints the area should not be too far, neither too close and not too big\dots

\relSection{Application}

\relSubsection{Measurements}
\label{sec:EFCmeasurements}
The measurements for Electric Field Conjugation consist in doing the following for each probe,
\begin{enumerate}
\item Apply probe positively and negatively to the Deformable Mirror and take an image for both.
\item Extract the pixels of the area of interest from both images.
\item Reshape the pixels in vector form.
\item Subtract the vector of the positive probe to the vector of the negative probe and divide by two.
\end{enumerate}
Then one needs to concatenate the resulting vectors of each probe for building the measurement vector called $\delta I_\Phi$ in \eqref{eq:EFCproblem}. \\
It is important that the centroid of the Point Spread Function doesn't move during the measurements and it has to be at the same relative sub-pixel position than when building the Interaction Matrix $G$. \\
In practice a dark exposure was also taken before and after the acquisition of the image. Besides each image where taken twice in order to estimate the photon noise in the correction vector. \marginpar{A reference image with no applied probe can also be taken for measuring the intensity in the dark area.}
For SPHERE a template was implemented on the instrument to run this sequence of image acquisition. The template instructions are given in \citeauthor{kasperNCPATemplate2014} \citep{kasperNCPATemplate2014}.

\begin{figure}[bth]
\myfloatalign
\subfloat[Interaction Matrix.]
{\includegraphics[width=.45\linewidth]{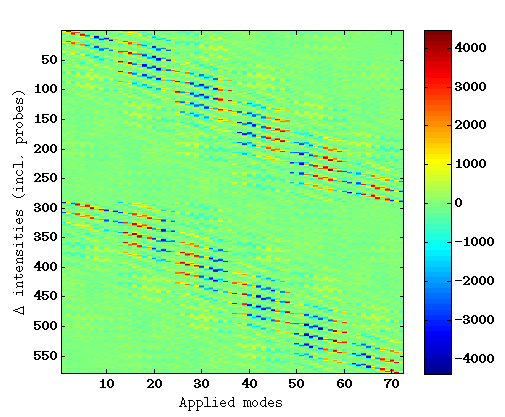}} \quad
\subfloat[Reconstruction Matrix]
{\includegraphics[width=.45\linewidth]{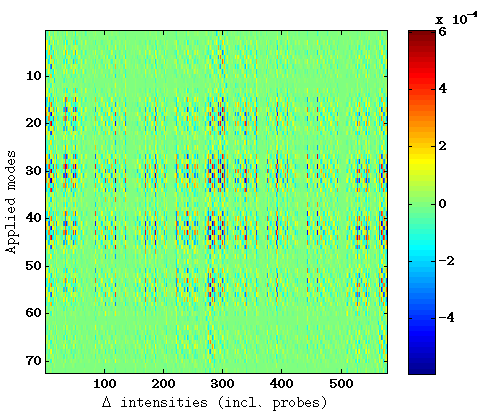}}
\caption[Interaction and Reconstruction Matrix.]{Example of Interaction Matrix and its inverse the Reconstruction Matrix for Electric Field Conjugation using Fourier modes spaced by $1(\lambda/d)$ in a rectangle area. The two diagonal shapes in the Interaction Matrix correspond to the two probes.}
\label{fig:EFCIMRM}
\end{figure}

\relSubsection{Interaction Matrix}
\label{sec:EFCIM}
The Interaction Matrix can be either simulated or directly measured. An example of synthetic matrix is given in \autoref{fig:EFCIMRM}.\marginpar{It is easier and more secure to measure it. However this calibration takes a long time when performed on a real system. For example building a Interaction Matrix for SPHERE with two probes and a hundred modes takes roughly two hours.} Beforehand one needs to notice that the columns of $G$ are measurement vectors when a single mode is applied $\delta I_\Omega$. Therefore building the Interaction Matrix consists in applying each mode one by one and then construct the measurement vector as explained in \autoref{sec:EFCmeasurements}. This naive approach would work only if there were no existing aberration while doing it. The solution to get rid of the aberrations is to apply the modes positively and negatively and then subtracting the two resulting measurement vectors. The demonstration is given in appendix \autoref{app:EFCAppIM}.

To conclude the steps for building the Interaction Matrix consists in doing for each mode and for each probe,
\begin{enumerate}
\item Take an image for all possible combinations of positive and negative probes and positive and negative modes.
\item Extract the pixels of the area of interest from all images.
\item Reshape the pixels in vector form.
\item Subtract the vectors of the positive probe to the vectors of the negative probe  for each mode and divide by two.
\item Subtract the difference vectors from last step of the positive and the negative mode.
\end{enumerate}
Then one needs to concatenate the vectors of the different probes for building the measurement vector. To finish the resulting vectors should be concatenated together as column vectors for building the Interaction Matrix.

\relSubsection{Iterations}
\label{sec:EFCAppIt}
Once the measurement vector $\delta I_\Phi$ is build one just needs to multiply it with the Reconstruction Matrix,
\begin{equation}
A = -R \delta I_\Phi = -G^\dagger \delta I_\Phi.
\end{equation}

\marginpar{Note that even if a Pseudo-Inverse is used here the author uses in practice a filtered Singular Value Decomposition.}
Again $A$ is the coordinate vector of the correction phase in the modal base. The command vector $\alpha$ for the Deformable Mirror can then be computed from $A$ using the influence functions using,
\begin{equation}
\alpha = I_f^{\dagger}ZA,
\end{equation}
With $I_f$ the influence function matrix and $Z$ the matrix to transform modal base coordinates into pupil phase. The computation of the commands can also be done by adding the commands of each mode with the correct amplitude of $A$. \\

Once the correction commands are applied another iteration can be performed. Electric Field Conjugation assumes the linearity of the system so it needs a few steps to reach the best correction. \marginpar{For an example of simulated EFC iterations see \autoref{sec:EFCsimu}.}\\

In order to measure the performance of the iterations one can measure the standard deviation of the intensity inside the area of interest. The ratio between the standard deviation of the last step and the first image corresponds to the contrast improvement.

\relChapter{Results} 
This chapter exposes the simulated or experimental results of Electric Field Conjugation applied on SPHERE.

\paragraph{Author's Contributions} This whole section is author's contribution. The simulation were done using Matlab. The author developed a set of functions for computing Fourier optics optionally including perturbations or coronagraphs for instance. An overview of the code is available in \citeauthor{EFCPDCodes} \citep{EFCPDCodes}. When experimenting on SPHERE the author did also all the data reduction and the computation of the correction vectors with tools he implemented.

\relSection{Simulations}
\label{sec:EFCsimu}

For the author the first goal of the simulations were to check his understanding of the method. The simulation are indeed not required for applying Electric Field Conjugation on a real system\marginpar{This is true only if one measures the Interaction Matrix instead of using a synthetic one.}. However a few tools of the simulations were still used on the real system. For example the definition of the probes and the modes commands for feeding the Deformable Mirror are the same. Besides the image reduction is identical in simulation and on the instrument.

Then the simulations were used to give an idea of what could be expected and how to reach the best performance. This aspect has a limited range because the model is too perfect compared to the real system.
\paragraph{Note} Because of the number of pages requirement almost all the simulations can be found in \autoref{sec:EFCAppSimu}. \marginpar{In \autoref{sec:EFCAppSimu} one can also find the improvement with singular modes, the effect of the distance of the area and the combined effect of the number of modes and the number of pixels of the area.}

\relSubsection{Simulation Principles}
\label{sec:EFCsimuPrinc}
The principle of the simulation is to do the following actions, \marginpar{Note that an how-to tutorial is available in \citep{EFCPDCodes}.}
\begin{enumerate}
\item Define the optics,
\item Load the influence functions of the Deformable Mirror,
\item Construct the probes and the modes,
\item Construct the Interaction Matrix by simulating images with the probes and the modes,
\item Invert the Interaction Matrix,
\item Apply a few iterations of Electric Field Conjugation,
\item And analyze the results.
\end{enumerate}

\relSubsection{General Parameters}
\label{sec:EFCsimuPrincPara}
The coronagraph of SPHERE is an Apodized-Lyot-Coronagraph which is simply modeled by its different planes: apodization function, focal plane mask and Lyot-Stop \marginpar{For more about the Apodized-Lyot-Coronagraph see \autoref{app:APLCdef}.}. The masks correspond to the configuration used on SPHERE and especially in \autoref{sec:EFCresSpherePara}.

If nothing is specified Fourier modes refer to modes centred on the unit grid. It means they are spaced by $1(\lambda/d)$\footnote{This is roughly $3$ pixels for IRDIS SPHERE.} in both axes in the detector. Singular modes are here defined from the Singular Value Decomposition of an Interaction Matrix built with a pixel based set of modes.\marginpar{When simulating with singular modes a new Interaction Matrix with only the singular modes is built. However the result by just filtering the pixel defined Interaction Matrix to keep the same number of singular modes is almost equivalent in simulation.} The modes respectively the probes have an amplitude of $0.01\text{rad}$ respectively $0.1\text{rad}$. The probe needs higher amplitude because it is spread on a bigger area. To finish the noise in the phase for creating the speckles has a standard deviation of $0.01\text{rad}$.

\relSubsection{Simulated Iterations}
\label{sec:EFCsimuIt}
The first test is to verify that running \autoref{sec:EFCsimuPrinc} indeed creates a dark hole in the speckles of the image.

\begin{figure}[bth]
\myfloatalign
\subfloat[Coronagraphic Point Spread Functions]
{\label{fig:EFCsimuMedRect-a} \includegraphics[width=\linewidth]{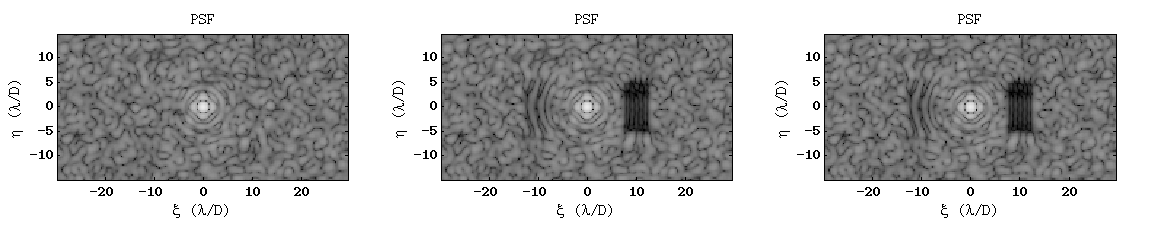}} \\
\subfloat[Dark Hole Speckle Intensity]
{\label{fig:EFCsimuMedRect-b} \includegraphics[width=0.5\linewidth]{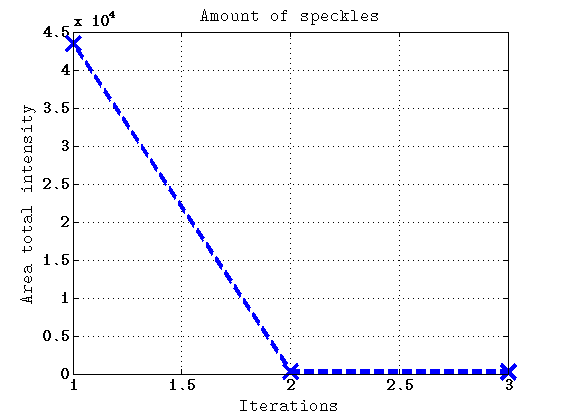}}
\caption[Simulated EFC iterations on a rectangle.]{Results of two iterations of Electric Field Conjugation using simulated data on a rectangle of dimension $5\times 10 (\lambda/d)$ at a distance of $10 (\lambda/d)$ of the center of the Point Spread Function. This corresponds to an area of $561$ pixels and $132$ Fourier modes were used. The two probes are defined as slightly bigger rectangles $0.5 (\lambda/d)$ on each side to avoid edge effect. No singular mode has been filtered as a Pseudo-Inverse is used for inverting the Interaction Matrix. \autoref{fig:EFCsimuMedRect-a} diplays the Coronagraphic Point Spread Function images after each iteration. \autoref{fig:EFCsimuMedRect-b} shows the mean over ten simulations of the total intensity in the area. The contrast improvement is given in \autoref{tab:EFCsimuMedRect}}
\label{fig:EFCsimuMedRect}
\end{figure}

 For a simple rectangle case the result of two iterations is given in \autoref{fig:EFCsimuMedRect} and the mean performance values over ten simulations is given in \autoref{tab:EFCsimuMedRect}.

\begin{table}
\myfloatalign
\begin{tabularx}{\textwidth}{Xcc}
\toprule
\tableheadline{Modes} & \tableheadline{RMS} & \tableheadline{Intensity} \\ \midrule
$132$ Fourier & $140$ &  $183$ \\
\bottomrule
\end{tabularx}
\caption[Simulated EFC contrast improvement on a rectangle with Fourier modes.]{Contrast improvement for the study case of \autoref{fig:EFCsimuMedRect} with a $5\times 10 (\lambda/d)$ rectangle $10(\lambda/d)$ away from the center. The gain is computed in term of the Root-Mean-Square value and the total intensity in the dark hole area. The values are the result of the mean over ten simulations.}  
\label{tab:EFCsimuMedRect}
\end{table}

\relSection{SPHERE}
The team for the tests included Markus Kasper, Christophe V\'erinaud and the author himself.

\relSubsection{General Parameters}
\label{sec:EFCresSpherePara}
The experiments on SPHERE were done using an Apodized-Lyot-Coronograph \marginpar{The APO1 apodizer, the ALC2 mask (4 $\lambda/d$ in diameter) and ST\_ALC Lyot-Stop were used. See the technical documentation \citep{SPHEREcoro2012}.} configuration and the internal fibre as a infrared light source. A narrow-band filter in H-band at $1.6\mu m$ was used.

The first attempts were realized with the Adaptive Optics in open-loop. The tip-tilt correction was however in closed-loop so that the image doesn't move too much on the detector. Open-loop is indeed slightly easier to manage because applying a shape to the Deformable Mirror corresponds simply to the addition of a voltage vector to the commands. When doing so in closed-loop the feedback would automatically kill the shape for going back to a flat wavefront. However the assumption of negligible internal convection in SPHERE\marginpar{SPHERE is closed and cooled environment.} was too optimistic so that the experiment had to be done in closed-loop. In that case the reference slopes of the wavefront sensor need to be used for applying shapes to the Deformable Mirror instead of the voltages. When doing so a problem of synchronization between the image acquisition and the Deformable Mirror occurred. Indeed loading reference slopes needs time but the delay is random.\marginpar{The delay is up to two seconds.} Therefore when acquiring images the right shape of the mirror was sometimes applied at the middle of the integration time. The solution found was to pause the system for two seconds between the sending of the reference slopes and the image acquisition. \marginpar{This is not optimized and lot of time is wasted because of that. However the system does not have a proper synchronization with a flag telling when it is ready to go. So there is no choice until it is implemented.} Including overheads one image requires in total about ten seconds. An example of a mode and an example of a probe are given in Appendix in \autoref{fig:EFCSPHEREprobeMode}.

All the image acquisition part was done using on-board SPHERE software but all the data reduction and the computation of the commands or slopes vectors were done using external Matlab codes.

The amplitudes of the probes and of the modes were adjusted by hand. The used amplitudes were not always recorded so that the author can't give their value. Still it can be noticed that the amplitude of the probes was decreased\marginpar{However this was not proven to have an effect\dots} at each iteration to follow the intensity drop in the dark hole. The criteria for a fine amplitude were that one could visually see the effect of a probe or a mode but without being brighter than the speckles.

\paragraph{Warning} A mistake was made in the scripts when iterating with simulated interaction matrices so that the author can't be exactly sure that the right matrices were used. This is particularly relevant for the first rectangle case where the simulated Interaction Matrix works surprisingly well. It could be that in fact it was the measured matrix which was used. In order to validate this result one should try it again.

\begin{figure}[p] 
\myfloatalign
\subfloat[Initial image.]
{\label{fig:EFCmedRectImage132Four-a} \includegraphics[width=.45\linewidth]{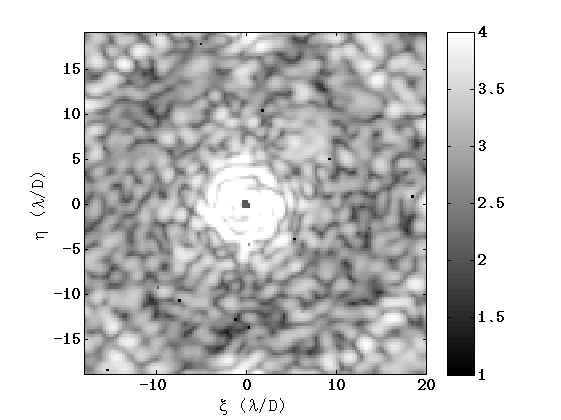}} \quad
\subfloat[Corrected area.]
{\label{fig:EFCmedRectImage132Four-b} \includegraphics[width=.45\linewidth]{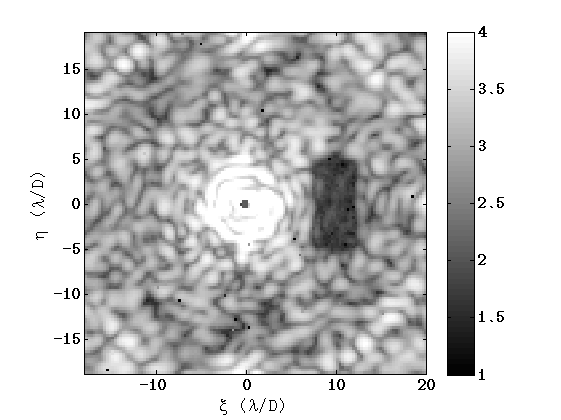}}
\caption[Rectangle dark hole with EFC on IRDIS SPHERE.]{Rectangle dark hole created by after seven iterations of Electric Field Conjugation on IRDIS SPHERE with a measured $132$ Fourier modes Interaction Matrix. Both images use logarithmic scale.}
\label{fig:EFCmedRectImage132Four}
\end{figure}

\relSubsection{Rectangle}
The first successful attempt of Electric Field Conjugation iterations was performed on a $5\times 10 (\lambda/d)$ rectangle at a distance of $10 (\lambda/d)$ of the center. Fourier modes and singular modes were tested as well as a synthetic Interaction Matrix. The resulting corrected area for the best case with $132$ Fourier modes is shown in \autoref{fig:EFCmedRectImage132Four}. The performance for all the tests is given in \autoref{tab:EFCgainMedRect}. One can find the complete results in Appendix in \autoref{fig:IMmedRect}, \autoref{fig:EFCitMedRect} and \autoref{fig:EFCmedRectImages}.

\relSubsection{Arc}
The second experiment was done on a quarter of a ring with a width of $5 (\lambda/d)$ and a mean radius of $10 (\lambda/d)$. $125$ singular modes were used. Again a comparison between a synthetic and a measured Interaction Matrix was performed. However in this case the synthetic matrix didn't performed well. The results are given in \autoref{tab:EFCgainArc} and the resulting dark hole for the measured Interaction Matrix is shown in \autoref{fig:EFCarcImageSing125}. One can find the complete results in Appendix in \autoref{fig:IMarc}, \autoref{fig:EFCitArc} and \autoref{fig:EFCarcImages}.

\relSubsection{Big Rectangle}

The last test was done on a big area formed by a $8\times 16 (\lambda/d)$ rectangle at a distance of $11 (\lambda/d)$ of the center. A synthetic Interaction Matrix with 306 Fourier modes was used. The results are given in \autoref{tab:EFCgainBigRect} and the images are shown in \autoref{fig:EFCbigRectImageChris306}. One can find the complete results in Appendix in \autoref{fig:EFCitBigRect} and \autoref{fig:EFCbigRectImages}.

\begin{figure}[p]
\myfloatalign
\subfloat[Initial image.]
{\label{fig:EFCarcImageSing125-a} \includegraphics[width=.45\linewidth]{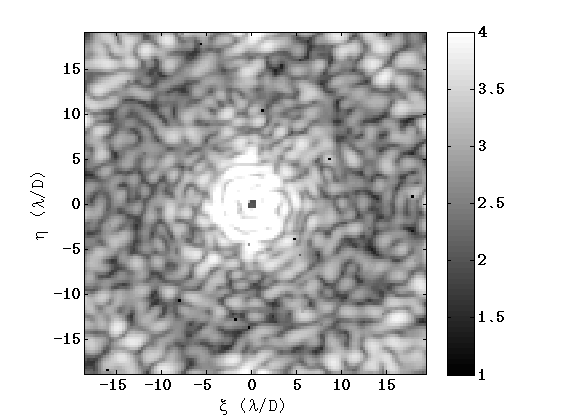}} \quad
\subfloat[Corrected area.]
{\label{fig:EFCarcImageSing125-b} \includegraphics[width=.45\linewidth]{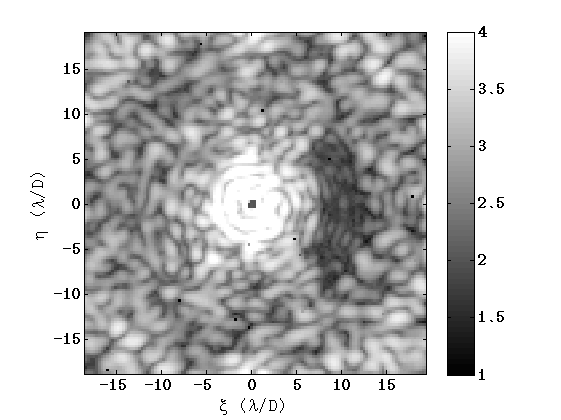}}
\caption[Quarter of ring dark hole with EFC on IRDIS SPHERE.]{Quarter ring dark hole created after four iterations of Electric Field Conjugation on IRDIS SPHERE with a measured $125$ singular modes Interaction Matrix. Both images use logarithmic scale.}
\label{fig:EFCarcImageSing125}
\end{figure}

\begin{figure}[p]
\myfloatalign
\subfloat[Initial image.]
{\label{fig:EFCbigRectImageChris306-a} \includegraphics[width=.45\linewidth]{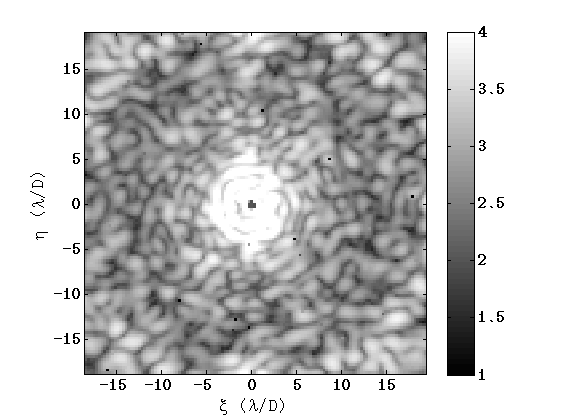}} \quad
\subfloat[Corrected area.]
{\label{fig:EFCbigRectImageChris306-b} \includegraphics[width=.45\linewidth]{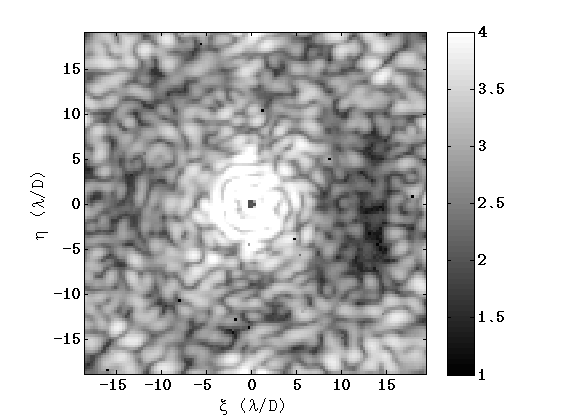}}
\caption[Big rectangle area before and after correction.]{Big rectangle dark hole created after four iterations of Electric Field Conjugation on IRDIS SPHERE with a synthetic $306$ Fourier modes Interaction Matrix. Both images use logarithmic scale.}
\label{fig:EFCbigRectImageChris306}
\end{figure}

\FloatBarrier

\begin{table}
\myfloatalign
\begin{tabularx}{\textwidth}{Xcccc}
\toprule
\tableheadline{Modes}& \tableheadline{IM} & \tableheadline{Cond.} & \tableheadline{RMS} & \tableheadline{RMS (small)} \\ \midrule
$132$ Fourier & Measured & $87$ &  $10.1$ & $13.3$ \\
$109$ Singular & Measured & $33$ &  $9.1$ & $12.5$ \\
$109$ Singular & Synthetic & $36$ &  $7.1$ & $7.8$ \\
\bottomrule
\end{tabularx}
\caption[EFC contrast improvement for the rectangle.]{Maximum contrast gain when iterating on a $5\times 10 (\lambda/d)$ rectangle at a distance of $10 (\lambda/d)$ of the center with different sets of modes. The columns are respectively indicating from left to right the set of modes, the type of Interaction Matrix, its conditioning number, the Root-Mean-Square contrast improvement in the area respectively in a smaller area without the edges.}  
\label{tab:EFCgainMedRect}
\end{table}

\begin{table}
\myfloatalign
\begin{tabularx}{\textwidth}{Xcccc}
\toprule
\tableheadline{Modes}& \tableheadline{IM} & \tableheadline{Cond.} & \tableheadline{RMS} & \tableheadline{RMS (small)} \\ \midrule
$125$ Singular & Measured & $10$ &  $6.1$ & $7.2$ \\
$125$ Singular & Synthetic & $5$ &  $1.3$ & $1.3$ \\
\bottomrule
\end{tabularx}
\caption[EFC contrast improvement for the quarter of ring.]{Maximum contrast gain when iterating on a quarter of a ring with a width of $5 (\lambda/d)$ and a mean radius of $10 (\lambda/d)$ using $125$ singular modes. The columns are respectively indicating from left to right the set of modes, the type of Interaction Matrix, its conditioning number, the Root-Mean-Square contrast improvement in the area respectively in a smaller area without the edges. }  
\label{tab:EFCgainArc}
\end{table}

\begin{table}
\myfloatalign
\begin{tabularx}{\textwidth}{Xcccc}
\toprule
\tableheadline{Modes}& \tableheadline{IM} & \tableheadline{Cond.} & \tableheadline{RMS} & \tableheadline{RMS (small)} \\ \midrule
$306$ Fourier & Synthetic & $36$ &  $2.0$ & $2.3$ \\
\bottomrule
\end{tabularx}
\caption[EFC contrast improvement for the big rectangle.]{Maximum contrast gain when iterating on a $8\times 16 (\lambda/d)$ rectangle at a distance of $11 (\lambda/d)$ of the center with $306$ Fourier modes and a synthetic Interaction Matrix. The fourth respectively the last columns indicates the Root-Mean-Square contrast improvement in the area   a smaller area without the edges. }  
\label{tab:EFCgainBigRect}
\end{table}

\FloatBarrier

\relChapter{Discussion and Perspective}
\label{cha:EFCresDiscu}
The ultimate goal of high contrast imaging is the observation of an earth like planet in the habitable zone of its host star. In the case of $G2$-type star like the Sun with an habitable zone around $1\text{AU}$\marginpar{$1\text{AU}\approx 1.5\,10^{11}\text{m}$} the required contrast is $2\, 10^{-10}$. Although it depends slightly on the albedo and the orbital position of the planet. This contrast is still clearly not achievable yet. The first step would probably be to look around an $M$-type star because it requires a contrast of about only $10^{-8}$. However the habitable zone is situated much closer to the star around $0.1\text{AU}$ and it becomes challenging to have a good contrast at this kind of separations. By definition of the $\text{parsec}$\marginpar{$1\text{parsec}\approx 3.1\,10^{16}\text{m}$} the angular separation between a planet and its star separated by $1\text{AU}$ and $1 \text{parsec}$ away from Earth is $0.1\text{arcsec}$. This corresponds to $2.4(\lambda/d)$\marginpar{It is about $8$ pixels on IRDIS SPHERE.} for an instrument on the Very Large Telescope at a wavelength of $1.6\mu m$. In the case of SPHERE it already falls in the Coronagraphic halo. Besides this value is inversely  proportional to the distance of the stars. \marginpar{The closest star Proxima Centauri is at $1.3 \text{parsec}$.} This is therefore a job for Extremely Large Telescopes with resolution of around $0.006$ to $0.007\text{arcsec}$ in $J$ band. They would hopefully be able to image the habitable zone of nearby stars between $1$ to $10\text{parsec}$. SPHERE was never meant to observe Earth-like planets anyway but to observe giant extra-solar-planets at a distance of $1$ to $100 \text{AU}$ of their star and in a range of $5$ to $15 (\lambda/d)$ on the detector. A fine contrast is possible either with a relatively faint star or with a really bright young planet still emitting heat from its formation.

SPHERE has a raw coronagraphic speckle contrast of about $8\, 10^{-6}$. It might be possible to gain a factor $7$ with Spectral Differential Imaging \citeauthor{marois2006} \citep{marois2006} and another factor $10$ with Angular Differential Imaging. However this stays to be demonstrated during the next SPHERE commissionings\marginpar{For now the integration time previously used was too short.}. So far with the previous commissionings a contrast of $5\, 10^{-7}$ was demonstrated. The highest demonstrated gains with Spectral respectively Differential Imaging are $30$ respectively $10$. These techniques brings the contrast of SPHERE up to $1.1\, 10^{-7}$.

Electric Field Conjugation has just proven to be able to divide another factor $10$ bringing the contrast to $10^{-8}$. It is really getting closer to the ideal value. Although it is really a net gain only with very long exposure observations. Indeed the photon noise of the averaged speckles from the atmosphere would otherwise be dominant. The caveat of Electric Field Conjugation is that the position of the planet has to be known in advance for defining the dark hole. This method can therefore most efficiently be used for follow up studies once the planet has been discovered.\marginpar{Or one could be very lucky\dots} However Electric Field Conjugation has proven to be very efficient and it will very likely become a key element for improving the contrast on SPHERE.

Spectral or Angular Differential Imaging and Electric Field Conjugation are all three dealing with the same speckles. One could find surprising that the contrast improvement of each method can be multiplied when applied together. This is however validated by the experience. Even if it is not really intellectually convincing the reason is probably that each uses a different property of the speckles.

The size of the main mirror of the telescope is also important for improving the contrast. The gain corresponds to the square ratio of the size of the mirrors while keeping the quasi-static aberrations at a constant level. For example the gain of the Extremely Large Telescope in comparison with the Very Large Telescope is $\left(\frac{39\text{m}}{8\text{m}}\right)^2s=25$.

The last option to gain contrast is to go in space but then the costs have no equivalent. It is interesting to remember that the price of the James Webb Telescope is estimate at about $\$9$ billions in 2013\marginpar{When it was estimated at $\$0.5$ billions in 2007\dots}. The price of an Extremely Large Telescope is only estimated to $\$1$ billion even if it is true that these projects are hardly beginning to be built\dots This makes a perfect transition for another master thesis realised by Anthony Berdeu at the Institut de Recherche en Astrophysique et Plan\'{e}tologie, Toulouse, France where he studied the possibilities of Fresnel Lenses for observing from space in ultraviolet light. Reducing the wavelength would indeed improve the resolution of the instrument.

On the technical side of Electric Field Conjugation, the singular modes were not proven to be more efficient than the Fourier modes. This is probably due to the existence of other limitations such as the centring of the image. Indeed the experiments showed a saturation of the contrast gain when increasing the number of modes. The next step would be to verify the performance of the synthetic Interaction Matrix because of the mistake in the Matlab scripts. Then it would be interesting to understand better the limitations of the algorithm like the importance of the centring\marginpar{The centring was proven to have a great impact on \autoref{fig:EFCitMedRect-a}.}, the amplitude of the modes and the probes or the integration time. These would probably help to apply a good correction on larger area in order to maybe allow planets detection and not only characterization. To finish the whole algorithm should be hard-coded in SPHERE software for an easy use on sky.

\cleardoublepage

\label{app:bibliography} 


\addcontentsline{toc}{chapter}{\tocEntry{\bibname}}

\bibliographystyle{plainnat}

\bibliography{D:/latex/EFC/bibliography/bib_EFC} 

\chapter*{Conclusion}

\manualmark
\markboth{\spacedlowsmallcaps{Conclusion}}{\spacedlowsmallcaps{Conclusion}} 
\refstepcounter{dummy}

\addtocontents{toc}{\protect\vspace{\beforebibskip}} 
\addcontentsline{toc}{chapter}{\tocEntry{Conclusion}}

As a first conclusion Electric Field Conjugation was proven to be able to bring another order of magnitude to the contrast of SPHERE in a $5\times 10 (\lambda/d)$ rectangle area. This fulfils the primary objective of the thesis. Besides Electric Field Conjugation appeared to be a very simple method that can be applied on virtually any coronagraphic instruments with an Adaptive Optics system. Indeed the idea of measuring the Interaction Matrix removes the need of a model that can be tricky to parametrized even if it requires several hours to calibrate. Unexpectedly the model developed by the author for simulation purposes appeared to give very reasonable performance as well. These encouraging results advocate for further studies and automation of the method on SPHERE. It is very likely to become in the future a key feature on SPHERE for follow-up observations.

The second conclusion concerns the Adaptive Optics Facility. Its problem of low Strehl Point Spread Function has been answered by the negative: it is not caused by low order phase aberration in the optical path otherwise Phase Diversity would have somehow better worked.

\cleardoublepage 


\appendix

\part{Appendix Introduction} 
\relChapter{Context}
\relSection{Exo-planets Detection Methods}
\label{app:ExoDetecMeth}
Description of exoplanets detection method other than direct imaging.

\begin{figure}[tbp]
  \begin{center}
    \includegraphics[width=0.5\linewidth]{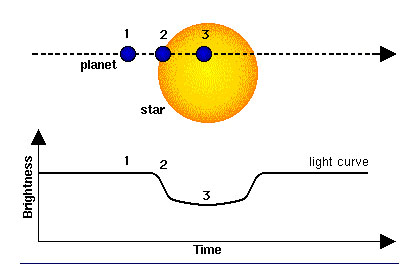}
    \caption[The transits Exo-Planets detection method.]{Description of the transits method for detecting exo-planets candidates. The planet goes through the line of sight of the observer with the star and the apparent star brightness drops. Source: Centre National d'\'{E}tudes spatiales (CNES).}
    \label{fig:transits}
  \end{center}
\end{figure}

\relSubsection{Transits}
 The most prolific method to detect exoplanets has clearly been the transits method. The principle is to monitor the star light intensity during a long period of time and then to detect periodic very faint drops of intensity or eclipses when a planet cross the line of sight as shown in \autoref{fig:transits}. The proportion of exoplanet orbits crossing the line of sight is very small but when observing a large number of stars it becomes significant. Besides the bigger the orbit radius, the lower the ratio of planets causing eclipses but also the further the planet the deeper is the intensity drop. These criteria introduce a bias in the observations.\marginpar{However this biais is known and can be corrected.} In addition one can only detect planets with orbital period a couple of times less than the total observation period. Several surveys have successfully discovered thousands of planets using space telescopes like Kepler or Corot. For instance Kepler identified over 2300 planet candidates in its first sixteen months of operation \citeauthor{batalhaKepler2013} \citep{batalhaKepler2013}. However transits are not considered has a proof of existence and other techniques like radial velocity have to be used to confirm any planet candidate.
 
 \relSubsection{Radial velocity}
 
\begin{figure}[tbp]
  \begin{center}
    \includegraphics[width=0.5\linewidth]{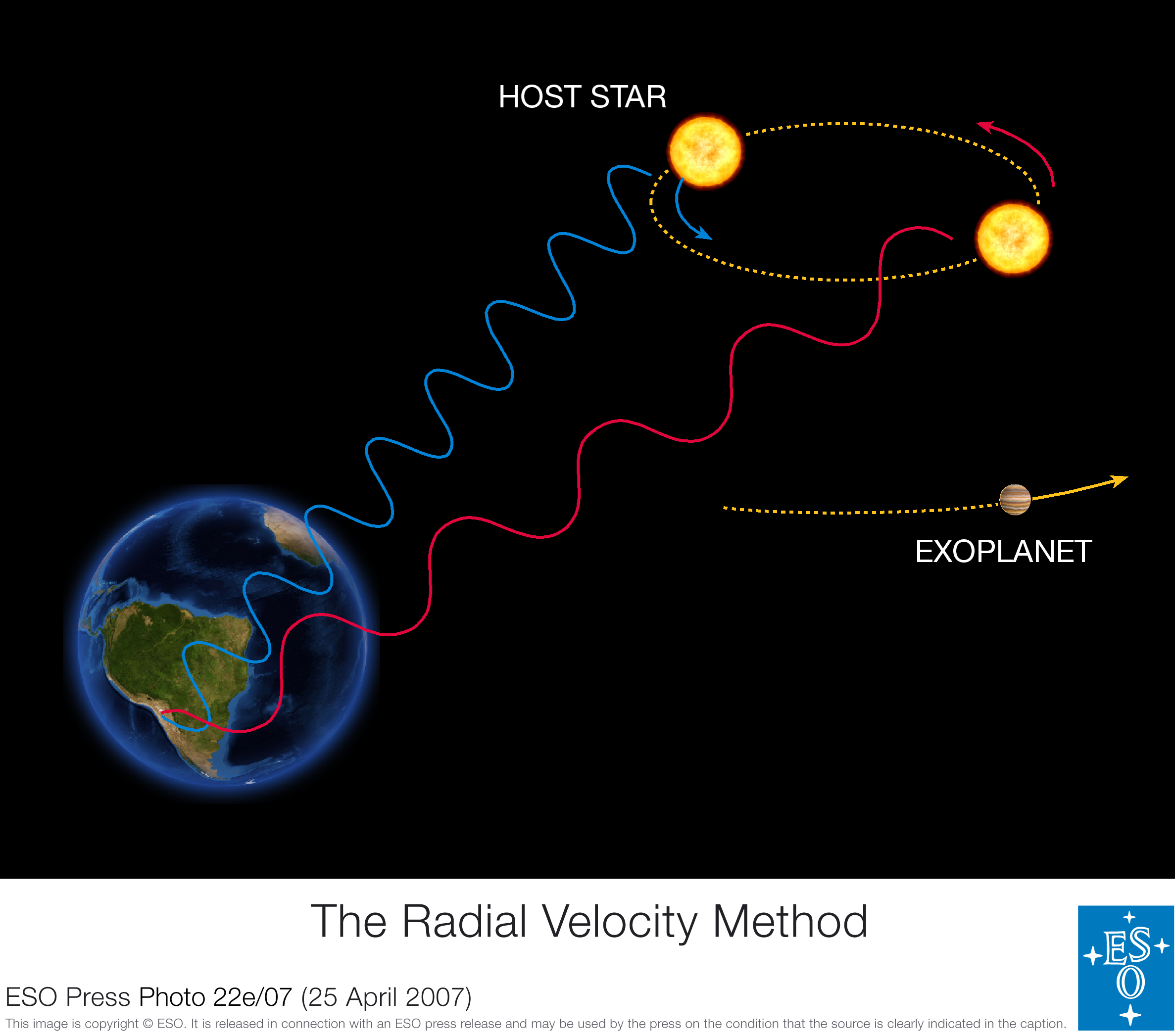}
    \caption[The Radial Velocity Exo-PLanets detection technique.]{Description of the radial velocity method for detecting exo-planets candidates. The planet gravitational force pulls the star backward and forward relative to the observer. This changes the relative velocity of the star which can be measured using Doppler effect. Source: European Southern Observatory (ESO).}
    \label{fig:Radial_Velocity}
  \end{center}
\end{figure}

Radial velocity studies the variation in Doppler shift of the star light spectrum as shown in \autoref{fig:Radial_Velocity}. The host star as well as the planet orbits around the center of mass of the binary system. Therefore an Earth observer will detect slight radial motion of the star if the orbit plane is not too inclined in regards to the line of sight. When the star moves away from the Earth the light shifts to longer wavelength while it shifts to shorter one if it moves toward us. The required instrument is a spectrograph with a stable reference light source. However only the radial velocity can be measured and that's why transit can be useful to constraint the orbit inclination. In the case of radial velocity it is easier to detect planets with higher orbit radii and bigger masses. The involved velocities of the stars lie around $12 \text{ms}^{-1}$ for a Jupiter-Sun system and only $1 \text{cm s}^{-1}$ for an Earth-Sun system. An example of such instrument is HARPS for High Accuracy Radial Velocity Planet Searcher mounted on ESO's La Silla $3.6\text{m}$ telescope. HARPS can detect velocities of $10 \text{cm s}^{-1}$ over one night and only $1 \text{m s}^{-1}$ over 60 days \citeauthor{wildiHARPS2011} \citep{wildiHARPS2011}.\marginpar{It is more difficult to be precise on long period because the calibrator needs to be stable enough.}
  
 \relSubsection{Others}
 Other less common methods exist like astrometry, timing or gravitational microlensing. 

\begin{description}
\item[Astrometry]  measures the movement of the centroid of the star in the sky. It has globally the same limitations as radial velocity. However it has the advantage to give the inclination and the eccentricity of the orbit which is directly related to the planet mass. As an indication $\alpha$ Centauri is at a distance of $1.34\text{pc}$ would move by only $7.34\text{mas}$ if a Jupiter were to orbit around it.
\item[Timing] measures the variation of a periodic signal emitting from the star for example pulsars. The required conditions are quite rare.
\item[Gravitational Microlensing] measures the light curve of star passing behind another star with a planet. The gravitational lensing increases the amount of light received and the presence of a planet appears as a very narrow secondary maximum. Such events are extremely rare.
\end{description}

For other clearer and extensive description Wikipedia stays your best friend:
 \begin{center}
 \url{http://en.wikipedia.org/wiki/Methods_of_detecting_exoplanets}
 \end{center}

\relSection{State of the Art: Phase Aberrations Correction}
\label{app:aberrCorrRefs}
A few references to different methods for correcting phase aberration are given here.

\marginpar{Thanks to Markus Kasper for giving this list!} One can cite COFFEE\footnote{COFFEE stands for COronagraphic PHase diversitY.} in \citeauthor{paulCOFFEE2013} \citep{paulCOFFEE2013}. COFFEE is an adaptation of Phase Diversity for coronagraphic systems. Speckle Nulling in \citeauthor{borde2006} \citep{borde2006} which is very similar to Electric Field Conjugation. The Phase Sorting Technique (PSI) \citeauthor{CodonaPSI2008} \citep{CodonaPSI2008} uses the random noise of the phase as interferometer to probe the quasi-static speckles. The random noise is measured with the Wavefront Sensor and this knowledge combined with short exposure frames of a science camera allows one to infer the speckles electric field. \citeauthor{codonaDOTF2013} \citep{codonaDOTF2013} developped a technique for non coronagraphic system called DOTF for Differential optical transfer function. It is meant to measure the complex field in the pupil using a couple of images differing by an artificial modification of the pupil. There is also \citeauthor{martinache2011} \citep{martinache2011} and many others\dots\marginpar{The author read only the abstracts of these papers which are not really self-explaining\dots}

\relChapter{Optics} 

\relSection{Fourier Optics}
According to Wikipedia because only Wikipedia gives clear and short definition of things.\marginpar{The author tried for days to understand a convincing demonstration of Fourier Optics without success. His problem is that Huygens-Fresnel principle is something that needs to be proven. And the demonstration if it exists might be too long or too complicated for the author\dots}
\begin{center}
\url{http://en.wikipedia.org/wiki/Fourier_optics}
\end{center}
\begin{quotation}
Fourier optics is the study of classical optics using Fourier transforms and can be seen as the dual of the Huygens-Fresnel principle.
\end{quotation}

The most famous reference book for optical science is \citeauthor{Born1980} \citep{Born1980}. From this book page 370 Huygens-Fresnel principle states that,
\begin{quotation}
Every point of a wave-front may be considered as a centre of a secondary disturbance which gives rise to spherical wavelets, and the wave-front at any later instant may be regarded as the envelope of these wavelets.
\end{quotation}

Huygens-Fresnel principle is an extension of the purely geometrical Huygens' Construction with the postulate of interfering secondary wavelets.

Fourier Optics states that the complex amplitude of the focal plane can be reconstructed from the complex amplitude of the pupil plane through a Fourier Transform. This is the Fraunhofer diffraction formula\marginpar{The fundamental formula of Fourier Optics is given page 385 equation (38).} applicable in the far field approximation. Far field approximation is equivalent to being at the focal plane of convergent lens put after the pupil aperture. Indeed any parallel rays before the lens have equal light path at the convergence point so it is like being infinitely far from the pupil. See Fresnel diffraction for near field. The Fraunhofer formula is defined as,
 
\begin{equation}
  \mathcal{D}(\xi,\eta)=C\iint_{\mathbb{R}^{2}} \! \mathcal{P}(x,y) e^{-2i\pi(x\xi+y\eta)} \, \mathrm{d} x \mathrm{d} y.
\end{equation}
 With $\mathcal{P}(x,y)$ the complex amplitude in the Pupil plane and $\mathcal{D}(\xi,\eta)$ the complex amplitude in the Detector plane. Besides $\mathcal{P}$ should be null outside the pupil mask defined by $\mathcal{P}_{m}(x,y)$. $x$ and $y$ are the spatial coordinates in the pupil normalized by the pupil mask diameter, $d$. $\xi$ and $\eta$ are the spatial coordinates in the focal plane normalized by $\lambda / d$\marginpar{The unit of $\xi$ and $\eta$ is sometimes called resel.}.
 
 Complete diffraction formula might include other terms especially wavelength dependency but things are kept simple here. For instance one can consider monochromatic light.

\relSection{Strehl Ratio}
\label{sec:strehlRatio}
The definition of the Strehl Ratio is the fraction between the peak intensity of the noisy Point Spread Function and the prefect one.
\begin{equation}
s=\frac{\mathcal{I}_\phi(0,0)}{\mathcal{I}_{\phi=0}(0,0)}
\end{equation}
With $s$ the Strehl Ratio, $\mathcal{I}(0,0)$ the peak intensity in the focal plane and $\phi$ the phase aberration.
The term perfect refers here to a Point Spread Function without speckles. \marginpar{A perfect Point Spread Function can have spider for instance.} Speckles indeed scatter light of the perfect Point Spread Function over the detector and tend to lower its peak value as the overall energy is conserved. However in practice it would be really difficult to know the absolute peak value to expect in the detector when there is no speckle. That's why the computation of the Strehl Ratio is usually performed by fitting a model to the image. The fitted model is considered to be the perfect Point Spread Function. There is no such thing as a standardized method to compute Strehl Ratio which makes it very difficult to compare Strehl Ratio from different sources.

The pupil complex amplitude is defined as,
\begin{equation}
\mathcal{P} = \mathcal{P}_m e^{i\phi},
\end{equation}
With $\mathcal{P}_m$ the pupil mask including the constant real amplitude. \marginpar{It means $\mathcal{P}_m$ does not necessarily equal $1$.}

It is also possible to compute the Strehl Ratio from the aberrations in the pupil phase. The most known expression is the approximated Marechal formula,
\begin{equation}
s=1-\sigma_{\phi}^2,
\end{equation}
With $\sigma_{\phi}$ the standard deviation of the phase aberration $ \sqrt{\mean{(\phi-\mean{\phi})^2}}$. \marginpar{$\mean{.}$ is the notation for the mean.}
Other related expressions are developed in \citeauthor{mahajan1982} \citep{mahajan1982} and summarized in \citep{mahajan1983}.
The demonstration is interesting and reproduced below. At least it allows to know exactly the approximations made for each formula.

First the Strehl Ratio can be expressed with a sole dependence to the phase aberration,
\begin{align}
s&= \frac{\mathcal{I}_\phi(0,0)}{\mathcal{I}_{\phi=0}(0,0)} = \abs{\frac{\mathcal{F}(P_\phi)(0,0)}{\mathcal{F}(P_{\phi=0})(0,0)}}^2,& \nonumber \\
 &= \abs{\frac{\int_{\mathbb{R}^2}\! \mathcal{P}_m(x,y) e^{i\phi(x,y)}e^{-2i\pi(0.x + 0.y)}\, \mathrm{d} x \mathrm{d} y}{\int_{\mathbb{R}^2}\! \mathcal{P}_m(x,y) e^{i0}e^{-2i\pi(0.x + 0.y)}\, \mathrm{d} x \mathrm{d} y}}^2,&\nonumber \\
 &= \abs{\frac{\int_{\mathbb{R}^2}\! \mathcal{P}_m(x,y) e^{i\phi(x,y)}\, \mathrm{d} x \mathrm{d} y}{\int_{\mathbb{R}^2}\! \mathcal{P}_m(x,y)\, \mathrm{d} x \mathrm{d} y}}^2,&  \equiv \abs{\mean{e^{i\phi}}}^2,  \nonumber \\
 &= \abs{e{-i\mean{\phi}}}^2 \abs{\frac{\int_{\mathbb{R}^2}\! \mathcal{P}_m(x,y) e^{i\phi(x,y)}\, \mathrm{d} x \mathrm{d} y}{\int_{\mathbb{R}^2}\! \mathcal{P}_m(x,y)\, \mathrm{d} x \mathrm{d} y}}^2 ,& \abs{e^{-i\mean{\phi}}}=1, \nonumber \\
 &= \abs{\frac{\int_{\mathbb{R}^2}\! \mathcal{P}_m(x,y) e^{i(\phi(x,y)-\mean{\phi})}\, \mathrm{d} x \mathrm{d} y}{\int_{\mathbb{R}^2}\! \mathcal{P}_m(x,y)\, \mathrm{d} x \mathrm{d} y}}^2 ,& \nonumber \\
s&= \abs{\mean{e^{i(\phi-\mean{\phi})}}}^2.&
\end{align}
Besides,
\begin{equation}
e^{i(\phi-\mean{\phi})} = \cos(\phi-\mean{\phi})+i\sin(\phi-\mean{\phi}),
\end{equation}
So,
\begin{equation}
s=\abs{\mean{\cos(\phi-\mean{\phi})}}^2 + \abs{\mean{\sin(\phi-\mean{\phi})}}^2
\end{equation}
However $(\phi-\mean{\phi})$ can be assumed small so that $\cos(\phi-\mean{\phi})\simeq1$ and $\sin(\phi-\mean{\phi})\simeq0$

Therefore one gets,
\begin{align}
s & \geq \abs{\mean{\cos(\phi-\mean{\phi})}}^2 \nonumber \\
  & \geq \abs{\mean{1-\frac{(\phi-\mean{\phi})^2}{2} + o((\phi-\mean{\phi})^3)}}^2 \nonumber \\
  & \geq \abs{1-\frac{1}{2}\mean{(\phi-\mean{\phi})^2} + \mean{o((\phi-\mean{\phi})^3)}}^2 \nonumber \\
s & \geq \abs{1-\frac{1}{2}\mean{(\phi-\mean{\phi})^2}}^2 + \mean{o((\phi-\mean{\phi})^3)}
\end{align}

Which gives directly the Marechal equation,
\begin{equation}
s_1 \approx \left(1-\frac{1}{2}\sigma_{\phi}^2\right)^2
\label{eq:marechal}
\end{equation}
Then a Taylor expansion gives,
\begin{equation}
s_2 \approx 1-\sigma_{\phi}^2
\label{eq:marechalApprox}
\end{equation}
Which is also the Taylor expansion of the exponential so one could also tell,
\begin{equation}
s_3 \approx e^{-\sigma_{\phi}^2}
\label{eq:marechalExp}
\end{equation}

\relSection{Adaptive Optics}
\label{app:AO}
Description of Adaptive Optics System.

\relSubsection{Overview-Speckles}
The theoretical width of the impulse response \marginpar{Also called PSF for Point Spread Function.} of the optical system is proportional to $\frac{\lambda}{D}$ where $\lambda$ is the wavelength and $D$ is the diameter of the mirror. However other perturbations tend to spread the light around a bigger area. For example the atmospheric turbulence lowers the resolution from at least an order of magnitude and it can be a lot more. This is due to the presence of so called-speckles. Adaptive Optics is a technology allowing the real time correction of the wavefront distortion. The most recent Adaptive Optics system allows a resolution close to the limit of diffraction. 
\paragraph{High Contrast Imaging} Even with that correction the difference in brightness of the two objects makes the detection challenging and this is independent from the turbulence. The solution is to use a coronagraph which blocks the light of the star allowing longer exposure time without saturation of the detector. A coronagraph and a powerful Adaptive Optics system are what characterize an high contrast imaging instrument. However another type of speckles remains called quasi-static speckles.

\relSubsection{Wavefront Sensing}

\begin{figure}[tbp]
  \begin{center}
    \includegraphics[width=\linewidth]{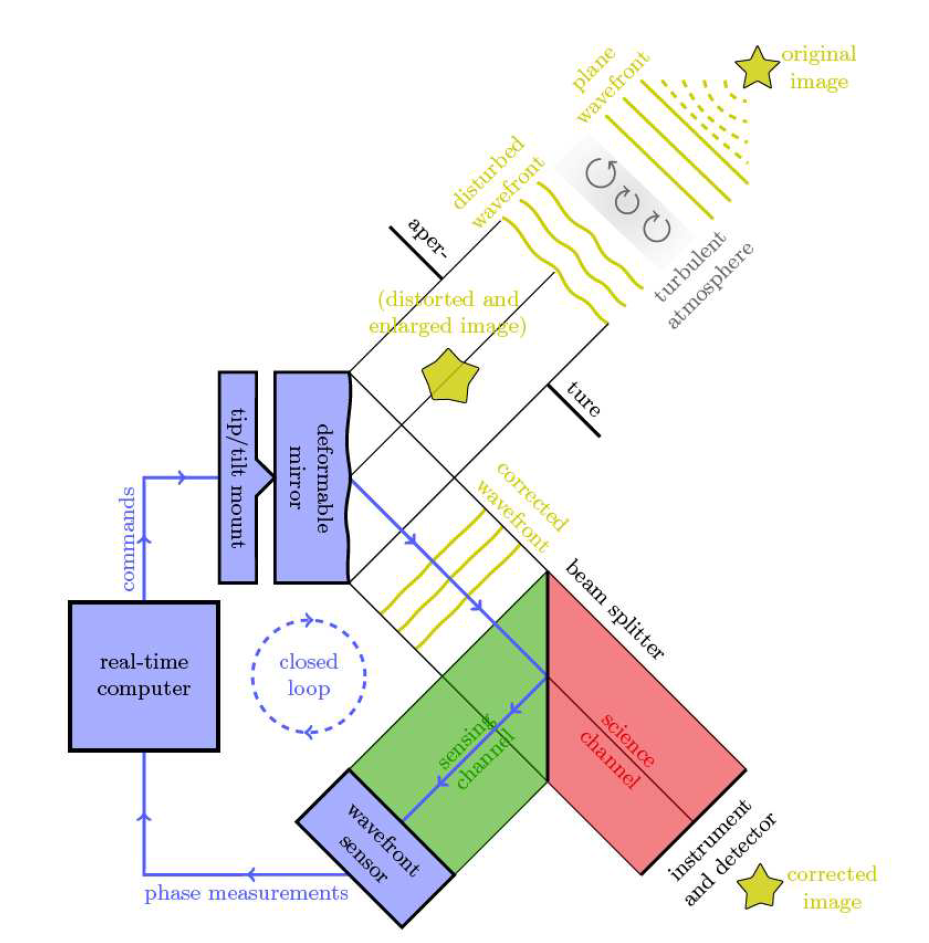}
    \caption[Basic architecture of an Adaptive Optics system.]{Working principle of an Adaptive Optics system. Source: \citeauthor{loose2011} \citep{loose2011}}
    \label{fig:AO}
  \end{center}
\end{figure}

An Adaptive Optics system is basically composed of a Deformable Mirror followed by a Wavefront Sensor as shown in figure \autoref{fig:AO}. The light beam is split into a science path and a sensing path where the Wavefront Sensor is positioned. The Wavefront Sensor measures the slopes of the wavefront and then the control loop translates these slopes into shapes for the DM. The frequency at which the correction works is of the order of the $kHz$.

\begin{figure}[tbp]
  \begin{center}
    \includegraphics[width=0.5\linewidth]{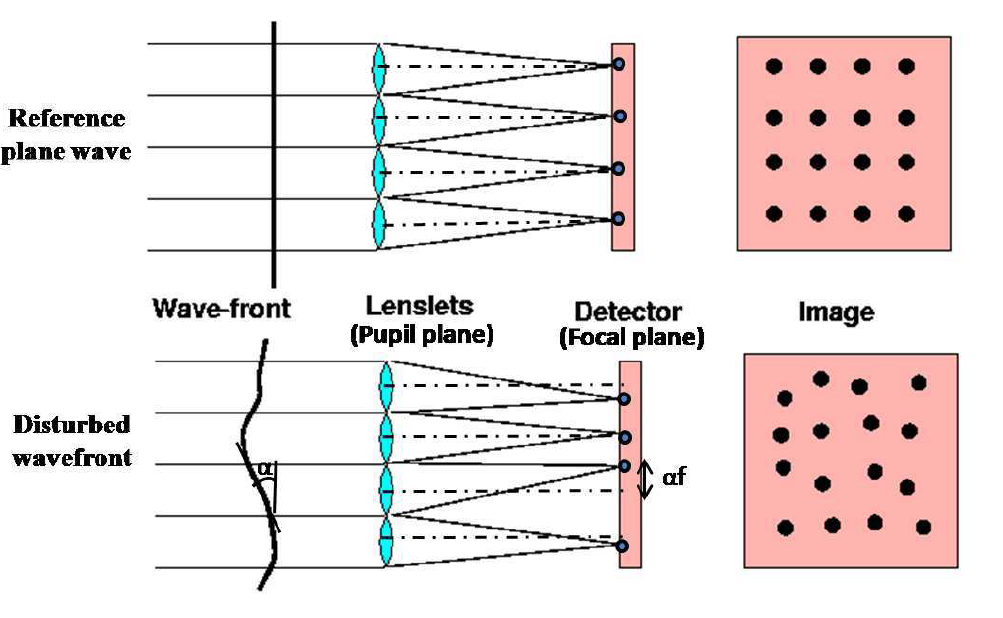}
    \caption[Shack-Hartmann sensor working principle.]{Working principle of a Shack-Hartmann sensor. Source: \citeauthor{allercarpentierPHD2011} \citep{allercarpentierPHD2011}}
    \label{fig:SHsensor}
  \end{center}
\end{figure}

The Schack-Hartmann sensor is the most common type of Wavefront Sensor currently used. Its principle is described in figure \autoref{fig:SHsensor}. The main component is an array of small lenslets which image the pupil plane. Each lenslet will focus on a detector creating a regular grid if the wavefront is flat. The direction of propagation of a wave is perpendicular to the wavefront so if some deformations are induced the lenslet will focus at a shifted point. If $\alpha$ is the slope of the wavefront, the distance of the focal  point to its nominal position in the detector will be $\delta = \alpha f$ where f is the focal distance of the lenslet. By measuring the shifts it is possible to infer the slope of the wavefront in front of each lenslet.

\relSubsection{Non-Common-Path}
\label{app:NCPA}
The Non-Common-Path corresponds to the sensing and the science channel represented in \autoref{fig:AO}. The Non-Common-Path-Aberrations can't be corrected by the closed-loop of the Adaptive Optics System. Indeed all aberrations included in the sensing path will be corrected however these aberrations were not part of the path taken by the light going to the science camera. Therefore the correction is actually adding non existing aberrations in the science detector. Regarding the science channel it is more obviously not seen by the Wavefront Sensor and therefore not corrected.

\relSubsection{Calibration and correction}
A calibration is required in order to know what shape to apply for correcting a known deformation. The calibration product is an Interaction Matrix formed by several measurements vectors $C_{n}$ of the slopes resulting from a given shape of the Deformable Mirror. Considering a set of different shapes of the Deformable Mirror one can record the measurements in a matrix $C$ so that,

\begin{equation}
C=(C_{1},C_{2},C_{3},\dots,C_{N}).
\label{eq:interact_matrix}
\end{equation}

A reference $Z_{n}$ with an undisturbed wavefront is also taken for each measurement and subtracted from it. The corrected Interaction Matrix is $I$ where,

\begin{equation}
I=C-Z.
\label{eq:correc_interact_matrix}
\end{equation}

This matrix relates a shape of the Deformable Mirror with the resulting deformation of the wavefront. The shape of the Deformable Mirror is expressed in a setting vector $S$ and the slopes of the wavefront are recorded in the vector $M$.

\begin{equation}
IS=M.
\label{eq:ao_correction}
\end{equation}

An Adaptive Optics correction works the other way around and equation \autoref{eq:ao_correction} needs to be inverted. We measure the slopes $M$ and we want to infer the Deformable Mirror setting $S$. The system is not necessarily square so that a Pseudo-Inverse can be used. $R=I^{-1}$ is called the Reconstruction Matrix.

\relSection{Coronagraph}

\relSubsection{Perfect Coronagraph}
A perfect coronagraph is able to remove all the light of the perfect on-axis Point Spread Function. However it would not be able to remove the coherent speckles of this Point Spread Function.\marginpar{Some people call coherent light the perfect Point Spread Function to oppose with the speckles. However it is very confusing because the speckles have coherent light with the Point Spread Function so the author is not going to use this term.} \citeauthor{cavarroc2005} \citep{cavarroc2005} uses a formula based on the Strehl Ratio to compute the effect of a perfect coronagraph in the pupil plane.
\begin{equation}
\mathcal{P}_C = \mathcal{P}_m \left( \sqrt{s}-e^{i\phi} \right)
\label{eq:coroCavarroc}
\end{equation}
With $\mathcal{P}_C$ the complex amplitude in the pupil plane after the coronagraph, $\mathcal{P}=\mathcal{P}_m e^{i\phi}$ the complex amplitude in the pupil before the coronagraph and $s$ the Strehl Ratio. The Strehl Ratio can be computed using the standard deviation of the phase in the pupil as shown in \autoref{sec:strehlRatio}.

The Strehl Ratio can be seen as the ratio of two perfect Point Spread Functions. The peak value of the one on top of the fraction would be equal to the peak value of the Point Spread Function with speckles. The Strehl Ratio can therefore be seen as a fraction of the pupil real amplitude,
\begin{equation}
\sqrt{s} = \frac{\mathcal{P}_{m,phi}}{\mathcal{P}_{m,phi=0}}
\end{equation}
Then \eqref{eq:coroCavarroc} can be written as,
\begin{equation}
\mathcal{P}_C = \mathcal{P}_{m,phi} - \mathcal{P}_m e^{i\phi}
\end{equation}
where the subtraction of the perfect Point Spread Function from the noisy one is clear. \marginpar{However the author doesn't understand why one doesn't simply write $ \mathcal{P}_{m} - \mathcal{P}_m e^{i\phi}$}

\begin{figure}[bth]
	\centering	
   	\includegraphics[width=\linewidth]{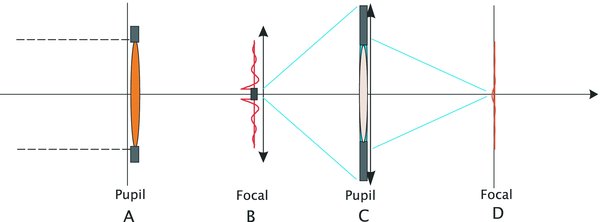}
   	\caption[Architecture of Apodized-Lyot-Coronagraphs.]{\label{fig:APLCArch} Architecture of Apodized-Lyot-Coronagraphs as a series of pupil and focal planes. Source: \citeauthor{soummer2009} \citep{soummer2009}.}
\end{figure}

\begin{figure}[bth]
	\centering	
   	\includegraphics[width=\linewidth]{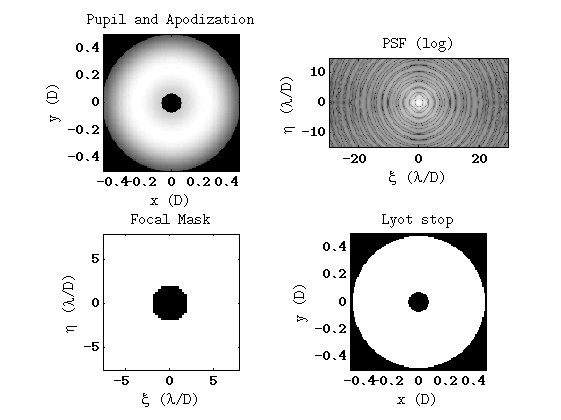}
   	\caption[Example of Apodized-Lyot-Coronagraph of SPHERE.]{\label{fig:APLC} Apodized-Lyot-Coronagraph of SPHERE for the $4(\lambda/d)$ diameter focal mask and the $1.6\mu m$ wavelength.}
\end{figure}

\relSubsection{Apodized-Lyot-Coronagraph}
\label{app:APLCdef}
The Apodized-Lyot-Coronagraph is the most advanced type of coronagraph at the time of writing. It consists in a series of focal and pupil planes like in \autoref{fig:APLCArch}. \autoref{fig:APLC} shows the different planes of an example of Apodized-Lyot-Coronagraph.

The complex amplitude at the entrance of the telescope in the first pupil plane is defined as $\mathcal{P}_0$.

\relSubsubsection{Apodizer}
The First pupil plane  called $A$ in \autoref{fig:APLCArch} includes the apodizer which helps removing the rings of the Point Spread Function. It modifies the amplitude profile of the pupil from a hat function to something closer to a Gaussian. \marginpar{The Fourier Transform of a Gaussian is a Gaussian.} Usually it is a mask with a variable density of black dots to partially absorb the light.

The complex amplitude in the pupil plane becomes,
 \begin{equation}
 \mathcal{P} = \mathcal{A}*\mathcal{P}_0,
 \end{equation}
 With $\mathcal{A}$ the apodization function.

The apodizer is not the key feature of the coronagraph but it improves the efficiency of the Lyot-Coronograph.

For example \citeauthor{Soummer2004} \citep{Soummer2004} gives a method to optimize an apodizer for a given pupil even with a central obscuration.

\relSubsubsection{Focal Mask}
The focal mask is an obstruction disk at the center of the focal plane $B$ in \autoref{fig:APLCArch}. If one wants only to separate the light it can also be a mirror with hole at the center.

The complex amplitude in the focal plane becomes here,
\begin{equation}
\mathcal{D}=\mathcal{M}_F\mathcal{F}(\mathcal{A}*\mathcal{P}_0),
\end{equation}
With $\mathcal{M}_F$ the focal mask function.

The size of the disk is wavelength dependent. The bigger the wavelength the bigger the width of the Point Spread function and the bigger the focal mask.

\relSubsubsection{Lyot Stop}
The Lyot stop is a mask in the pupil plane which blocks the light at the edges of the pupil $C$ in \autoref{fig:APLCArch}. It helps reducing the diffracted light from the Focal Mask. Indeed the presence of the focal mask tend to push the light at the edge of the pupil. It can be demonstrated using Fourier Transforms.

\cleardoublepage

\label{app:bibliography} 


\addcontentsline{toc}{chapter}{\tocEntry{\bibname}}

\bibliographystyle{plainnat}

\bibliography{D:/latex/ESO_PFE/masterThesis/bibliography/bib_intro} 

\part{Appendix Phase Diversity}

\chapter{PD Theory}

\relSection{Model}
\label{app:Model} 

\relSubsection{Intensity}
 The aberrations and the defocus are modelled thanks to an additional phase in the complex amplitude of the pupil.
 
 The focused and defocused images will be computed using Fourier Optics,\marginpar{Curly font is used for continuous variables and straight font will be used after discretization.}
\begin{equation}
  \begin{dcases*}
        \mathcal{H}^{f}(\xi,\eta)= \abs*{\mathcal{F}\left( \mathcal{P}_{m}(x,y) e^{i\phi(x,y)}\right)}^{2}   & focused,\\
        \mathcal{H}^{d}(\xi,\eta)= \abs*{\mathcal{F}\left( \mathcal{P}_{m}(x,y) e^{i(\phi(x,y)+\delta(x,y))}\right)}^{2} & defocused.
  \end{dcases*}
  \label{eq:PSF}
\end{equation}
 With $\mathcal{H}$ the intensity in the focal plane, $\mathcal{F}$ the Fourier Transform, $\phi$ the phase of the aberrations and $\delta$ the defocused expressed in radian.

\relSubsection{Extended source}
\label{sec:extendedSource}
Equation~\eqref{eq:PSF} is expressing the Point Spread Function of the system but it was stated that Phase Diversity models extended light source which is called object. Assuming that the object $\mathcal{O}(\xi,\eta)$ is not coherent the real image results from a convolution between the Point Spread Function and the object. The convolution of two functions is noted $f\ast g$.

\begin{equation}
  \begin{dcases*}%
        \mathcal{G}^{f}(\xi,\eta)= \left( \mathcal{H}^{f}\ast\mathcal{O}\right)(\xi,\eta)  = \abs*{\mathcal{F}\left( \mathcal{P}_{m} e^{i\phi}\right)}^{2}\ast\mathcal{O}   & focused,\\
        \mathcal{G}^{d}(\xi,\eta)= \left( \mathcal{H}^{d}\ast\mathcal{O}\right)(\xi,\eta) = \abs*{\mathcal{F}\left( \mathcal{P}_{m} e^{i(\phi+\delta)}\right)}^{2}\ast\mathcal{O} & defocused.
  \end{dcases*}
  \label{eq:PSFobj}
\end{equation}

\relSubsection{Complete model}
 The measured intensities will be noted $\mathcal{I}^{f}$ and $\mathcal{I}^{d}$ for the focused respectively defocused images. Here the spatial dependency is forgotten in the notation and replaced by the phase dependency one gets,

\begin{equation}
  \begin{dcases*}%
        \mathcal{I}^{f} = \mathcal{G}^{f}(\Phi) + n^{f} = \mathcal{H}^{f}(\phi)\ast\mathcal{O} + n^{f} & focused,\\
        \mathcal{I}^{d} = \mathcal{G}^{d}(\Phi) + n^{d} = \mathcal{H}^{d}(\phi)\ast\mathcal{O} + n^{d}& defocused,
  \end{dcases*}
  \label{eq:modelCont}
\end{equation}

With $n^{f}$ and $n^{d}$ two random variables. Their laws are assumed to be centered gaussians so it is assumed that the background has been subtracted. Besides the photon noise is also ignored.

\relSubsection{Discrete form}
The expressions below \eqref{eq:modelCont} can be expressed in discrete form. Assuming periodicity of the images the convolution can take the form of a matrix multiplication,\marginpar{In order to see what error these approximations create in Fourier space one could use the spectral window of the windowing with a hat function and of the discretization with a dirac comb.}
 
\begin{equation}
  \begin{dcases*}%
        I^{f} = H^{f}(\Phi)\times O + N^{f} & focused,\\
        I^{d} = H^{d}(\Phi)\times O + N^{d}& defocused,
  \end{dcases*}
  \label{eq:modelCont}
\end{equation}
In this expression $O$ and $I$ are image-size matrices put in vector form $N^2\times1$. $H$ is however a real matrix but with a width equal to the total number of pixel of the image $N^2\times N^2$. $N^{f}$ and $N^{d}$ are random Gaussian vectors $N^2\times N^2$.

Besides the matrices $H^{f}$ and $H^{d}$ are Circulant-bloc-Circulant\footnote{Circulant means Toeplitz where each row is a circular permutation of the previous one. A bloc-Circulant matrix is a bloc-matrix where the blocs of the matrix follow the 
circular permutation rule. Circulant-bloc-Circulant matrix is a bloc-circulant matrix where the blocs are also circulant. Toeplitz means a matrix with constant diagonals. It can be also the matrix of a convolution however without the periodicity approximation.}\marginpar{The expression of $H$ is given in \autoref{app:convoMat}} as the periodicity in space of the images is assumed. This is going to be useful when going in Fourier space. The assumption of periodicity is reasonable if the images are big enough and tend toward zero near the edges. See \autoref{app:circulDiag} for interesting properties of the circulant matrices especially regarding Fourier Transform.

\relSubsection{Zernike Polynomials}
\label{app:EFCzerPol}
For the estimation the aberration phase is expended on modes and these modes are Zernike polynomials in this case. Zernike polynomials are here indexed by a single variable. The phase gives,
\marginpar{The coefficients $a_{1}$, $a_{2}$ and $a_{3}$ have been ignored as they are respectively the piston, the tip and the tilt.}
\begin{equation}
\phi(x,y) = \sum_{k=4}^{M+3}a_{k}\mathcal{Z}_{k}(x,y).
\end{equation}

In discrete form it takes the form of a matrix multiplication where $Z$ transforms a vector of coefficients into the discrete phase matrix,
\begin{equation}
\Phi = Z\times A,
\end{equation}
With $A=(a_{4},\dots,a_{M+3})$ the coefficient vector.

\relSection{Joint Maximum \textit{A Posteriori}}
The definition of the problem can be found in \autoref{sec:JMAPprobDef}.

\relSubsection{Bayesian Approach}
\label{app:JMAPBaysApproach}
The likelihood of the measurements is defined by,\marginpar{This is the probability of the noise and the author sees the problem as what value of the noise is needed to give the right measurements for given parameters.}
\begin{equation}
  \begin{dcases*}
        f\left(I^{f}|A,O\right) = \frac{1}{\sqrt{2\pi}^{N^2}\sigma^{N^2}}e^{-\frac{1}{2\sigma^{2}}(I^{f}-H^{f}O)^\mathsf{T}(I^{f}-H^{f}O)},\\
        f\left(I^{d}|A,O\right) = \frac{1}{\sqrt{2\pi}^{N^2}\sigma^{N^2}}e^{-\frac{1}{2\sigma^{2}}(I^{d}-H^{d}O)^\mathsf{T}(I^{d}-H^{d}O)}.
  \end{dcases*}
  \label{eq:probaMeasurements}
\end{equation}

The Maximum \textit{A Posteriori} estimator is a Bayesian approach with \textit{a priori} information on the unknowns. The \textit{a priori} laws are also gaussians with,
\begin{equation}
  \begin{dcases*}
        f\left(A\right) = \frac{1}{\sqrt{2\pi}^{M}\sqrt{\abs*{R_{a}}}}e^{-\frac{1}{2}A^\mathsf{T}R_{a}^{-1}A},\\
        f\left(O\right) = \frac{1}{\sqrt{2\pi}^{N^2}\sqrt{\abs*{R_{o}}}}e^{-\frac{1}{2}(O-O_m)^\mathsf{T}R_{o}^{-1}(O-O_m))},
  \end{dcases*}
  \label{eq:probaApriori}
\end{equation}
With $R_{a}$ respectively $R_{o}$ the covariance matrix of the random vector $A$ respectively $O$ and $O_m$ the mean object.

All the tools are now defined to pose the real joint estimation. In a Bayesian approach the idea is to maximize the probability of the parameters knowing the results of the measurements. The density of probability of the parameters is,
\begin{equation}
f\left(A,O|I^{f},I^{d}\right) = \frac{f\left(A,O,I^{f},I^{d}\right)}{f\left(I^{f},I^{d}\right)} = \frac{f\left(I^{f},I^{d}|A,O\right)f\left(A,O\right)}{f\left(I^{f},I^{d}\right)}.
\end{equation}
Besides $N^{f}$ and $N^{d}$ (therefore $I^{f}|A,O$ and $I^{d}|A,O$) respectively $A$ and $O$ are independent so one can write,
\begin{equation}
f\left(A,O,I^{f},I^{d}\right) = f\left(I^{f}|A,O\right)f\left(I^{d}|A,O\right)f\left(A\right)f\left(O\right).
\end{equation}
The expression of each probability can found in equations \eqref{eq:probaMeasurements} and \eqref{eq:probaApriori} which give,
\begin{align}
 f\left(A,O,I^{f},I^{d}\right)&= \frac{1}{\sqrt{2\pi}^{N^2}\sigma^{N^2}}e^{-\frac{1}{2\sigma^{2}}(I^{f}-H^{f}O)^\mathsf{T}(I^{f}-H^{f}O)} \nonumber \\
 &* \frac{1}{\sqrt{2\pi}^{N^2}\sigma^{N^2}}e^{-\frac{1}{2\sigma^{2}}(I^{d}-H^{d}O)^\mathsf{T}(I^{d}-H^{d}O)} \nonumber \\
 &* \frac{1}{\sqrt{2\pi}^{M}\sqrt{\abs*{R_{a}}}}e^{-\frac{1}{2}A^\mathsf{T}R_{a}^{-1}A} \nonumber \\
 &* \frac{1}{\sqrt{2\pi}^{N^2}\sqrt{\abs*{R_{o}}}}e^{-\frac{1}{2}(O-O_m)^\mathsf{T}R_{o}^{-1}(O-O_m))}
 \label{ref:probaParameters}
\end{align}

The Joint Maximum \textit{A Posteriori} is given by the maximization of \eqref{ref:probaParameters}. The estimation will be written with a tilde,
\begin{align}
\left(\tilde{A},\tilde{O}\right) &= \underset{A,O}{\text{argmax}}\, f\left(A,O,I^{f},I^{d}\right) \nonumber \\
\left(\tilde{A},\tilde{O}\right) &= \underset{A,O}{\text{argmin}} \underbrace{-\text{Ln}\left(f\left(A,O,I^{f},I^{d}\right)\right)}_{L_{JMAP}}.
\end{align}

The complete expression of $J_{JMAP}$ is given by,\marginpar{The expression for the Maximum Likelihood is simply the third line if ignoring the constants}
\begin{align}
L_{JMAP} &= \frac{N^2}{2}\text{Ln}2\pi + \frac{N^2}{2}\text{Ln}2\pi + \frac{M}{2}\text{Ln}2\pi + \frac{N}{2}\text{Ln}2\pi & \text{Big constant} \nonumber \\
 &+ N^2\text{Ln}\sigma^2 + \frac{1}{2}\text{Ln}\abs*{R_a} + \frac{1}{2}\text{Ln}\abs*{R_o} & \text{Uncertainties} \nonumber \\
 &+ \frac{1}{2\sigma^{2}}(I^{f}-H^{f}O)^\mathsf{T}(I^{f}-H^{f}O) & \dots \nonumber \\
 &+  \frac{1}{2\sigma^{2}}(I^{d}-H^{d}O)^\mathsf{T}(I^{d}-H^{d}O) & \dots\text{Model} \nonumber \\
 &+ \frac{1}{2}A^\mathsf{T}R_{a}^{-1}A + \frac{1}{2}(O-O_m)^\mathsf{T}R_{o}^{-1}(O-O_m)) & \text{\textit{A priori}}
 \label{eq:LJMAP}
\end{align}
Only the last two lines depend on the unknowns. The second line is constant unless one wants to do hierarchical Bayesian approach.\marginpar{Hierarchical Bayesian approach means "also estimating the hyper-parameters" which are the parameters defining the \textit{a priori} knowledge} However it seems that hierarchical approach is not possible in the joint estimation but possible in the Maximum \textit{A Posteriori} approach defined in \autoref{app:MAP}. It is also possible to get a closed-form expression of the derivative of the criterion but is is a bit more complicated to infer. However the expression of the derivative is given in \citeauthor{blancJOSAA2003} \citep{blancJOSAA2003}.

\paragraph{Point Source} Considering a point source in \eqref{eq:LJMAP} would mean setting $O=(1\, 0\, \dots\, 0)^\mathsf{T}$ and removing the terms with covariance matrix $R_o$.

\relSubsection{Object Estimation}
\label{app:JMAPobjEsti}
In order to get the closed-form expression of the estimated object when all the other parameters are constant one can find the point where the derivative of $L_{JMAP}$ is null.\marginpar{This could be done also by expressing the problem in canonic form and then computing the pseudo-inverse using matrix multiplication only.} Remembering that if $\alpha = X^\mathsf{T}AX$ with $A$ symmetric then  $\frac{\mathrm{d}\alpha}{\mathrm{d}x}=2AX$ and $\frac{\mathrm{d}AX}{\mathrm{d}x}=A$ one gets,
\begin{align}
\frac{\mathrm{d}L_{JMAP}}{\mathrm{d}O} &= \frac{1}{\sigma^2} F^{f\mathsf{T}}H^f O - \frac{1}{\sigma^2} H^{f\mathsf{T}}I^{f} + \frac{1}{\sigma^2} H^{d\mathsf{T}}H^{d}O \nonumber \\
&- \frac{1}{\sigma^2} H^{d\mathsf{T}}I^{d} + R_{o}^{-1}O - R_{o}^{-1}O_{m}.
\end{align}
Then,
\begin{align}
& \frac{\mathrm{d}L_{JMAP}}{\mathrm{d}O}\left(\tilde{O}\right) = 0  \nonumber \\
\Leftrightarrow &  \tilde{O}\left(H^{f\mathsf{T}}H^{f} + H^{d\mathsf{T}}H^{d} + \sigma^2 R_o^{-1} \right) = H^{f\mathsf{T}}I^{f} + H^{d\mathsf{T}}I^{d} + \sigma^2 R_o^{-1}O_m \nonumber \\
\Leftrightarrow & \tilde{O} = \left(H^{f\mathsf{T}}H^{f} + H^{d\mathsf{T}}H^{d} + \sigma^2 R_o^{-1} \right)^{-1} \left(  H^{f\mathsf{T}}I^{f} + H^{d\mathsf{T}}I^{d} + \sigma^2 R_o^{-1}O_m \right),
\label{eq:dLJMAPdOnull}
\end{align}
gives the estimated object in closed-form. Therefore instead of minimizing $L_{JMAP}$ through two variables it is now possible to replace $O$ by its estimate $\tilde{O}$ and minimizing only on the aberration $A$.

\relSubsection{Fourier Space}
\label{app:JMAPfour}

\relSubsubsection{Criterion}
It was mentioned in \autoref{sec:extendedSource} that the matrices of the convolution $H^{f}$ and $H^d$ are Circulant-bloc-Circulant. It is here also assumed that the covariance matrix of the object $R_o$ is Circulant-block-Circulant.\marginpar{This assumption means that the correlation between two pixels depends on the relative position of the pixels. It means that the correlation pattern with the neighbouring pixels is the same for all pixels.} Therefore all these matrices can be diagonalized in Fourier space as,

\begin{equation}
  \begin{dcases*}
        H^f = F^{-1}\text{diag}\left( \widehat{h^f}_{kl} \right)F   & Convolution of the focused image,\\
        H^d = F^{-1}\text{diag}\left( \widehat{h^d}_{kl} \right)F   & Convolution of the defocused image,\\
        R_o = F^{-1}\text{diag}\left( s_{o,kl} \right)F   & \textit{a priori} on the object,
  \end{dcases*}
  \label{eq:fourDecompo}
\end{equation}
With $F$ the Discrete Fourier Transform matrix and $\widehat{h}_{kl}$ the coefficients of the Discrete Fourier Transform of the simulated image formed by the pixels $h_{ij}$ where $ij$ are the indices of the pixel. Besides $\text{diag}\left( \widehat{h^f}_{kl} \right)$ stands for the diagonal matrix formed by the values $\widehat{h^f}_{kl}$ when the indices $h$ and $k$ vary. Again it is considered as a vector even if there are two indices. Note by the way that the indices are still written here but they are going to be forgotten soon. Any time a lower case $h$,$i$,$o$ or $s$ is used the indexation with $k,l$ is assumed. Besides the letters $k,l$ are usually used in Fourier space while $i,j$ should be used in real space. The proof of \eqref{eq:fourDecompo} can be found in the appendix \autoref{app:circulDiag}.

Most of the term of \eqref{eq:LJMAP} can now be simplified using the decomposition in \eqref{eq:fourDecompo}.
Noting that the model model can be written,
\begin{align}
\left(I^f-H^f O\right) &= F^{-1} \left(F i^f-\text{diag}\left(\widehat{h^f}_{kl}\right)F O\right)  \nonumber \\
\left(I^f-H^f O\right) &= F^{-1}\left(\widehat{i^f}-\text{diag}\left(\widehat{h^f}_{kl}\right)\widehat{O}\right),
\end{align}
and remembering that $F^{-1}=F^\mathsf{T}$ with $\mathsf{T}$ the conjugate transposition one gets,
\begin{align}
\left(I^f-H^f O\right)^\mathsf{T}\left(I^f-H^f O\right) &= \left(I^f-\text{diag}\left(\widehat{h^f}_{kl}\right)\widehat{O}\right)^\mathsf{T} \underbrace{F^{-1\mathsf{T}}F^{-1}}_{I_{N}} \left(I^f-\text{diag}\left(\widehat{h^f}_{kl}\right)\widehat{O}\right), \nonumber \\
\left(I^f-H^f O\right)^\mathsf{T}\left(I^f-H^f O\right) &= \sum_{k,l=1}^{N} \abs*{\widehat{i^f}_{kl}-\widehat{h^f}_{kl}\widehat{o}_{kl}}^2,
\end{align}
With $\widehat{O} = \left[\widehat{o}_{kl}\right]$ and $\widehat{I} = \left[\widehat{i}_{kl}\right]$ both in vector form even if there are two indices.
It was done for the focused image but it is exactly the same for the defocused one.

Then the term of the object,
\begin{align}
\left(O-O_m\right)^\mathsf{T}R_o^{-1}\left(O-O_m\right) &= \left(O-O_m\right)^\mathsf{T}F^{-1}\text{diag}\left( \frac{1}{s_{o,kl}} \right)F\left(O-O_m\right) \nonumber \\
&= \left(F\left(O-O_m\right)\right)^\mathsf{T}\text{diag}\left( \frac{1}{s_{o,kl}} \right)F\left(O-O_m\right) \nonumber \\
&= \left(\widehat{O}-\widehat{O_m}\right)^\mathsf{T}\text{diag}\left( \frac{1}{S_{O,kl}} \right)\left(\widehat{O}-\widehat{O_m}\right) \nonumber \\
\left(O-O_m\right)^\mathsf{T}R_o^{-1}\left(O-O_m\right) &= \sum_{k,l=1}^{N} \frac{\abs*{\widehat{o}_{kl}-\widehat{o_m}_{kl}}^2}{s_{o,kl}}.
\end{align}

The criterion $L_{JMAP}$ of \eqref{eq:LJMAP} can now take the simpler form of,
\begin{align}
L_{JMAP}(O,A) &= \text{cst}+ N^2\text{Ln}\sigma^2 + \frac{1}{2}\text{Ln}\abs*{R_a} + \frac{1}{2}\text{Ln}\abs*{R_o} \nonumber \\
&+ \sum_{k,l=1}^{N} \frac{1}{2\sigma^{2}}  \abs*{\widehat{i^f}_{kl}-\widehat{h^f}_{kl}\widehat{o}_{kl}}^2  \nonumber \\
&+ \sum_{k,l=1}^{N} \frac{1}{2\sigma^{2}}  \abs*{\widehat{i^d}_{kl}-\widehat{h^d}_{kl}\widehat{o}_{kl}}^2  \nonumber \\
&+ \sum_{k,l=1}^{N} \frac{1}{2\sigma^{2}}  \frac{\abs*{\widehat{o}_{kl}-\widehat{o_m}_{kl}}^2}{s_{o,kl}}  \nonumber \\
&+ \frac{1}{2}A^\mathsf{T}R_{a}^{-1}A.
\end{align}

\relSubsubsection{Object}
However it has to be remembered that it is $L_{JMAP}(\tilde{O},A)$  which is actually minimized in practice so let's compute $\tilde{O}$ in Fourier space.\marginpar{$X^\ast$ is the conjugation.}
\begin{align}
\tilde{O} &= \left(H^{f\mathsf{T}}H^{f} + H^{d\mathsf{T}}H^{d} + \sigma^2 R_o^{-1} \right)^{-1}\dots  \nonumber \\
&\dots \times \left(  H^{f\mathsf{T}}I^{f} + H^{d\mathsf{T}}I^{d} + \sigma^2 R_o^{-1}O_m \right) \nonumber \\
\tilde{O} &= F^{-1} \left( \text{diag}\left( \abs*{\widehat{h^f}}^2 \right) + \text{diag}\left( \abs*{\widehat{h^d}}^2 \right)  + \text{diag}\left( \frac{\sigma^2}{s_{o}} \right)\right)^{-1} F \dots  \nonumber \\
&\dots  \times F^{-1} \left( \text{diag}\left(\widehat{h^f}_{kl}^{\ast} \right) \widehat{I^f} + \text{diag}\left(\widehat{h^d}^{\ast} \right) \widehat{I^d}  +\sigma^2 \text{diag}\left(\frac{1}{s_{o}}\right)\widehat{O_m} \right) \nonumber \\
\widehat{\tilde{O}} &= \text{diag}\left( \frac{1}{\abs*{\widehat{h^f}}^2+\abs*{\widehat{h^d}}^2+\frac{\sigma^2}{s_{o}}}\right) \dots  \nonumber \\
&\dots \times \left( \text{diag}\left(\widehat{h^f}^{\ast} \right) \widehat{I^f} + \text{diag}\left(\widehat{h^d}^{\ast} \right) \widehat{I^d}  +\sigma^2 \text{diag}\left(\frac{1}{s_{o}}\right)\widehat{o_m} \right). \nonumber \\
\end{align}
So,\marginpar{Note that it is possible to add or remove terms if one adds a third image or removes the \textit{a priori} on the object.}
\begin{equation}
\widehat{\tilde{o}}_{kl} = \frac{\widehat{h^f}_{kl}^{\ast}\widehat{i^f}_{kl} + \widehat{h^d}_{kl}^{\ast}\widehat{i^d}_{kl} + \sigma^2\frac{\widehat{o_m}_{kl}}{s_{o,kl}}}{\abs*{\widehat{h^f}_{kl}}^2+\abs*{\widehat{h^d}_{kl}}^2+\frac{\sigma^2}{s_{o,kl}}}.
\end{equation}

\relSection{Maximum \textit{A Posteriori}}
\label{app:MAP}

In \autoref{sec:JMAP} the idea was to maximized the probability $f(A,O|I^f,I^d)$. But there is no real need for the estimation of the object itself because it doesn't help correcting the aberrations. However the object takes a lot of unknowns to describe. So the idea here is to maximize the probability of the aberration vector whatever the object is. This "whatever" is actually translated in a integral of the probability function over the object.\marginpar{Note that the Gaussian law of the object is still assumed.} Reducing the number of unknowns improves what is called the asymptotic property of the method. Good asymptotic property means that the more the measurements the better the result. If the number of unknowns grows proportionally with the measurements then having more data is not helping the estimation which is the case for the joint estimation. Once the object is out of the problem the number of unknowns is constant and adding measurements will help a lot the estimation. Another advantage of the method is apparently \marginpar{"Apparently" because the author didn't take the time to prove this point and he doesn't understand it completely.}to allow the hierarchical estimation or also called unsupervised estimation.

\relSubsection{Bayesian Approach}
The estimator is therefore defined as follow by integrating the object out of $f(A,O|I^f,I^d)$,
\begin{align}
\tilde{A} &=  \underset{A}{\text{argmax}}\, f(A|I^f,I^d) \nonumber \\
&= \underset{A}{\text{argmax}}\, \int_{\mathbb{R}^{N^2}} f(A,O|I^f,I^d) \mathrm{d} O \nonumber \\
&= \underset{A}{\text{argmax}}\, \int_{\mathbb{R}^{N^2}} f(A|I^f,I^d,O) f(O) \mathrm{d} O \nonumber \\
&= \underset{A}{\text{argmax}}\, \int_{\mathbb{R}^{N^2}} \frac{f(I^f,I^d|A,O)f(A)}{f(I^f,I^d)} f(O) \mathrm{d} O \nonumber \\
&= \underset{A}{\text{argmax}}\, \int_{\mathbb{R}^{N^2}} f(I^f,I^d|A,O) f(A) f(O) \mathrm{d} O \nonumber \\
&= \underset{A}{\text{argmax}}\, f(I^f,I^d) f(A) \nonumber \\
\tilde{A} &= \underset{A}{\text{argmax}}\, f\left(I=\vecTwoD{I^f}{I^d}\right) f(A)
\end{align}

\relSubsection{Likelihood Closed-Form}
From \eqref{eq:modelCont} it can be seen that $f(I|A)$ is a combination of Gaussian laws as the noise and the object are both Gaussian laws. So it means that $f(I|A)$ follows also a Gaussian law which is entirely defined by its mean and its covariance matrix.
The computation of the mean is straightforward,
\begin{equation}
I_m = \left[
\begin{array}{c}
\mean{H^f O} \\
\mean{H^d O}
\end{array}
\right] + \left[
\begin{array}{c}
\mean{N^f} \\
\mean{N^d}
\end{array}
\right] =\left[
\begin{array}{c}
H^f O_m \\
H^d O_m
\end{array}
\right]. 
\end{equation}

The computation of the covariance matrix $R_I$ is a bit trickier and gives, \marginpar{$[M,M]$ is a block matrix as well as $\vecTwoD{M}{M}$.}
\begin{align}
R_I &= \mean{II^\mathsf{T}}- \mean{I}\mean{I}^\mathsf{T} \nonumber \\
R_I &= \mean{ \left( \vecTwoD{H^f}{H^d}  O + \left[ \begin{array}{c} N^f\\N^d \end{array} \right] \right) \left( O^\mathsf{T} [H^{f\mathsf{T}}, H^{d\mathsf{T}}] + [N^{f\mathsf{T}}, N^{d\mathsf{T}}] \right) } \dots  \nonumber \\
&\dots - \vecTwoD{H^f}{H^d} O_m O_m^\mathsf{T} [H^{f\mathsf{T}}, H^{d\mathsf{T}}] \nonumber \\
R_I &= \vecTwoD{H^f}{H^d} \mean{O O^\mathsf{T}} [H^{f\mathsf{T}}, H^{d\mathsf{T}}] + \vecTwoD{H^f}{H^d} \mean{O_m [N^{f\mathsf{T}}, N^{d\mathsf{T}}]} \dots  \nonumber \\
&\dots + \mean{\left[ \begin{array}{c} N^f\\N^d \end{array} \right] O_m^\mathsf{T}} [H^{f\mathsf{T}}, H^{d\mathsf{T}}] +  \mean{\left[ \begin{array}{c} N^f\\N^d \end{array} \right] [N^{f\mathsf{T}}, N^{d\mathsf{T}}]} \nonumber \\
&\dots - \vecTwoD{H^f}{H^d} O_m O_m^\mathsf{T} [H^{f\mathsf{T}}, H^{d\mathsf{T}}] \nonumber \\
R_I &= \vecTwoD{H^f}{H^d} R_o [H^{f\mathsf{T}}, H^{d\mathsf{T}}] + \left[ \begin{array}{cc}
\text{cov}(N^f) & \text{cov}(N^f,N^d) \\
\text{cov}(N^f,N^d) & \text{cov}(N^d)
\end{array} \right] \nonumber \\
R_I &= \vecTwoD{H^f}{H^d} R_o [H^{f\mathsf{T}}, H^{d\mathsf{T}}] + \sigma^2 I_{2N^2}
\end{align}

Finally $R_I$ is a bloc-matrix constructed as,
\begin{equation}
R_I = \left[ \begin{array}{cc}
\widehat{h^f} R_o H^{f\mathsf{T}} + \sigma^2 I_N & \widehat{h^f} R_o H^{d\mathsf{T}} \\
\widehat{h^d} R_o H^{f\mathsf{T}} & \widehat{h^d} R_o H^{d\mathsf{T}} + \sigma^2 I_N
\end{array} \right].
\end{equation}
 
So to conclude,
\begin{equation}
f(I|A) = \frac{1}{\sqrt{2\pi}^{2N^2}\sqrt{\abs{R_I}}} e^{\frac{-1}{2}(I-I_m)^\mathsf{T}R_I^{-1}(I-I_m)}.
\end{equation}

And $\tilde{A}$ is given by solving,
\begin{align}
\tilde{A} &= \underset{A}{\text{argmax}} f\left(I | A\right) f(A) \nonumber \\
\tilde{A} &= \underset{A,O}{\text{argmin}} \underbrace{-\text{Ln}\left(f\left(I | A\right) f(A)\right)}_{L_{MAP}}.
\end{align}

The closed-form expression of $L_{MAP}$ is, \marginpar{There is a relation between $L_{MAP}$ and $L_{JMAP}(\tilde{O},.)$ that can be computed but the author didn't succeed yet in the demonstration.}
\begin{align}
L_{MAP} &= N^2 \text{Ln}(2\pi) + \frac{1}{2} \text{Ln}\abs{R_I}+\frac{1}{2}(I-I_m)^\mathsf{T} R_I^{-1} (I-I_m) \nonumber \\
& + \frac{1}{2} A^\mathsf{T} R_a^{-1} A+\frac{1}{2} \text{Ln}\abs{R_a} + \frac{1}{2} M \text{Ln}(2\pi).
\label{eq:LMAP}
\end{align}

\relSubsection{Criterion Closed-Form}
The criterion $L_{MAP}$ \eqref{eq:LMAP} depends on the determinant and the inverse of the covariance matrix of $I|A$ and both have a tractable nice closed-form expression. In both cases a formula for bloc-matrices is used so let's define the matrices $A$,$B$,$C$ and $D$ as
\begin{equation}
R_I = \left[ \begin{array}{cc}
A & B \\
C & D
\end{array} \right]
\end{equation}

\relSubsubsection{Determinant of $R_I$}
The determinant of a bloc matrix where the blocs are squared is given by
\begin{equation}
\abs{R_I} = \abs{A}\abs{D-C A^{-1} B}.
\end{equation}

\marginpar{Reminder: The indices $kl$ for the lower case letters $k$,$i$,$o$ and $s$ like in $\widehat{h^f}_{kl}$ have been forgotten for readability.} The determinant is invariant by change of base so computing it in Fourier space or real space doesn't matter. Therefore one gets,
\begin{align}
\abs{R_I} &= \abs{H^f R_o H^{f\mathsf{T}} + \sigma^2 I_N} \dots \nonumber \\
&\dots * \abs{ H^d R_o H^{d\mathsf{T}} + \sigma^2 I_N - \left(H^d R_o H^{f\mathsf{T}}\right)\left(H^f R_o H^{f\mathsf{T}} + \sigma^2 I_N\right)^{-1} \left(\widehat{h^f} R_o H^{d\mathsf{T}}\right) } \nonumber \\
&= \abs{\diag{\widehat{h^f} s_o \widehat{h^{f\ast}} + \sigma^2}} \dots \nonumber \\
&\dots * \abs{\diag{\widehat{h^d} s_o \widehat{h^{d\ast}} + \sigma^2 - \left(\widehat{h^d} s_o \widehat{h^{f\ast}}\right)\frac{1}{\left(\widehat{h^f} s_o \widehat{h^{f\ast}} + \sigma^2\right)} \left(\widehat{h^f} s_o \widehat{h^{d\ast}}\right) }} \nonumber \\
&= \prod_{k,l=1}^{N} \left( \widehat{h^f} s_{o} \widehat{h^{f\ast}} + \sigma^2 \right) \dots \nonumber \\
&\dots * \prod_{k,l=1}^{N} \left( \widehat{h^d} s_{o} \widehat{h^{d\ast}} + \sigma^2  - \frac{\left(\widehat{h^d} s_o \widehat{h^{f\ast}}\right)\left(\widehat{h^f} s_o \widehat{h^{d\ast}}\right)}{\left(\widehat{h^f} s_o \widehat{h^{f\ast}} + \sigma^2\right)}\right) \nonumber \\
&= \prod_{k,l=1}^{N} \left( \abs{\widehat{h^f}}^2 s_{o} + \sigma^2 \right) \dots \nonumber \\
&\dots * \prod_{k,l=1}^{N} \left(  \abs{\widehat{h^d}}^2 s_{o} + \sigma^2  - \frac{ \abs{\widehat{h^f}}^2 \abs{\widehat{h^d}}^2 s_{o}^2}{\left(\abs{\widehat{h^f}}^2 s_{o} + \sigma^2\right)}\right)  \nonumber \\
&= \prod_{k,l=1}^{N} \left( \abs{\widehat{h^f}}^2 s_{o} + \sigma^2 \right) \dots \nonumber \\
&\dots * \prod_{k,l=1}^{N} \left( \frac{ \abs{\widehat{h^f}}^2\abs{\widehat{h^d}}^2 s_{o}^2 + \sigma^4 + \sigma^2 s_{o} \left( \abs{\widehat{h^f}}^2 + \abs{\widehat{h^d}}^2 \right) - \abs{\widehat{h^f}}^2 \abs{\widehat{h^d}}^2 s_{o}^2}{\left(\abs{\widehat{h^f}}^2 s_{o} + \sigma^2\right)}\right)  \nonumber \\
\abs{R_I} &= \prod_{k,l=1}^{N} \sigma^2 \prod_{k,l=1}^{N} s_{o} \prod_{k,l=1}^{N} \left( \abs{\widehat{h^f}}^2 + \abs{\widehat{h^d}}^2 + \frac{\sigma^2}{s_{o}} \right) .
\end{align}

\relSubsubsection{Inverse of $R_I$}
Let's define the inverse of $R_I$ by
\begin{equation}
R_I^{-1} = \left[ \begin{array}{cc}
E & Q \\
G & H
\end{array} \right] .
\end{equation}

The bloc-matrix inversion lemma proves that the previous blocs are equal to,
\begin{subequations}
\begin{align}
E &= \left( A-BD^{-1}C\right)^{-1} \label{eq:defE} \\
Q &= - A^{-1}B\left( D-CA^{-1}B\right)^{-1} \label{eq:defQ} \\
G &= - D^{-1}C\left( A-BD^{-1}C\right)^{-1} \label{eq:defG} \\
H &= \left( D-CA^{-1}B\right)^{-1} \label{eq:defH}
\end{align}
\end{subequations}

With,
\begin{align}
A &= H^f R_o H^{f\mathsf{T}} + \sigma^2 I_N \nonumber  \\
B &= H^f R_o H^{d\mathsf{T}} \nonumber \\
C &= H^d R_o H^{f\mathsf{T}} \nonumber \\
D &= H^d R_o H^{d\mathsf{T}} + \sigma^2 I_N,
\end{align}

And using the diagonalization in Fourier space it gives,
\begin{align}
A &= F^{-1}\diag{\widehat{h^f} s_o \widehat{h^{f}}^\ast + \sigma^2}F \nonumber \\
B &= F^{-1}\diag{\widehat{h^f} s_o \widehat{h^{d}}^\ast}F \nonumber \\
C &= F^{-1}\diag{\widehat{h^d} s_o \widehat{h^{f}}^\ast}F \nonumber \\
D &= F^{-1}\diag{\widehat{h^d} s_o \widehat{h^{d}}^\ast + \sigma^2}F .
\end{align}

Then each bloc of the inverse can be computed. \\
First \eqref{eq:defE} gives,
\begin{align}
E &= F^{-1} \diag{\abs{\widehat{h^f}}^2 s_{o} + \sigma^2  - \frac{ \abs{\widehat{h^f}}^2 \abs{\widehat{h^d}}^2 s_{o}^2}{\left(\abs{\widehat{h^d}}^2 s_{o} + \sigma^2\right)}}^{-1} F \nonumber \\
E &= F^{-1} \diag{\frac{ \abs{\widehat{h^f}}^2\abs{\widehat{h^d}}^2 s_{o}^2 + \sigma^4 + \sigma^2 s_{o} \left( \abs{\widehat{h^f}}^2 + \abs{\widehat{h^d}}^2 \right) - \abs{\widehat{h^f}}^2 \abs{\widehat{h^d}}^2 s_{o}^2}{\left(\abs{\widehat{h^d}}^2 s_{o} + \sigma^2\right)}}^{-1} F \nonumber \\
E&= \frac{1}{\sigma^2 s_{o}} F^{-1} \diag{ \frac{\abs{\widehat{h^d}}^2 s_{o} + \sigma^2}{\abs{\widehat{h^f}}^2 + \abs{\widehat{h^d}}^2 + \frac{\sigma^2}{s_{o}}} } F ,
\end{align}

Similarly \eqref{eq:defH} gives,
\begin{equation}
H = \frac{1}{\sigma^2 s_{o}} F^{-1} \diag{ \frac{\abs{\widehat{h^f}}^2 s_{o} + \sigma^2}{\abs{\widehat{h^f}}^2 + \abs{\widehat{h^d}}^2 + \frac{\sigma^2}{s_{o}}} } F .
\end{equation}

Then \eqref{eq:defQ} gives,
\begin{align}
Q &= F^{-1} \diag{\frac{\widehat{h^f} s_{o} \widehat{h^d}}{\abs{\widehat{h^f}}^2 s_{o} + \sigma^2}} \frac{1}{\sigma^2 s_{o}} \diag{\frac{\abs{\widehat{h^f}}^2 s_{o} + \sigma^2}{\abs{\widehat{h^f}}^2 + \abs{\widehat{h^d}}^2 + \frac{\sigma^2}{s_{o}}} }  F \nonumber \\
Q &= F^{-1} \frac{-1}{\sigma^2} \diag{\frac{\widehat{h^f} \widehat{h^{d}}^\ast}{\abs{\widehat{h^f}}^2 + \abs{\widehat{h^d}}^2 + \frac{\sigma^2}{s_{o}}}}  F ,
\end{align}

Similarly \eqref{eq:defG} gives,
\begin{equation}
G = F^{-1} \frac{-1}{\sigma^2} \diag{\frac{\widehat{h^{f}}^\ast h^{d}}{\abs{\widehat{h^f}}^2 + \abs{\widehat{h^d}}^2 + \frac{\sigma^2}{s_{o}}}}  F .
\end{equation}

\relSubsubsection{Quadratic Form}
In order to obtain a quite simple expression of the criterion $L_{MAP}$ a last thing needs to be computed. It is the quadratic form of the gaussian law that appears in \eqref{eq:LMAP}. This calculations will be done in Fourier space as the bloc-matrices of $R_I^{-1}$ are diagonal in that space,
\begin{subequations}
\begin{align}
(I-I_m)^\mathsf{T} R_I^{-1} (I-I_m) &= \vecTwoD{I^f - H^f \widehat{o_m}}{I^d - H^d \widehat{o_m}}^{\mathsf{T}} \matTwoD{E}{Q}{G}{H} \vecTwoD{I^f - H^f \widehat{o_m}}{I^d - H^d \widehat{o_m}} \nonumber \\
&= (I^f - H^f \widehat{o_m})^{\mathsf{T}} E (I^f - H^f \widehat{o_m}) \label{eq:quadE} \\
&+ (I^f - H^f \widehat{o_m})^{\mathsf{T}} Q (I^d - H^d \widehat{o_m}) \label{eq:quadQ} \\
&+ (I^d - H^d \widehat{o_m})^{\mathsf{T}} G (I^f - H^f \widehat{o_m}) \label{eq:quadG} \\
&+ (I^d - H^d \widehat{o_m})^{\mathsf{T}} H (I^d - H^d \widehat{o_m}) \label{eq:quadH} .
\end{align}
\end{subequations}

In Fourier space all the terms \eqref{eq:quadE},\eqref{eq:quadQ},\eqref{eq:quadG} and \eqref{eq:quadH} will be transformed in sums as the bloc-matrices are diagonal. Besides as a quadratic form the expresion in Fourier space are equal to the one in real space. \\
Therefore term \eqref{eq:quadE} becomes,
\begin{equation}
(I^f - H^f \widehat{o_m})^{\mathsf{T}} E (I^f - H^f \widehat{o_m}) = \sum_{k,l=1}^{N} \frac{1}{\sigma^2 s_{o}} \frac{\abs{\widehat{h^d}}^2 s_{o} + \sigma^2}{\abs{\widehat{h^f}}^2 + \abs{\widehat{h^d}}^2 + \frac{\sigma^2}{s_{o}}} \abs{\widehat{i^f} - \widehat{h^f} \widehat{o_m}}^2,
\end{equation}

\eqref{eq:quadQ} becomes,
\begin{equation}
(I^f - H^f \widehat{o_m})^{\mathsf{T}} Q (I^d - H^d \widehat{o_m}) = \sum_{k,l=1}^{N} \frac{-1}{\sigma^2} \frac{\widehat{h^f} \widehat{h^{d}}^\ast}{\abs{\widehat{h^f}}^2 + \abs{\widehat{h^d}}^2 + \frac{\sigma^2}{s_{o}}} (\widehat{i^f} - \widehat{h^f} \widehat{o_m})^{\ast}(\widehat{i^d} - \widehat{h^d} \widehat{o_m}),
\end{equation}

\eqref{eq:quadG} becomes,
\begin{equation}
(I^d - H^d \widehat{o_m})^{\mathsf{T}} G (I^f - H^f \widehat{o_m}) = \sum_{k,l=1}^{N} \frac{-1}{\sigma^2} \frac{\widehat{h^{f}}^\ast \widehat{h^d}}{\abs{\widehat{h^f}}^2 + \abs{\widehat{h^d}}^2 + \frac{\sigma^2}{s_{o}}} (\widehat{i^d} - \widehat{h^d} \widehat{o_m})^{\ast}(\widehat{i^f} - \widehat{h^f} \widehat{o_m}),
\end{equation}

\eqref{eq:quadH} becomes,
\begin{equation}
(I^d - H^d \widehat{o_m})^{\mathsf{T}} H (I^d - H^d \widehat{o_m}) = \sum_{k,l=1}^{N} \frac{1}{\sigma^2 s_{o}} \frac{\abs{\widehat{h^f}}^2 s_{o} + \sigma^2}{\abs{\widehat{h^f}}^2 + \abs{\widehat{h^d}}^2 + \frac{\sigma^2}{s_{o}}} \abs{\widehat{i^d} - \widehat{h^d} \widehat{o_m}}^2.
\end{equation}

Then it is only a work of factorisation to get to,
\begin{align}
&(I-I_m)^\mathsf{T} R_I^{-1} (I-I_m) \nonumber \\
=& \sum_{k,l=1}^{N} \frac{1}{\abs{\widehat{h^f}}^2 + \abs{\widehat{h^d}}^2 + \frac{\sigma^2}{s_{o}}}
\left(
 \frac{1}{\widehat{s_o}} \left( \abs{\widehat{i^f} - \widehat{h^f} \widehat{o_m}}^2 + \abs{\widehat{i^d} - \widehat{h^d} \widehat{o_m}}^2 \right)  \right. \dots \nonumber \\
 \dots +&\left.  \frac{1}{\sigma^2}\abs{\widehat{i^f}\widehat{h^d} - \widehat{i^d}\widehat{h^f}} \right).
\end{align}

\relSection{Application}
\relSubsection{Measurements}
\label{app:PDappMeas}

Different methods can be used to measure focused and defocused images. Using the Deformable Mirror has been described in \autoref{sec:measurements}. The displacement of the source or the detector and the beam splitter methods are described right below.

\begin{description}
\item[Displacement] The easiest way to acquire focused and defocused images is to move the light source or the detector by a constant amount between each image. The displacement can be computed from the equivalent pupil phase using the following equations,
\begin{equation}
  \begin{cases}
\delta_{\text{mm}} &= \delta_{\text{p2v,rad}}\frac{4.10^{-3}\lambda_{\mu m}r^2}{\pi}, \\
r &= \frac{f}{d}, \\
\delta_{\text{p2v,rad}} &= \delta_{\text{p2v,nm}} \frac{2\pi 10^{-3}}{\lambda_{\mu m}}, \\
\delta_{\text{p2v,nm}} &= \delta_{\text{rms,nm}} 2\sqrt{3}, 
  \end{cases}
  \label{eq:DisplDefoc}
\end{equation}
With $\delta_{mm}$ the displacement of the source in millimetres, $\delta_{p2v,rad}$ respectively $\delta_{p2v,nm}$ the peak-to-valley value of the defocus phase function in radian respectively in nanometre, $\delta_{rms,nm}$ the Root-Mean-Square value of the defocus phase function in nanometre, $f$ the focal distance of the optical system in meter, $d$ the diameter of the pupil in meter and $\lambda_{\mu m}$ the wavelength in micrometre.\\
This method is easy but it is also the less precise because of hysteresis in the back and forth displacements.

\item[Beam splitter] Using the same idea of a physical longer distance between the source and the detector it is possible to use a beam splitter. Half of the light will go to the regular science detector and the other half would go to another detector  defocus by with displacement equal to \eqref{eq:DisplDefoc}. The advantage here is that there are no movements of the optics and that both images are taken at the same time. However the light path for the two detectors are not identical.
\end{description}

\chapter{PD Mathematics}

\relSection{Convolution Matrix}
\label{app:convoMat}

Just for fun the matrices of the convolution will be described here. The variables noted with a lower case are the elements of the matrices. It means that $i_{nm}$, $h_{kl}$ and $o_{kl}$ are respectively the pixels of the measured image, of the simulated image and of the object.
\begin{equation}
i_{n,m} = \sum_{k,l=1}^{K,L} h_{k,l} o_{n-k,m-l} = \sum_{k,l=1}^{K,L} o_{k,l} h_{n-k,m-l}
\end{equation}

\marginpar{Let's have fun with indices!!} Now the convolution matrix H is given by,
\begin{equation}
H =
\left[
\begin{array}{ccccc}
h_{1,\star} & h_{n,\star}& h_{(n-1),\star}& \ddots & h_{2,\star} \\
h_{2,\star} & h_{1,\star}& h_{n,\star}& \ddots & \ddots \\
h_{3,\star} & h_{1,\star}& h_{1,\star}& \ddots & h_{(n-1),\star} \\
\ddots & \ddots & \ddots& \ddots & h_{n,\star} \\
h_{n,\star} & \ddots& h_{3,\star}& h_{2,\star} & h_{1,\star}
\end{array}
\right],
\end{equation}

With $h_{k,\star}$ bloc-matrices defined as,
\begin{equation}
h_{k,\star} =
\left[
\begin{array}{ccccc}
h_{k,1} & h_{k,n}& h_{k,(n-1)}& \ddots & h_{k,2} \\
h_{k,2} & h_{k,1}& h_{k,n}& \ddots & \ddots \\
h_{k,3} & h_{k,2}& h_{k,1}& \ddots & h_{k,(n-1)} \\
\ddots & \ddots & \ddots& \ddots & h_{(k,n} \\
h_{k,n} & \ddots& h_{k,3}& h_{k,2} & h_{k,1}
\end{array}
\right].
\end{equation}
One sees that $H$ is indeed circulant-block-circulant.

\relSection{Circulant Matrix}

\relSubsection{Definition}
Here are going to be developed interesting mathematical properties of the circulant matrices. It is true that Phase Diversity deals with Circulant-block-Circulant matrices but let's assume these results can be generalized from the simple Circulant matrix. Let's define the Circulant matrix $C$ as,
\begin{equation}
C =
\left[
\begin{array}{ccccc}
a_{N-1} & a_{N-2} & a_{N-3} & \cdots & a_{0} \\
a_{0} & a_{N-1} & a_{N-2} & \ddots& \vdots \\
\vdots & a_{0} & \ddots & \ddots & 0 \\
a_{N-3} & \ddots & \ddots & \ddots & a_{N-2} \\
a_{N-2} & a_{N-3} & \cdots & a_{0} & a_{N-1}
\end{array}
\right].
\end{equation}

\relSubsection{Base}
$C$ can be expanded on a base composed of the powers of the matrix $J$ defined by,
\begin{equation}
J =
\left[
\begin{array}{ccccc}
0 & 1 & 0 & \cdots & 0 \\
0 & 0 & 1 & \ddots& \vdots \\
\vdots & \ddots & \ddots & \ddots & 0 \\
0 & \ddots & \ddots & \ddots & 1 \\
1 & 0 & \cdots & 0 & 0
\end{array}
\right].
\end{equation}

$C$ can be written as,\marginpar{Note that $J^{N}=I_{N}$, the identity matrix in dimension $N$}
\begin{equation}
C=\sum_{n=0}^{N-1} a_{0}J^{n} = \mathcal{P}_{C}(J),
\end{equation}
With $\mathcal{P}$ a polynomial of coefficients the elements of the vector $A=(a_0, \dots, a_{N-1})^\mathsf{T}$.

\relSubsection{Eigen-values and -vectors}
Looking for the eigen-values of the matrix $J^{n}$ is equivalent to search the roots of unity, \marginpar{Solving $det(J^{n}-\lambda . I_{N})=0$}
\begin{equation}
  \begin{dcases*}%
        \lambda_{k} = \omega^{k} & eigen-values with $\omega=e^{2i\pi/N}$,\\
        v_{k} = (1,\omega^{k},\dots,\omega^{(N-1)k})^\mathsf{T}& eigen-vectors.
  \end{dcases*}
\end{equation}

The eigen-values of $C$ are then given by $P_{C}(\omega^{k})=\sum_{n=0}^{N-1} a_{0}\omega^{kn} $ with the same eigen-vectors $v_{k}$.\marginpar{One can just verify it.} Besides it appears that all the $P_{C}(\omega^{k})$ are the coefficients of the Fourier expansion of $A$ so one can write,
\begin{equation}
\hat{a}_{k} = P_{C}(\omega^{k}).
\end{equation}

\relSubsection{Diagonalization: Fourier space}
\label{app:circulDiag}
The matrix $P$ for changing bases between $C$ and its diagonal matrix $D=diag(\hat{a}_{1},\hat{a}_{2},\dots,\hat{a}_{K})$ is built using the eigen-vectors as columns of the matrix. Therefore,
\begin{equation}
P =\left([v_{1}][v_{2}]\dots[v_{N}]\right),
\end{equation}
which is no other than the matrix of the inverse Discrete Fourier Transform so $P=F^{-1}$ and,
\begin{equation}
D = P^{-1}\times C \times P = F\times C \times F^{-1}
\end{equation}

Now is time to go back to the convolution of $HO$. As it was said it is assumed than the Circulant matrix results can be applied to the Circulant-block-Circulant matrix with a two-dimensional Fourier Transform instead.
\begin{equation}
HO = F^{-1}$diag$(\hat{A})FO,
\end{equation}
or it can be written as,
\begin{equation}
\widehat{HO} = $diag$(\hat{A})\hat{O} = \left(\hat{A}_{nm} \hat{O}_{nm}\right)_{n,m}.
\label{eq:FourConv}
\end{equation}

\relSection{Convolution and Fourier}
This parenthesis with circulant matrices was fun but actually it could have gone much faster. The author did this digression only because the paper of \citeauthor{blancJOSAA2003} \citep{blancJOSAA2003} speaks about Toeplitz matrices with Circulant approximation. The other way to demonstrate \eqref{eq:FourConv} is to prove directly that the convolution in real space is equivalent to a multiplication in Fourier space. This can be done in a few lines even for the two-dimensional convolution with two matrices $A$ and $B$,
\begin{align}
    \left( \widehat{A \ast B}\right)_{k,l} &= \sum_{n,m=0}^{N-1,M-1}\left(\sum_{i,j=0}^{N-1,M-1}a_{i,j}b_{n-i,m-j}\right)e^{-2i\pi\left(\frac{kn}{N}+\frac{lm}{M}\right)} \nonumber \\
      &= \left(\sum_{i,j=0}^{I-1,J-1} a_{i,j} e^{-2i\pi\left(\frac{ki}{N}+\frac{lj}{M}\right)} \right) \dots \nonumber  \\
      &\dots *\left(\sum_{n,m=0}^{N-1,M-1}b_{n-i,m-j} e^{-2i\pi\left(\frac{k(n-i)}{N}+\frac{l(m-j)}{M}\right)}\right) \nonumber \\
    \left( \widehat{A \ast B}\right)_{k,l} &= \hat{A}_{k,l}\hat{B}_{k,l}
\end{align}

\relChapter{PD Results} 

\relSection{AOF Iterations}
\label{app:PDresAOFit}

\begin{figure}[bth]
\myfloatalign
\subfloat[Projection Matrix.]
{\label{fig:PDAOFiterPDPM-a} \includegraphics[width=0.45\linewidth]{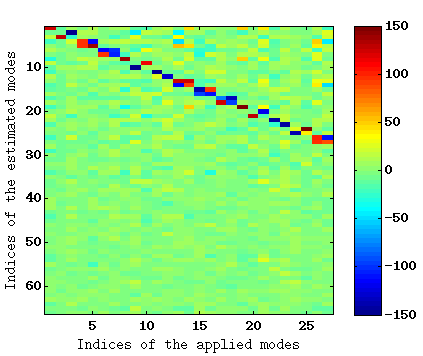}} \quad
\subfloat[Sole aberrations.]
{\label{fig:PDAOFiterPDPM-b} \includegraphics[width=0.45\linewidth]{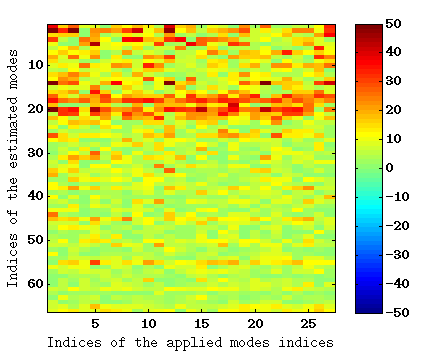}}
\caption[Phase Diversity Projection Matrix for AOF.]{Phase Diversity Projection Matrix for AOF iterations with $66$ estimated modes. \autoref{fig:PDAOFiterPDPM-a} contains the sole mode contribution by half subtraction of the estimation of the positive respectively negative offset of the modes applied during calibration. \autoref{fig:PDAOFiterPDPM-b} includes the sole aberration which is the result of the mean instead.}.
\label{fig:PDAOFiterPDPM}
\end{figure}

\cleardoublepage

\label{app:bibliography} 


\addcontentsline{toc}{chapter}{\tocEntry{\bibname}}

\bibliographystyle{plainnat}

\bibliography{D:/latex/phaseDiversity/bibliography/bib_phase_diversty} 

\part{Appendix Electric Field Conjugation}

\chapter{EFC Theory}

\relSection{Model}
\label{app:EFCAppModel}
The complex amplitude in the focal plane is given by,\marginpar{This is \eqref{eq:EFCpup2foc} of \autoref{sec:EFCModel}.}
\begin{equation}
  \mathcal{D}(\xi,\eta)=\mathcal{C}\left( \mathcal{P} \right)=\mathcal{C}\left( \mathcal{P}_m e^{i(\phi+\psi+\omega)} \right) ,
\end{equation}
With $\mathcal{C}$ the linear operator modelling the coronagraph. $\omega$ is forgotten for now.

A second order Taylor expansion gives in addition with the linearity of $\mathcal{C}$,
\begin{align}
	\mathcal{D}(\xi,\eta)&=\mathcal{C}(\mathcal{P}_m)  & 0\,\text{Order,} \nonumber \\
	&+ i\mathcal{C}(\mathcal{P}_m \phi)+ i\mathcal{C}(\mathcal{P}_m \psi)) & 1^{st}\, \text{Order,} \nonumber \\
	&- \mathcal{C}(\mathcal{P}_m \frac{\phi^2}{2}) - \mathcal{C}(\mathcal{P}_m \frac{\psi^2}{2})) - \mathcal{C}(\mathcal{P}_m \psi\omega) & 2^{st}\, \text{Order.} 
	\label{eq:EFCcomplFocal}
\end{align}

\paragraph{Demonstration comments}
\citeauthor{giveon2011} \citep{giveon2011} used only a first order Taylor expansion of $\mathcal{D}$. The problem doing so is that one doesn't get a real linear relation between the electric field in the focal plane and the intensity of the detector. In this case the matrix relating both is indeed itself dependent on the electric field in the focal plane. Therefore it requires a fuzzy approximation expressed in equation (7) of \citep{giveon2011}. The author solved this problem by pushing the expansion to the second order. It still requires an assumption but the assumption is the use of a coronagraph which is not incredible. In practice it requires that the rings below the speckles are less bright than the speckles themselves. In the opinion of the author this demonstration is better because the linear relation appears as a second order term when expanding the intensity so keeping a first order expansion at the beginning doesn't make sense.\\

The operator $\mathcal{C}$ is noted with an upper bar and the conjugate of $z$ is noted $z^\ast$.
The intensity in the focal plane is given by the square of the absolute value of the complex amplitude $\mathcal{I} = \mathcal{D} \mathcal{D}^{\ast}$. Using the second order Taylor expansion of \eqref{eq:EFCcomplFocal} on gets, 
\begin{align}
 \mathcal{I} &= \abs*{\overline{\mathcal{P}_m}}^2 & 0\, \text{Order,} \nonumber \\
 &- i \overline{\mathcal{P}_m} \left( \overline{\mathcal{P}_m \phi}+ \overline{\mathcal{P}_m \psi} \right)^{\ast} & 1^{st}\, \text{Order,} \nonumber \\
 &+ i \left(\overline{\mathcal{P}_m}\right)^{\ast} \left( \overline{\mathcal{P}_m \phi}+ \overline{\mathcal{P}_m \psi} \right) & \dots \nonumber \\
 &- \overline{\mathcal{P}_m} \left( \overline{\mathcal{P}_m \frac{\phi^2}{2}} + \overline{\mathcal{P}_m \frac{\psi^2}{2}} + \overline{\mathcal{P}_m \psi\phi} \right)^{\ast} & 2^{nd}\, \text{Order,} \nonumber \\
 &- \overline{\mathcal{P}_m}^{\ast} \left( \overline{\mathcal{P}_m \frac{\phi^2}{2}} + \overline{\mathcal{P}_m \frac{\psi^2}{2}} + \overline{\mathcal{P}_m \psi\phi} \right) & \dots  \nonumber \\
 &+ \abs*{\overline{\mathcal{P}_m \phi}}^2 + \abs*{\overline{\mathcal{P}_m \psi}}^2 + \left(\overline{\mathcal{P}_m \phi}\right)^{\ast}\overline{\mathcal{P}_m \psi} +  \overline{\mathcal{P}_m \phi}\left(\overline{\mathcal{P}_m \psi}\right)^{\ast}& \dots
\end{align}

Which can be simplified using $z\overline{w}+\overline{z}w = 2\text{Re} \left[z\overline{w}\right]$, \marginpar{If time it would be interesting to see what happens with amplitude errors.}
\begin{align}
 \mathcal{I} &= \abs*{\overline{\mathcal{P}_m}}^2 & 0\, \text{Order,} \nonumber \\
 &+ 2\text{Re} \left[ i \left(\overline{\mathcal{P}_m}\right)^{\ast} \left( \overline{\mathcal{P}_m \phi}+ \overline{\mathcal{P}_m \psi} \right) \right] & 1^{st}\, \text{Order,} \nonumber \\
 &- 2\text{Re} \left[ \overline{\mathcal{P}_m} \left( \overline{\mathcal{P}_m \frac{\phi^2}{2}} + \overline{\mathcal{P}_m \frac{\psi^2}{2}} + \overline{\mathcal{P}_m \psi\phi} \right)^{\ast} \right] & 2^{nd}\, \text{Order,} \nonumber \\
 &+ \abs*{\overline{\mathcal{P}_m \phi}}^2 + \abs*{\overline{\mathcal{P}_m \psi}}^2 + 2\text{Re} \left[ \left(\overline{\mathcal{P}_m \phi}\right)^{\ast}\overline{\mathcal{P}_m \psi} \right]& \dots
 \label{eq:EFCintensAppen}
\end{align}

\relSection{Electric Field Estimation}
\label{app:EFCAppEFEst}
When subtracting the intensity of a positive probe with the same probe but applied negatively the only terms remaining in \eqref{eq:EFCintens}\marginpar{\eqref{eq:EFCintensAppen} is identical to \eqref{eq:EFCintens}.} are the one proportional to $\psi$. The other terms of the equation are cancelled out so that,
\begin{equation}
\frac{1}{2}\left( \mathcal{I}^+ - \mathcal{I}^-\right) = 2\text{Re} \left[ i \left(\overline{\mathcal{P}_m}\right)^{\ast} \overline{\mathcal{P}_m \psi} \right]
- 2\text{Re} \left[ \overline{\mathcal{P}_m} \left( \overline{\mathcal{P}_m \psi\phi} \right)^{\ast} \right]
+ 2\text{Re} \left[ \left(\overline{\mathcal{P}_m \phi}\right)^{\ast}\overline{\mathcal{P}_m \psi} \right],
\label{eq:EFCdeltaIraw}
\end{equation}
With $\mathcal{I}^+$ respectively $\mathcal{I}^-$ the intensity for the probe applied positively respectively negatively.
 
 The assumption of a perfect coronagraph gives $\overline{\mathcal{P}_m} = 0$ so that the first two terms become negligible. It means that the rings below the speckle should be negligible compared to the intensity of the speckles themselves. After expanding the remaining real part of the product in \eqref{eq:EFCdeltaIraw} one finds,
 \begin{equation}
\frac{1}{2}\left( \mathcal{I}^+ - \mathcal{I}^-\right) = 2\text{Re} \left[\overline{\mathcal{P}_m\phi}\right] \text{Re} \left[\overline{\mathcal{P}_m\psi}\right] + 2\text{Im} \left[\overline{\mathcal{P}_m\phi}\right] \text{Im} \left[\overline{\mathcal{P}_m\psi}\right].
 \label{eq:EFCdeltaI}
 \end{equation}
 
 The images are discrete so each function below can take the form of a vector in which both spatial dimensions are reshaped in a single dimension. The vector corresponding to each function is noted with a straight font for the Latin letters and with an upper case letter for the Greek letters. One also remembers that the upper bar is a linear operator modelling the coronagraph so it can be defined using a matrix $C$. Besides for clarity the phase vector will be assumed null outside the pupil mask so that $\Phi = \text{diag}(P_m)\Phi$. In bloc-matrix form it gives,
 \begin{equation}
 \frac{1}{2}(I^+ - I^-)= 2
 \left[ \text{Re} \left[C \Psi\right] \, \text{Im} \left[C \Psi\right] \right]
 \vecTwoD{\text{Re} \left[C \Phi\right]}{\text{Im} \left[C \Phi\right]}.
 \label{eq:EFCestiSingle}
 \end{equation}
 Note also that the matrix $C$ is defined so that the vectors $I^+$ and $I^-$ includes only the pixels of the area of interest. The area of interest is the area that one wants to darken.
 
\eqref{eq:EFCestiSingle} allows one to estimate the complex amplitude of the speckles \marginpar{$i\overline{\mathcal{P}_m\Phi}$ is indeed the first order contribution of the speckles to the electric field in the focal plane. See \eqref{eq:EFCcomplFocal}.} in the focal plane from a simple image subtraction. Besides this relation is linear so that a Pseudo-Inverse would solve the problem. However one single probe would not suffice to remove the degeneracy due to the intensity measurements. That's why at least two independent probes are necessary. The problem takes then the following form, \marginpar{Extensions of equation \eqref{eq:EFCesti} for pixel weighting, actuators regularization, multi Deformable Mirror or multi-wavelength correction are given in \citep{giveon2007}}
 \begin{equation}
 \vecTwoD{\frac{1}{2}(I_1^+ - I_1^-)}{\frac{1}{2}(I_2^+ - I_2^-)}= 2
 \underbrace{\left[ \begin{array}{cc}
 \text{Re} \left[C \Psi_1\right] & \text{Im} \left[C \Psi_1\right] \\
 \text{Re} \left[C \Psi_2\right] & \text{Im} \left[C \Psi_2\right] \\
 \end{array} \right]}_{E}
 \vecTwoD{\text{Re} \left[C \Phi \right]}{\text{Im} \left[C \Phi \right]},
 \label{eq:EFCestiApp}
 \end{equation}
With $\Psi_1$ and $\Psi_2$ the two different probe functions in vectors form and $I_1$ respectively $I_2$ the vector of the pixel intensities for the respective probes.

\relSection{Correction}
\label{app:EFCAppCorr}
The phase $\omega$ produced by the Deformable Mirror is related to the actuator commands by the influence functions. \marginpar{Commands are sometimes also called voltages.} If the pupil is discretized this can take the form of a matrix multiplication,
\begin{equation}
\Omega  = I_f A,
\label{eq:EFCinflFunc}
\end{equation}
With $\Omega$ the vector containing all the pixels of the pupil, $I_f$ the matrix containing the influence functions and $A$ the vector of commands to apply to each actuator. \marginpar{For exemple on SPHERE there are 1377 actuators.}
The influence function are also supposed to carry out the pupil mask so that the elements of $\Omega$ outside the pupil are null.\\

As $A$ is a real vector it can also be written as,
\begin{equation}
\vecTwoD{\text{Re}(C\Omega)}{\text{Im}(C\Omega)}=\vecTwoD{\text{Re}(CI_f)}{\text{Im}(CI_f)} A.
\end{equation}

Correcting the area of interest means setting all the intensities to zero. If the probes are well defined having $I_1^+-I_1^-=I_2^+-2_1^-=0$ is equivalent to $I=0$. If one adds the correcting term in \eqref{eq:EFCesti} $\Omega$ should verify, \marginpar{One just needs to replace $\Phi$ by $\Phi + \Omega$.}
\begin{equation}
\vecTwoD{0}{0}= 
 \underbrace{2E\vecTwoD{\text{Re} \left[C \Phi \right]}{\text{Im} \left[C \Phi \right]}}_{\delta I_\Phi}
 + 2E \vecTwoD{\text{Re} \left[C \Omega \right]}{\text{Im} \left[C \Omega \right]}
\end{equation}

To finish the commands for correcting the aberrations can then be obtained by inverting the following problem,
\begin{equation}
 \underbrace{2 E \vecTwoD{\text{Re}(CI_f)}{\text{Im}(CI_f)}}_{G} A
 = - \delta I_\Phi
\end{equation}
$G$ is called the Interaction Matrix. The inverse of $G$ noted $R$ is called the Reconstruction Matrix.

The final formulation of the problem is,
\begin{equation}
GA=-\delta I_\Phi\, \text{or}\, A=-R\delta I_\Phi,
\label{eq:EFCproblemApp}
\end{equation}
where $A$ is the unknown vector with the commands for the actuators and $\delta I$ is the measurement. 

\relSection{Modes}

\relSubsection{Commands Vector}
\label{app:EFCAppModesVolt}
As it was already explained in \autoref{sec:EFCAppIt} it is always possible to go from the modes coefficients to the actuators commands using the Pseudo-Inverse of $I_f$. The command vector $\alpha$ would indeed be equal to,
\begin{equation}
\alpha = I_f^{\dagger}ZA,
\label{eq:EFCmodeCommands}
\end{equation}
With $\dagger$ representing the Pseudo Inverse. This is by the way used to project the theoretical modes onto the influence functions to find the closest shape that can take the Deformable Mirror. $\alpha$ could also be computed by adding the commands vector of each mode of the base with the right amplitude.

\begin{figure}[h]
	\centering	
   	\includegraphics[width=\linewidth]{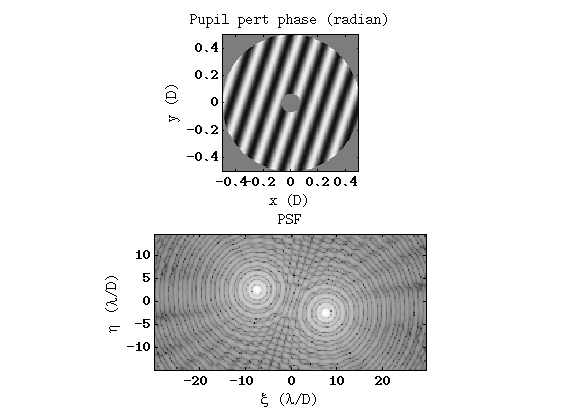}
   	\caption[Example of mode.]{\label{fig:modeSin} Example of Fourier mode with a perfect coronagraph. The upper figure shows the pupil pupil phase while the bottom one shows the detector intensities in logarithmic scale.}
\end{figure}

\relSubsection{Fourier Modes}
\label{app:EFCAppModesFour}
Sines and cosines functions have theoretically an effect on single pixels as their Fourier Transform is a pair of symmetric Dirac functions with opposite phase, \marginpar{$\mathcal{F}$ is the Fourier operator.}
\begin{equation}
\begin{cases}
\mathcal{F}\left(\sin(2\pi \xi_c .)\right)(\xi) &= \frac{\delta(\xi-\xi_c)-\delta(\xi+\xi_c)}{2i} \nonumber \\
\mathcal{F}\left(\cos(2\pi \xi_c .)\right)(\xi) &= \frac{\delta(\xi-\xi_c)+\delta(\xi+\xi_c)}{2}
\end{cases}.
\end{equation}
 However because of the finite size of the pupil the Fourier Transform is convolved with the Point Spread Function resulting in two symmetrical Point Spread Function. It can be seen in \autoref{fig:modeSin}. The presence of the coronagraph shouldn't affect too much the shape of the modes if they are far enough from the center. The idea is not to know the closed-form expression of the mode in the focal plane but only to know where it falls. The frequency of the sine and the coordinates of the Dirac in the focal plane in unit of $\lambda/d$ are the same. That's why it is possible to define a set of modes with sines and cosines which have an effect on each pixel of the area of interest. Both sines and cosines are needed to act on both dimension of a complex number. A set of Fourier modes would be defined as,
 \begin{equation}
 \forall k \in [1,K]
 \begin{cases}
 \mathcal{Z}_{2k-1}(x,y) &= a \sin(2\pi (\xi_k x + \eta_k y) \\
 \mathcal{Z}_{2k}(x,y) &= a \cos(2\pi (\xi_k x + \eta_k y)
 \end{cases},
 \end{equation}
With $(\xi_k,\eta_k)$ the set of coordinates of the modes in the focal plane and $a$ the amplitude of the modes\marginpar{\citeauthor{verinaudFREEEFC} advises to take probes and modes with the same intensity as the speckles in the image.} . The total number of modes is here $M=2K$.

For example $(\xi_k,\eta_k)$ can be the coordinates of each pixel of the area so that $K$ would be the number of pixels. Another solution is to regularly space the modes in the focal plane with a defined step. A step of $1\lambda/d$ would mean that each mode has an additional period in the phase of the pupil.

\relSubsection{Singular Modes}
\label{app:EFCAppModesSing}
The singular modes are modes computed after the Singular Value Decomposition of the Interaction Matrix. The Interaction Matrix is the matrix $G$ in $GA=-\delta I_\Phi$ \eqref{eq:EFCproblem}. As it was indicated in \autoref{sec:EFCmodes} the vector $A$ doesn't need to be the commands vector but it can be a coordinate vector in any modal basis.
The Singular Value Decomposition of $G$ gives three matrices $U$, $S$ and $V$ so that $G=USV^\mathsf{T}$ and $R=VSU^\mathsf{T}$. \marginpar{$V^\mathsf{T}$ is the conjugate transpose of $V$.} The columns of $V$ are called the right-singular vectors of $G$ and they correspond to the coordinates of some modes in the modal basis. These modes are called the singular modes. They are sorted by their efficiency in the area of interest. The efficiency can indeed be measured by the value of the singular value. The higher the singular value the greater the effect on the intensities in the area. \\
Obviously one needs a Interaction Matrix for computing the singular modes which requires the use of another modal base. A simple base with Fourier modes would work but it needs to be complete so that the Singular Value Decomposition has enough freedom to build the best modes. \marginpar{The author uses usually a pixel based set of modes. It means that there are one sine and one cosine mode centred on each pixel of the area of interest.}\\

\begin{figure}[h]
	\centering	
   	\includegraphics[width=\linewidth]{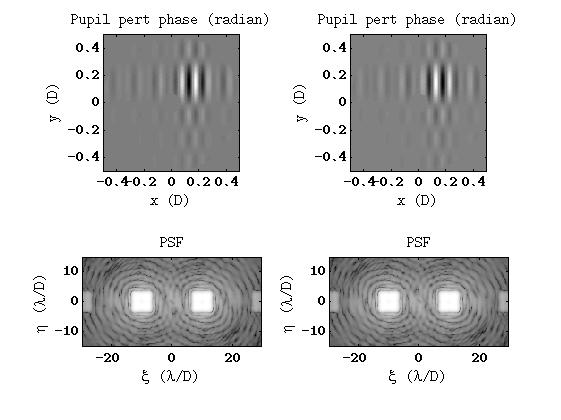}
   	\caption[Example of probe.]{\label{fig:probeSin} Example of probe with a perfect coronagraph. The probe is defined with a sine. The probe is applied positively on the left figures and negatively on the right figures. The upper figures include the pupil phase and the bottom one the detector intensities in logarithmic scale.}
\end{figure}

\relSection{Probes}
\label{app:EFCAppProbes}
As it was mentioned for the modes it is possible to know the effect of the probe in the focal plane by taking the Fourier Transform of the phase. The probes of \citeauthor{giveon2011} \citep{giveon2011} will create symmetrical rectangles horizontally shifted like in \autoref{fig:probeSin} and their phase functions are,
\begin{equation}
\begin{cases}
\psi_1(x,y) &= a\, \text{sinc}(\Delta \xi x)\text{sinc}(\Delta \eta y)\sin(2\pi \xi_c x) \nonumber \\
\psi_2(x,y) &= a\, \text{sinc}(\Delta \xi x)\text{sinc}(\Delta \eta y)\cos(2\pi \xi_c x)
\end{cases}
\end{equation}
With $a$ the amplitude of the probes, $\Delta \xi$ and $\Delta \eta$ the size of the rectangle in the detector and $\xi_c$ the distance of the center of the rectangle to the center of the image in the horizontal direction.\marginpar{All distances are given in unit of $\lambda/d$.} The sines cardinal are used to build the rectangle as the Fourier Transform of a sine cardinal is the hat function while the last sine or cosine is used to shift the rectangle from the center. \marginpar{One has to remember that the Fourier Transform of the product if the convolution of the Fourier Transforms.}

One can also rotate the probe from an angle $\theta$ using instead the more general formula,
\begin{equation}
\begin{cases}
\psi_1(x,y) &= a\, \text{sinc}\left(\Delta \xi (\cos(\theta)x+\sin(\theta)y)\right)\dots \nonumber \\
&\dots\times \text{sinc}\left(\Delta \eta (-\sin(\theta)x+\cos(\theta)y)\right)\sin\left(2\pi \xi_c (\cos(\theta)x+\sin(\theta)y)\right) \nonumber \\
\psi_2(x,y) &= a\, \text{sinc}\left(\Delta \xi (\cos(\theta)x+\sin(\theta)y)\right)\dots \nonumber \\
&\dots\times \text{sinc}\left(\Delta \eta (-\sin(\theta)x+\cos(\theta)y)\right)\cos\left(2\pi \xi_c (\cos(\theta)x+\sin(\theta)y)\right)
\end{cases}
\end{equation}

\relSection{Interaction Matrix}
\label{app:EFCAppIM}
Applying positively and negatively the modes when recording the Interaction Matrix allows one to get rid of the existing aberrations. The demonstration requires to expand the intensity like it was done for \eqref{eq:EFCintens} but this time including the phase function of the mode. For simplicity the negligible terms due to the coronagraph are not considered
\begin{align}
 \mathcal{I} &= \abs*{\overline{\mathcal{P}_m}}^2 & 0\, \text{Order,} \nonumber \\
 &+ \abs*{\overline{\mathcal{P}_m \phi}}^2 + \abs*{\overline{\mathcal{P}_m \psi}}^2 + \abs*{\overline{\mathcal{P}_m \omega}}^2 & 2^{nd}\, \text{Order,} \nonumber \\
 &+ 2\text{Re} \left[ \left(\overline{\mathcal{P}_m \phi}\right)^{\ast}\overline{\mathcal{P}_m \psi} \right]
  + 2\text{Re} \left[ \left(\overline{\mathcal{P}_m \phi}\right)^{\ast}\overline{\mathcal{P}_m \omega} \right]
  + 2\text{Re} \left[ \left(\overline{\mathcal{P}_m \psi}\right)^{\ast}\overline{\mathcal{P}_m \omega} \right]& \dots
 \label{eq:EFCintensMode}
\end{align}
Let's write $I_{\omega^{+/-}}^{\psi^+/-}$ the intensity corresponding to a positive $\omega^{+}$ or negative $\omega^{-}$ mode and a positive $\psi^{+}$ or negative $\psi^{-}$ probe.
When subtracting from each other the intensities of the positive and negative probe only the term proportional to the probe remains and similarly for the probe. Therefore the only term remaining is the one proportional to both the probe and the mode which gives,
\begin{equation}
\frac{1}{2}\left(\frac{1}{2}
(I_{\omega^{+}}^{\psi^+}-I_{\omega^{+}}^{\psi^-})-\frac{1}{2}(I_{\omega^{-}}^{\psi^+}-I_{\omega^{-}}^{\psi^-})
\right) = 2\text{Re} \left[ \overline{\mathcal{P}_m \psi}^{\ast}\overline{\mathcal{P}_m \omega} \right]
\end{equation}
This is exactly equivalent to \eqref{eq:EFCdeltaI} when replacing the aberration $\phi$ by the mode $\omega$ which exactly what is wanted.

\relChapter{EFC Results} 

\relSection{Simulations}
\label{sec:EFCAppSimu}

\relSubsection{Effect of the Modes}
\citeauthor{verinaudFREEEFC} \citep{verinaudFREEEFC} is using Fourier modes localized in the area of interest and spaced by $1 (\lambda/d)$. This base was thought to include all the frequencies for modelling the speckles inside the area of interest. The author however emphasized that this is not correct\marginpar{Even if right now the best results on SPHERE were achieved with these modes.}. Pixels are usually centred on non-integer frequencies \marginpar{Frequencies of the phase in the pupil plane are coordinates of the corresponding pixels in the detector in $\lambda/d$ are identical.} and all integer frequencies are used when projecting the corresponding phase onto the discrete array of actuators. This is an effect of the discretization of the frequencies. The solution suggested by the author was to not restraint the choice of the modes amongst the integer frequencies. The consequence is that one can actually use as many modes as wanted for any size of the area. However the number of modes is limited by the time it requires for measuring the Interaction Matrix. The problem is therefore to find a way to select the most efficient modes. This is the reason why the author suggested the use of the singular modes as described in \autoref{app:EFCAppModesSing}.

This test uses the same area and probes as in \autoref{sec:EFCsimuIt} and \autoref{fig:EFCsimuMedRect}.

First if one uses $132$ singular modes which is the same number of modes as before the results are already $5$ times better but using $300$ singular modes brings the area intensity to almost zero. The contrast improvement after two iterations is presented in \autoref{tab:EFCsimuMedRectsing}. These contrasts could not be reached in the existing simulation by \citeauthor{verinaudFREEEFC} because of the lack of modes.

\begin{table}
\myfloatalign
\begin{tabularx}{\textwidth}{Xcc}
\toprule
\tableheadline{Modes} & \tableheadline{RMS} & \tableheadline{Intensity} \\ \midrule
$132$ Singular ($5\times 10$ Rectangle) & $757$ &  $1063$ \\
$300$ Singular ($5\times 10$ Rectangle) & $3.8\, 10^4$ &  $3.8\, 10^4$ \\
\bottomrule
\end{tabularx}
\caption[Simulated EFC contrast improvement on a rectangle with singular modes.]{Contrast improvement with two iterations of EFC when using $132$ respectively $300$ singular modes. The area of interest is a $5\times 10 (\lambda/d)$ rectangle $10(\lambda/d)$ away from the center. This table should be compared with \autoref{tab:EFCsimuMedRect}. The gain is computed in term of the Root-Mean-Square value and the total intensity in the dark hole area. The values are the result of the mean over ten simulations.}  
\label{tab:EFCsimuMedRectsing}
\end{table}

\begin{figure}[h]
	\centering	
   	\includegraphics[width=0.66\linewidth]{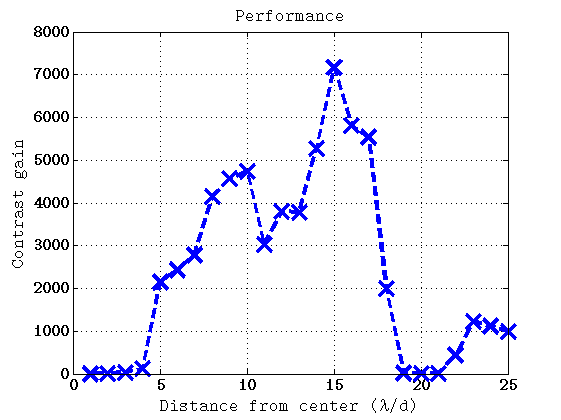}
   	\caption[Effect of the distance to the center on simulated EFC performance.]{Effect of the distance to the center of the dark hole on the contrast improvement after two iterations of Electric Field Conjugation. The gain is computed using the Root-Mean-Square of the speckles in the area of interest. The range is from $1$ to $25 (\lambda/d)$ with a step of $1 (\lambda/d)$.The algorithm is applied on a small $3\times 3 (\lambda/d)$ rectangle with $110$ pixels using $50$ singular modes. The probes are defined slightly bigger that the dark hole. The Interaction Matrix is inverted using a Pseudo-Inverse.}
   	\label{fig:EFCdistEffect} 
\end{figure}

\relSubsection{Effect of the Distance}
It is also interesting to check in simulation the ability of the Deformable Mirror to correct at certain distances of the center. Indeed the actuators grid doesn't sample well some frequencies and the performance can be diminished. Besides it emphasizes the limit distance of correction of $20 (\lambda/d)$ for a $40\times 40$ actuator grid. However one can see that after this limit the gain is still slightly positive. The author thinks that this correction is made thanks to folded frequencies.\marginpar{However the author didn't check yet if the frequencies were folded due to the discrete actuator grid or due to the sampling of the simulated pupil.} The results are shown in \autoref{fig:EFCdistEffect}.

\begin{figure}[h]
	\centering	
   	\includegraphics[width=\linewidth]{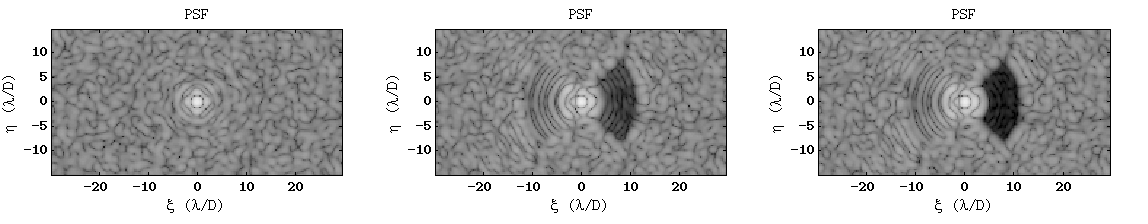}
   	\caption[Simulated EFC iterations on a quarter ring.]{Dark hole area for two iterations of Electric Field Conjugation on a quarter ring with a width $6 (\lambda/d)$ and a mean radius of $8 (\lambda/d)$. This corresponds to an area of $841$ pixels and $300$ singular modes were used. Four probes is used in this case with two in the vertical and two in the horizontal in order to cover the whole ring. The contrast improvement expressed in Root-Mean-Square is here $3.6\,10^3$.}
   	\label{fig:EFC90degArc} 
\end{figure}

\relSubsection{Effect of the Area Size}
The last simulation is meant to study the combined effect of the size of the area and the number of modes required for achieving a good darkening. The area is defined as a portion of ring with a width $6 (\lambda/d)$ and a mean radius of $8 (\lambda/d)$ in order to isolate the effect of the distance to the center. The angle defining the portion is proportional to the number of pixels. The dark hole for an area of $90^{o}$ is given in \autoref{fig:EFC90degArc}.

\begin{figure}[h]
	\centering	
   	\includegraphics[width=0.66\linewidth]{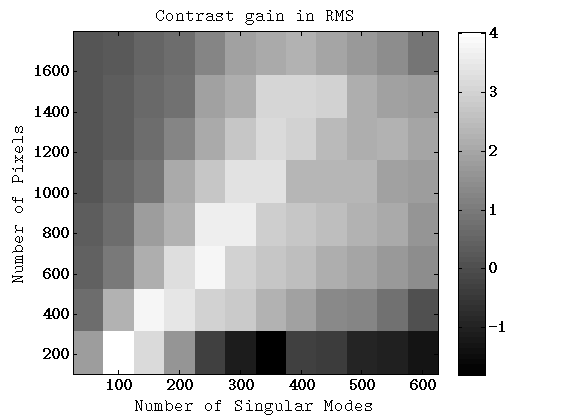}
   	\caption[Effect of the size of the area and the number of modes.]{Combined effect of the number of pixels and the number of singular modes on the contrast improvement. The gain is expressed as the Root-Mean-Square of the intensity in the dark hole after two iterations of Electric Field Conjugation. The area is a portion of a ring defined by an angle going from $\frac{\pi}{8}$ to $\pi$ by step of $\frac{\pi}{8}$. The number of singular modes goes from $50$ to $600$ by step of $50$. Like in \autoref{fig:EFC90degArc} four rectangular probes are used to cover the whole ring.}
   	\label{fig:EFCsizeEffect} 
\end{figure}

\autoref{fig:EFCsizeEffect} shows the contrast improvement depending on the number of pixels of the area and the number of singular modes used for iterating. One can extrapolate a linear law stating that a single mode is roughly enough for correcting two pixels. However this is simulation and it might not be completely applicable for a real case.\marginpar{And actually it is not\dots See \autoref{cha:EFCresDiscu}}

\relSection{SPHERE}
This appendix section contains the results of all the Electric Field Conjugation tests performed on SPHERE during the period of the $24-29$ of june 2013.

\begin{figure}[bth]
\myfloatalign
\subfloat[Fourier mode.]
{\label{fig:EFCSPHEREprobeMode-a} \includegraphics[width=.45\linewidth]{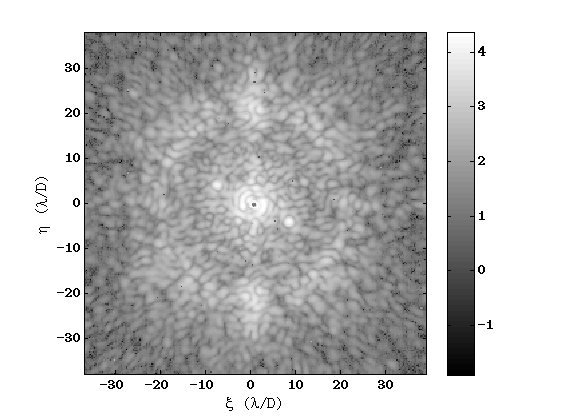}} \quad
\subfloat[Rectangular Probe.]
{\label{fig:EFCSPHEREprobeMode-b} \includegraphics[width=.45\linewidth]{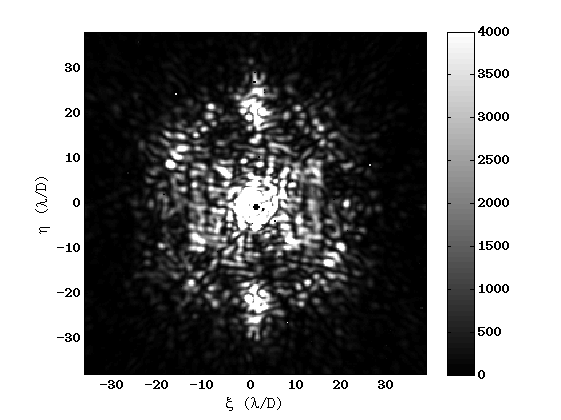}}
\caption[Example of Fourier mode and probe for EFC on SPHERE.]{Examples of SPHERE images with a high amplitude Fourier mode with coordinates $[8(\lambda/d),-4(\lambda/d)]$ and another one with a $10\times 20 (\lambda/d)$ rectangular probe $10(\lambda/d)$ away from the center. \autoref{fig:EFCSPHEREprobeMode-a} is using a logarithmic scale while \autoref{fig:EFCSPHEREprobeMode-b} is in linear scale with a cut at $4000$ data number for emphasizing the probe area.}
\label{fig:EFCSPHEREprobeMode}
\end{figure}

\begin{figure}[bth]
\myfloatalign
\subfloat[ 132 Fourier (Measured IM).]
{\label{fig:IMmedRect-a} \includegraphics[width=.45\linewidth]{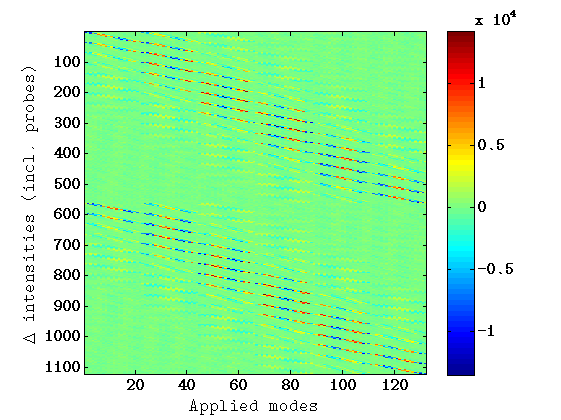}} \quad
\subfloat[109 Singular (Measured IM).]
{\label{fig:IMmedRect-b} \includegraphics[width=.45\linewidth]{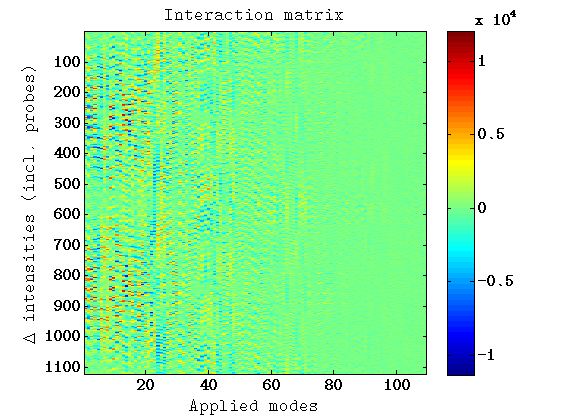}} \quad
\subfloat[109 Singular (Synthetic IM).]
{\label{fig:IMmedRect-c} \includegraphics[width=.45\linewidth]{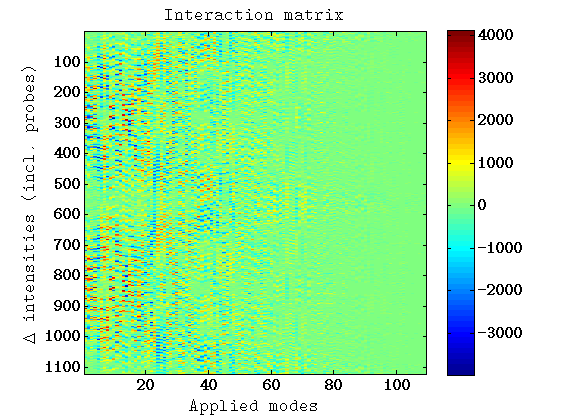}}
\caption[Electric Field Conjugation Interaction Matrices for the rectangle.]{Electric Field Conjugation Interaction Matrices for the rectangle with 132 Fourier modes or 109 singular modes. One can acknowledge the similarity between measured and the synthetic Interaction Matrix.}
\label{fig:IMmedRect}
\end{figure}

\begin{figure}[bth]
\myfloatalign
\subfloat[ 132 Fourier (Measured IM).]
{\label{fig:EFCitMedRect-a} \includegraphics[width=.45\linewidth]{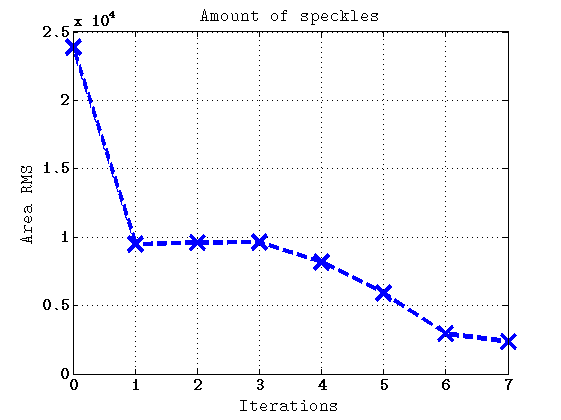}} \quad
\subfloat[109 Singular (Measured IM).]
{\label{fig:EFCitMedRect-b} \includegraphics[width=.45\linewidth]{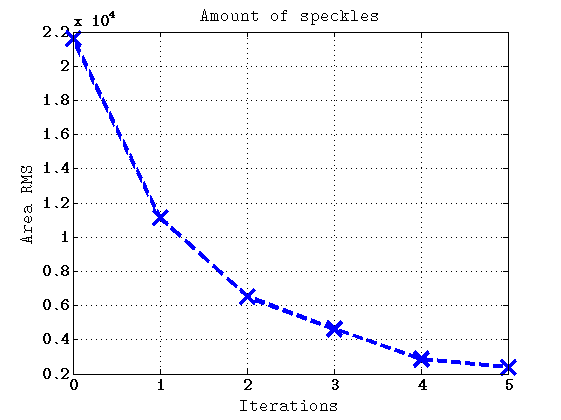}} \quad
\subfloat[109 Singular (Synthetic IM).]
{\label{fig:EFCitMedRect-c} \includegraphics[width=.45\linewidth]{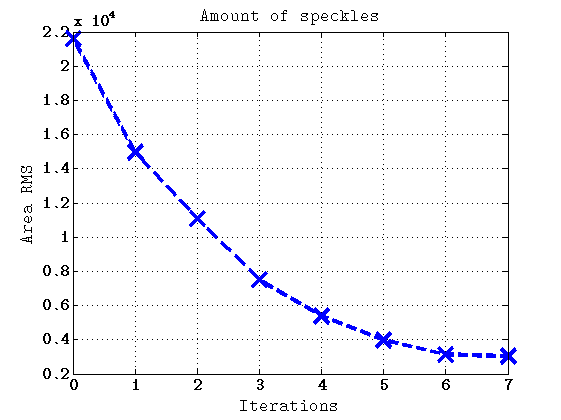}}
\caption[Speckles RMS during EFC iterations on the rectangle.]{Evolution of the speckles Root-Mean-Square in the rectangle dark hole. The strange curve of \autoref{fig:EFCitMedRect-a} is due to the fact the a re-centring of the image has been performed at the fourth iteration greatly improving the convergence.}
\label{fig:EFCitMedRect}
\end{figure}

\begin{figure}[bth]
\myfloatalign
\subfloat[132 Fourier modes (Measured IM).]
{\label{fig:EFCmedRectImages-a} \includegraphics[width=.45\linewidth]{imInitMedRectChris132} \quad
\includegraphics[width=.45\linewidth]{imCorrMedRectChris132}} \\
\subfloat[109 Singular modes (Measured IM).]
{\label{fig:EFCmedRectImages-b} \includegraphics[width=.45\linewidth]{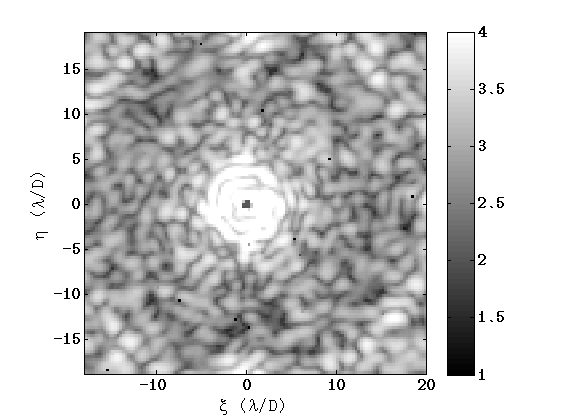} \quad
 \includegraphics[width=.45\linewidth]{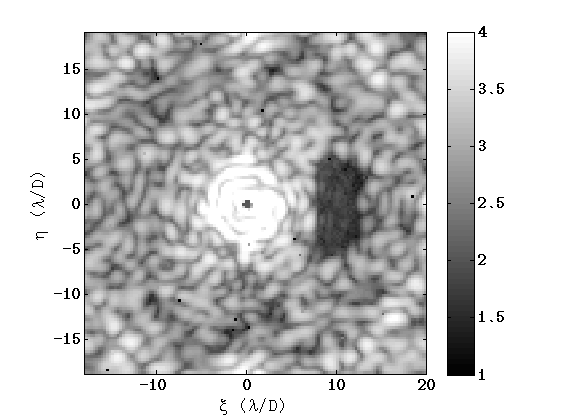}} \\
\subfloat[109 Singular modes (Synthetic IM).]
{\label{fig:EFCmedRectImages-c} \includegraphics[width=.45\linewidth]{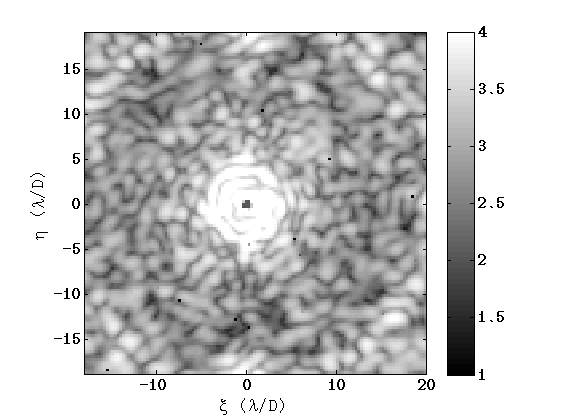} \quad
 \includegraphics[width=.45\linewidth]{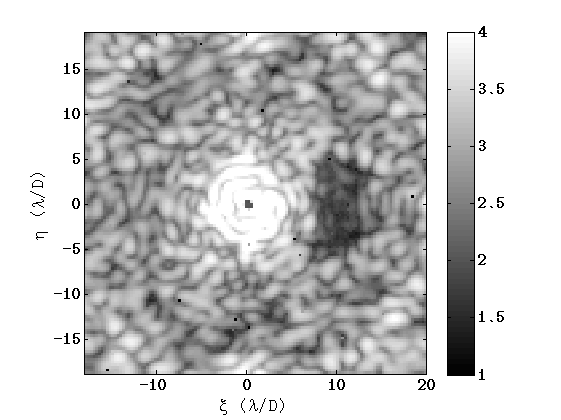}}
\caption[Rectangle dark holes before and after.]{Rectangle dark hole after Electric Field Conjugation iterations. On the left are the images before and on the right after the correction has been been applied.}
\label{fig:EFCmedRectImages}
\end{figure}

\relSubsection{Rectangle}
The first test was done on a $5\times 10 (\lambda/d)$ rectangle at a distance of $10 (\lambda/d)$ to the center. The first iterations were using $132$ Fourier modes  and a measured Interaction Matrix. Then $109$ singular modes with a measured Interaction Matrix and then with an synthetic one. The interaction matrices can be found in \autoref{fig:IMmedRect}. The convergence curves are plotted in \autoref{fig:EFCitMedRect}. The uncorrected and corrected images are shown in \autoref{fig:EFCmedRectImages}.

\begin{figure}[bth]
\myfloatalign
\subfloat[125 Singular modes (Measured IM).]
{\label{fig:IMarc-a} \includegraphics[width=.45\linewidth]{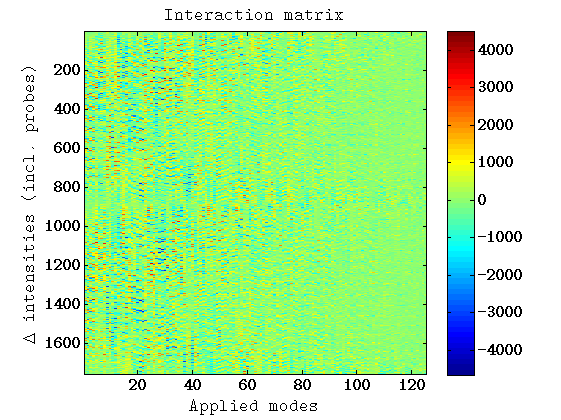}} \quad
\subfloat[125 Singular modes (Synthetic IM).]
{\label{fig:IMarc-b} \includegraphics[width=.45\linewidth]{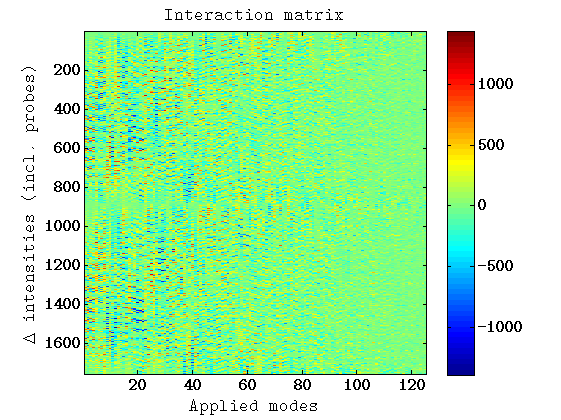}}
\caption[Electric Field Conjugation Interaction Matrices for the quarter ring.]{Electric Field Conjugation Interaction Matrices for the quarter ring with 125 singular modes.}
\label{fig:IMarc}
\end{figure}

\begin{figure}[bth]
\myfloatalign
\subfloat[125 Singular modes (Measured IM).]
{\label{fig:EFCitArc-a} \includegraphics[width=.45\linewidth]{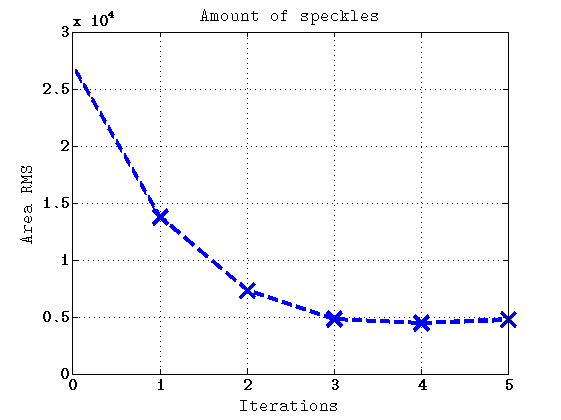}} \quad
\subfloat[125 Singular modes (Synthetic IM).]
{\label{fig:EFCitArc-b} \includegraphics[width=.45\linewidth]{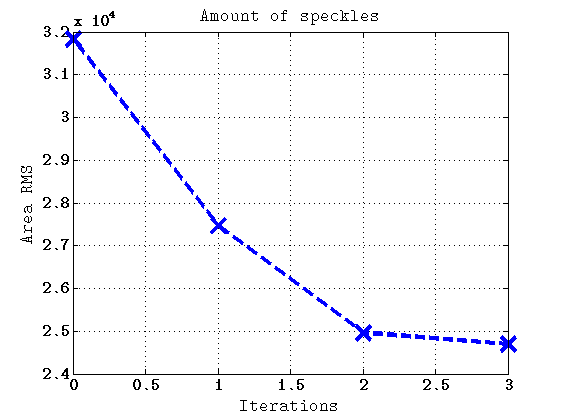}}
\caption[Speckles RMS during EFC iterations on the quarter ring.]{Evolution of the speckles Root-Mean-Square in the quarter ring dark hole. In \autoref{fig:EFCitArc-a} the amplitude of the probe was increased at the fourth iteration and then decreased a lot at the fifth iteration but the image was blurred.}
\label{fig:EFCitArc}
\end{figure}

\begin{figure}[bth]
\myfloatalign
\subfloat[125 Singular modes (Measured IM).]
{\label{fig:EFCarcImages-a} \includegraphics[width=.45\linewidth]{imInitArcSing125} \quad
 \includegraphics[width=.45\linewidth]{imCorrArcSing125}} \\
\subfloat[125 Singular modes (Synthetic IM).]
{\label{fig:EFCarcImages-b} \includegraphics[width=.45\linewidth]{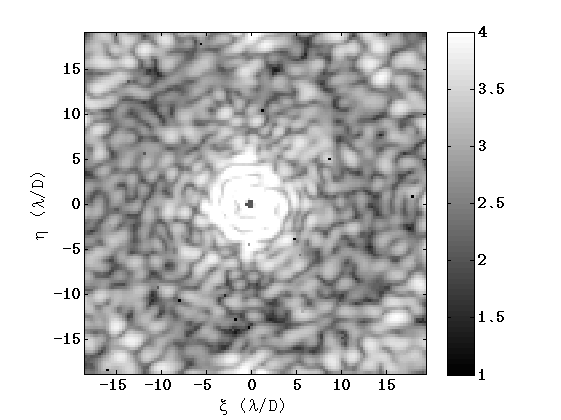} \quad
\includegraphics[width=.45\linewidth]{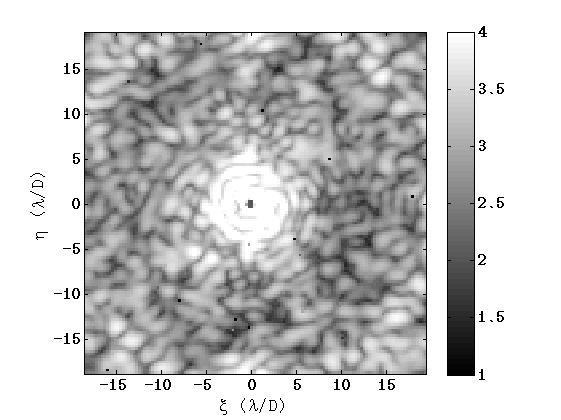}}
\caption[Quarter ring dark holes before and after.]{Quarter ring dark hole after Electric Field Conjugation iterations. On the left are the images before and on the right after the correction has been been applied.}
\label{fig:EFCarcImages}
\end{figure}

\relSubsection{Arc}
The second test was done on a quarter of a ring with a width of $5 (\lambda/d)$ and a mean radius of $10 (\lambda/d)$. $125$ singular modes are used firstly with a measured Interaction Matrix and then with a synthetic one. The interaction matrices can be found in \autoref{fig:IMarc}. The convergence curves are plotted in \autoref{fig:EFCitArc}. The uncorrected and corrected images are shown in \autoref{fig:EFCarcImages}.

\begin{figure}[bth]
\myfloatalign
\subfloat[306 Fourier Modes.]
{\includegraphics[width=.45\linewidth]{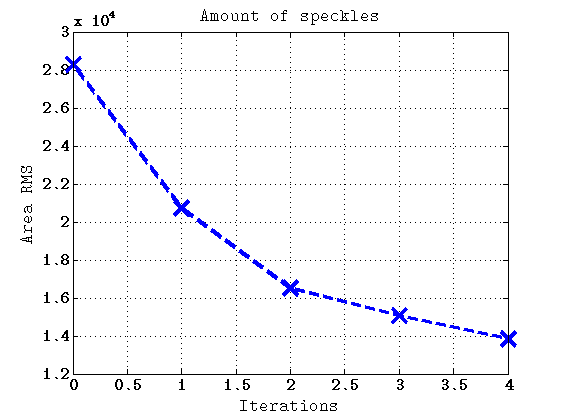}} 
\caption[Speckles RMS during EFC iterations on the big rectangle.]{Evolution of the speckles Root-Mean-Square in the big rectangle dark hole. }
\label{fig:EFCitBigRect}
\end{figure}

\begin{figure}[bth]
\myfloatalign
\subfloat[306 Fourier modes (Synthetic IM).]
{ \includegraphics[width=.45\linewidth]{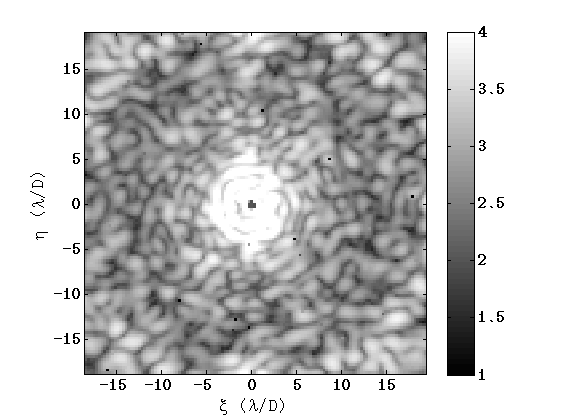} \quad
 \includegraphics[width=.45\linewidth]{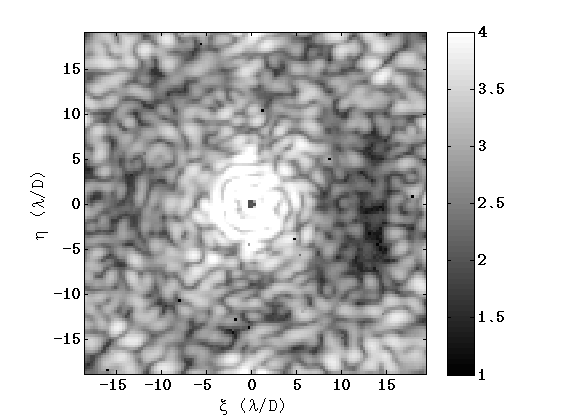}}
\caption[Big rectangle dark holes before and after.]{Big rectangle dark hole after Electric Field Conjugation iterations. On the left are the images before and on the right after the correction has been been applied.}
\label{fig:EFCbigRectImages}
\end{figure}

\relSubsection{Big Rectangle}
The third and last test was a $8\times 16 (\lambda/d)$ rectangle at a distance of $11 (\lambda/d)$ of the center. A synthetic Interaction Matrix with $306$ Fourier modes was used. The reconstruction matrix was built by filtering some singular modes but the number couldn't be found. The convergence curves are plotted in \autoref{fig:EFCitBigRect}. The uncorrected and corrected images are shown in \autoref{fig:EFCbigRectImages}.

\cleardoublepage

\label{app:bibliography} 


\addcontentsline{toc}{chapter}{\tocEntry{\bibname}}

\bibliographystyle{plainnat}

\bibliography{D:/latex/EFC/bibliography/bib_EFC} 



\cleardoublepage

\pagestyle{empty}

\hfill

\vfill

\pdfbookmark[0]{Colophon}{colophon}

\section*{Colophon}

This document was typeset using the typographical look-and-feel \texttt{classicthesis} developed by Andr\'e Miede. The style was inspired by Robert Bringhurst's seminal book on typography ``\emph{The Elements of Typographic Style}''. \texttt{classicthesis} is available for both \LaTeX\ and \mLyX: 

\begin{center}
\url{http://code.google.com/p/classicthesis/}
\end{center}

\noindent Happy users of \texttt{classicthesis} usually send a real postcard to the author, a collection of postcards received so far is featured here: 

\begin{center}
\url{http://postcards.miede.de/}
\end{center}
 
\bigskip

\noindent\finalVersionString 

\cleardoublepage

\refstepcounter{dummy}
\pdfbookmark[0]{Declaration}{declaration} 

\chapter*{Declaration} 

\thispagestyle{empty}

I declare that I have developed and writt
en the enclosed Master Thesis
completely
by
myself, and have not used sources or means without declaration in the text. Any
thoughts from others or literal quotations are clearly marked. The Master Thesis was
not used in the same or in a similar version to achieve an academic grading or is
being
published elsewhere.
\bigskip
 
\noindent\textit{\myLocation, \myTime}

\smallskip

\begin{flushright}
\begin{tabular}{m{8cm}}
\\ \hline
\centering\myName, \today \\
\end{tabular}
\end{flushright}


\end{document}